\documentclass[11pt]{article}

\usepackage[preprint]{acl}

\usepackage{times}
\usepackage{latexsym}

\usepackage[T1]{fontenc}

\usepackage[utf8]{inputenc}

\usepackage{microtype}

\usepackage{inconsolata}

\usepackage{graphicx}

\usepackage{algorithm}
\usepackage{algorithmic}
\usepackage{tabularx}
\usepackage{booktabs}
\usepackage{pifont}
\usepackage{xspace}
\usepackage[most]{tcolorbox}
\usepackage{amsmath}
\usepackage{amssymb}
\usepackage{subcaption}
\usepackage{marvosym}
\usepackage{fontawesome}
\usepackage{tikz}
\usetikzlibrary{arrows.meta, positioning, shapes.geometric}

\definecolor{accent}{HTML}{6B1D1D}
\definecolor{pteal}{HTML}{3D7068}
\definecolor{pamber}{HTML}{8D6A2B}
\definecolor{pwine}{HTML}{8A3B3B}

\newcommand{\cmark}{\ding{51}}  
\newcommand{\xmark}{\ding{55}}  
\newcommand{\agentframe}{\textsc{MoRE}\xspace}
\newcommand{\envframe}{\textsc{Social-Evol}\xspace}
\newcommand{\HP}{\mathrm{HP}}
\newcommand{\bound}[3]{\min(\max(#1, #2), #3)}

\title{Why Are We Moral? \\ An LLM-based Agent Simulation Approach to Study Moral Evolution
  \thanks{Accepted at ACL 2026 Main Conference.}}

\author{
Zhou Ziheng$^{1}$\textsuperscript{* \Letter},
Huacong Tang$^{1}$\textsuperscript{*},
Mingjie Bi$^{2}$\textsuperscript{*},
Yipeng Kang$^{2}$,
Wanying He$^{2}$, \\
\bf
Fang Sun$^{1}$,
Yizhou Sun$^{1}$,
Ying Nian Wu$^{1}$,
Demetri Terzopoulos$^{1}$,
Fangwei Zhong$^{3,2}$\textsuperscript{\Letter}
\\[0.3em]
$^{1}$University of California, Los Angeles  \quad
$^{2}$Beijing Institute for General Artificial Intelligence \quad \\
$^{3}$Beijing Normal University 
\\
\textsuperscript{*}Equal contribution.\quad
\textsuperscript{\Letter}Corresponding authors: josephziheng@ucla.edu, fangweizhong@bnu.edu.cn
\\[0.5em]
\href{https://MoralAgentSim.github.io}{\faGlobe~Project Page} \quad
\href{https://github.com/MoralAgentSim/social-evol-sim}{\faGithub~Code}
}

\begin{document}
\maketitle
\begin{abstract}
The evolution of morality presents a puzzle: natural selection should favor self-interest, yet humans developed moral systems promoting altruism. Traditional approaches must abstract away cognitive processes, leaving open how cognitive factors shape moral evolution. We introduce an LLM-based agent simulation framework that brings cognitive realism to this question: agents with varying moral dispositions perceive, remember, reason, and decide in a simulated prehistoric hunter-gatherer society. This enables us to manipulate factors that traditional models cannot represent---such as moral type observability and communication bandwidth---and to discover emergent cognitive mechanisms from agent interactions. Across 20 runs spanning four settings, we find that cooperation and mutual help are the central driver of evolutionary survival, with universal and reciprocal morality exhibiting the most stable outcomes across conditions while selfishness is strongly disfavoured. Beyond cooperation itself, we further identify cognition as a central mediator---most clearly through a \emph{cost of moral judgment} that shifts the winning moral type across settings, with a \emph{self-purging effect} among selfish agents as an additional cognitive pattern. We validate robustness across multiple LLM backbones, architecture ablations, and prompt sensitivity analyses. This work establishes LLM-based simulation as a powerful new paradigm to complement traditional research in evolutionary biology and anthropology, opening new avenues for investigating the complexities of moral and social evolution.
\end{abstract}

\section{Introduction}
\label{sec:introduction}
The emergence and evolution of morality represents one of the most enduring puzzles in evolutionary biology and social sciences~\citep{haidt2007moral, greene2013moral}. From an evolutionary standpoint, natural selection should favor individuals who maximize their reproductive success, often through selfish behaviors that increase resource acquisition at others' expense~\citep{dawkins1976selfish}. Yet, humans and some other species have evolved complex moral systems that frequently promote cooperation, altruism, and other prosocial behaviors that can seemingly contradict individual fitness maximization~\citep{tomasello2016natural}. This apparent contradiction presents a profound scientific question: Under what conditions does morality provide an evolutionary advantage?

Prior research has approached this question through evolutionary game theory~\citep{nowak2006five, axelrod1981evolution}, anthropological fieldwork on moral universals and cultural variation~\citep{henrich2015secret, curry2019good}, and biological mechanisms including kin selection~\citep{hamilton1964genetical}, reciprocal altruism~\citep{trivers1971evolution}, and group selection~\citep{wilson2007rethinking}. Moral frameworks have further identified patterns such as the expanding circle of moral concern~\citep{singer1981expanding} and fundamental moral dimensions~\citep{haidt2007moral}. While these approaches have yielded valuable insights, they share two simplifications (Fig.~\ref{fig:paradigm}, left): \emph{agents} are reduced to fixed strategies or payoff entries, and the \emph{world} is reduced to an abstract game. This prevents investigation of how cognitive factors---memory of past interactions, the ability to identify others' moral dispositions, communication bandwidth, reasoning under uncertainty---and richer environmental pressures jointly shape moral evolution.

Recent advances in LLMs present a novel methodological opportunity to address these limitations. LLM-based agent simulations can model entities with sophisticated cognitive architectures---including values, memory, perception, reasoning, and social dynamics---that generate emergent, complex behaviors~\citep{park2023generative, horton2023large}. This simulation paradigm allows us to observe interactions between moral cognition, behavior, and evolutionary outcomes under controlled conditions while providing rich, realistic psychological details that surpass traditional agent-based models~\citep{aher2023using}.
Nevertheless, the existing methods barely support research on morality evolution.

To address the gaps, our research advances the study of human evolution through three primary contributions:
1) Methodological: We propose a new research paradigm (Fig.~\ref{fig:paradigm}) that relaxes both simplifications of prior work---agents become cognitive and the world becomes a prehistoric hunter-gatherer ecology---turning cognition and environment into first-class experimental variables;
2) Programmatic: We release \agentframe, an extensible agent framework, and \envframe, a versatile environment platform. Together, they enable multifaceted inquiry into social evolution—from norm formation to inter-group conflict—through both long-term and scenario-based simulation;
3) Empirical: We conduct experiments across multiple settings showing that cooperation is consistently favoured by selection, with universal and reciprocal morality exhibiting the most stable survival. We further identify cognition itself as a central mediator---most clearly through the cost of moral judgment, which shifts the winning moral type across settings, with a self-purging effect among selfish agents as an additional cognitive pattern---both difficult to study with traditional approaches. We validate robustness across multiple LLM backbones, architecture ablations, and prompt sensitivity analyses.

\begin{figure}[t]
\centering
\begin{tikzpicture}[font=\scriptsize]
\node[font=\scriptsize\bfseries, color=gray!70, anchor=base] at (-1.75, 3.0) {TRADITIONAL};
\node[font=\scriptsize\bfseries, color=accent, anchor=base]  at ( 1.75, 3.0) {OUR PARADIGM};

\node[font=\tiny\bfseries, color=gray!65, rotate=90] at (-3.3, 1.75) {AGENTS};
\node[font=\tiny\bfseries, color=gray!65, rotate=90] at (-3.3, -0.25) {ENVIRON.};

\draw[gray!40, dashed, line width=0.3pt] (0, 2.85) -- (0, -1.3);
\draw[gray!40, dashed, line width=0.3pt] (-3.05, 0.35) -- (3.05, 0.35);

\node[circle, draw=gray!60, fill=gray!10, minimum size=5mm, inner sep=0pt, font=\tiny] at (-2.2, 1.75) {A};
\node[circle, draw=gray!60, fill=gray!10, minimum size=5mm, inner sep=0pt, font=\tiny] at (-1.3, 1.75) {B};
\node[font=\tiny\itshape, color=gray!70] at (-1.75, 0.6) {fixed strategies};

\node[circle, draw=accent, line width=0.7pt, minimum size=1.3cm, inner sep=0pt] at (1.75, 1.75) {};
\node[font=\tiny\bfseries, color=accent, align=center, inner sep=0pt] at (1.75, 1.75) {Moral\\Value};
\node[font=\tiny, anchor=south] at (1.75, 2.47) {Perceive};
\node[font=\tiny, anchor=north] at (1.75, 1.03) {Reflect};
\node[font=\tiny, anchor=east] at (0.95, 1.75) {Plan};
\node[font=\tiny, anchor=west] at (2.55, 1.75) {Reason};
\node[font=\tiny\itshape, color=accent] at (1.75, 0.6) {cognitive};

\draw[gray!60, line width=0.3pt] (-2.15, 0.05) rectangle (-1.35, -0.7);
\draw[gray!60, line width=0.3pt] (-1.75, 0.05) -- (-1.75, -0.7);
\draw[gray!60, line width=0.3pt] (-2.15, -0.325) -- (-1.35, -0.325);
\node[font=\tiny] at (-1.95, -0.13) {C,C};
\node[font=\tiny] at (-1.55, -0.13) {C,D};
\node[font=\tiny] at (-1.95, -0.52) {D,C};
\node[font=\tiny] at (-1.55, -0.52) {D,D};
\node[font=\tiny\itshape, color=gray!70] at (-1.75, -1.08) {$2{\times}2$ payoff};

\draw[color=pwine!55, line width=0.4pt] (0.35, -0.6) -- (3.0, -0.6);
\draw[color=pteal, line width=0.4pt, fill=pteal!30] (0.55, -0.6) -- (0.45, -0.3) -- (0.65, -0.3) -- cycle;
\draw[color=pamber, line width=0.4pt, fill=pamber!25] (0.95, -0.43) ellipse (0.11cm and 0.06cm);
\filldraw[color=pteal]  (1.25, -0.45) circle (0.06cm);
\filldraw[color=pamber] (1.40, -0.45) circle (0.06cm);
\filldraw[color=pwine]  (1.55, -0.45) circle (0.06cm);
\draw[color=accent, line width=0.35pt] (1.95, -0.45) circle (0.05cm);
\draw[color=accent, line width=0.35pt] (2.15, -0.45) circle (0.05cm);
\draw[->, color=accent, line width=0.3pt] (2.00, -0.45) -- (2.10, -0.45);
\filldraw[color=accent] (2.70, -0.37) circle (0.055cm);
\filldraw[color=accent] (2.70, -0.51) circle (0.035cm);
\node[font=\fontsize{4.5}{5}\selectfont, color=gray!70, anchor=north] at (0.55, -0.67) {plants};
\node[font=\fontsize{4.5}{5}\selectfont, color=gray!70, anchor=north] at (0.95, -0.67) {prey};
\node[font=\fontsize{4.5}{5}\selectfont, color=gray!70, anchor=north] at (1.40, -0.67) {peers};
\node[font=\fontsize{4.5}{5}\selectfont, color=gray!70, anchor=north] at (2.05, -0.67) {share};
\node[font=\fontsize{4.5}{5}\selectfont, color=gray!70, anchor=north] at (2.70, -0.67) {reproduction};
\node[font=\tiny\itshape, color=accent] at (1.75, -1.08) {hunter-gatherer};

\end{tikzpicture}
\caption{A new research paradigm. Prior approaches reduce agents to fixed strategies and the world to a $2{\times}2$ payoff matrix (\emph{left}). We relax both simplifications (\emph{right}): agents become cognitive (Section~\ref{sec:agent_architecture}) and the world becomes a prehistoric hunter-gatherer ecology (Section~\ref{sec:framework})---making cognition \emph{and} environment first-class experimental variables.}
\label{fig:paradigm}
\end{figure}
\vspace{-4mm}

\begin{table*}[t]
\setlength{\tabcolsep}{4pt}
\centering
\caption{Comparison of Existing LLM-based Simulation Frameworks for Social Science.}
\small
\begin{tabularx}{\textwidth}{@{}p{3cm} c c c c c c@{}}
\toprule
\textbf{Environment} & \textbf{Morality} & \textbf{Memory} & \textbf{Team Formation} & \textbf{Competition} & \textbf{Evolution} & \textbf{Engine} \\
\midrule
\citet{park2023generative} & \xmark & Short/Long-term &  \xmark & \xmark & \xmark & LLM \\
\citet{li2023camel} & \xmark & Undefined & Multi-round Negotiation & \xmark & \xmark & LLM \\
\citet{horton2023large} & \xmark & Undefined & \xmark & Explicit & \xmark & LLM \\
\citet{aher2023using} & \xmark & Short-term & \xmark & \xmark & \xmark & LLM \\
\citet{wang2023humanoid} & \xmark & Short-term & \xmark & \xmark & \xmark & LLM \\
\citet{huang2024adasociety} & \xmark & Short-term & Multi-round Negotiation & Implicit & \cmark & LLM/RL \\
\citet{wang2024simulating} & \xmark & Short-term & \xmark & \xmark & \xmark & LLM \\
\citet{piao2025agentsociety} & \xmark &  Short/Long-term & Task-dependent & Task-dependent & \cmark & LLM \\
\citet{guan2024richelieu} & \xmark & Short/Long-term & Multi-round Negotiation & Explicit & \cmark & LLM \\
\citet{dai2024artificial} & \xmark & Short/Long-term & Negotiation-based & Explicit & \cmark & LLM \\
\midrule
Ours & \cmark & Short/Long-term & Multi-round Negotiation & Explicit & \cmark & LLM \\
\bottomrule
\end{tabularx}
\label{tab:comparison}
\end{table*}

\section{Related Work}
\label{sec:related_work}
\paragraph{Evolutionary Origins of Morality}

Evolutionary biologists have proposed various mechanisms to explain how altruistic traits might evolve. Theories of kin selection~\citep{hamilton1964genetical} and reciprocal altruism~\citep{trivers1971evolution} show how limited forms of cooperation could evolve among relatives and repeated interaction partners. Recently, cultural group selection theories~\citep{boyd2011cultural, henrich2015secret} explain how groups with stronger moral norms outcompeted others, leading to the genetic evolution of psychological predispositions supporting moral behavior.
Evolutionary game theory provided mathematical frameworks demonstrating how cooperation can evolve under some strategies similar like previously mentioned mechanisms by showing strategies can yield higher payoffs than pure selfishness under specific conditions. \citet{nowak2006five}'s ``five rules for the evolution of cooperation'' identifies key mechanisms: kin selection, direct reciprocity, indirect reciprocity, network reciprocity, and group selection. 

Though these works provide great insights and a mathematical foundation for cooperation and moral evolution, they highly abstract away the rich complexity of human cognition and cooperation using mathematical models, blocking a full view of moral evolution dynamics.

\paragraph{Moral Frameworks}

Moral Foundations Theory identifies five moral dimensions: care/harm, loyalty/betrayal, authority/subversion, and sanctity/degradation~\citep{Haidt2007MFT}. The Theory of Dyadic Morality~\citep{Gray2012TDM} emphasizes harm as the root of morality, while Morality-as-Cooperation theory~\citep{Curry2019MAC} identifies seven cooperative behaviors as essential: helping kin, helping group members, reciprocating, being brave, deferring to superiors, dividing resources fairly, and respecting others' property. Other theories ground morality in distinctions between rules and conventions~\citep{Turiel1983MDT}, social-relational models~\citep{Fiske1992RMT, Rai2011RRT}, specific moral emotions~\citep{Rozin1999CAD}, or an ethic of care~\citep{Gilligan1982EOC}.

While existing theories offer rich descriptions of morality's central traits, our initial investigation desires a framework that organizes these traits into scalable ``levels'' for a more systematic analysis. The Expanding Circle Theory~\citep{singer1981expanding} provides an ideal structure for this purpose. Its concept of a ``circle of concern'' that expands from the self, to kin, and finally to society offers a clear, tiered structure for defining an agent's moral level. Furthermore, this progression provides a natural way to integrate key concepts of other theories: care and loyalty are paramount within the kin circle, while fairness and reciprocity are essential for the stability of the broader group circle. This concentric model has cross-cultural relevance due to its deep resonance with philosophical traditions like Confucianism~\citep{fei1992from}, which also defines morality through the proper ordering of relational duties. Given its implementable structure and integrative power, we adopt the Expanding Circle as the primary theoretical foundation for our simulation.

\paragraph{LLM-Based Agent Simulation}

Recent advances in LLMs provide the methodological foundation for our approach to studying morality's evolution. LLM-based agent simulations can model entities with sophisticated cognitive architectures—including values, memory, perception, reasoning, and social dynamics—that generate emergent, complex behaviors~\citep{park2023generative, horton2023large}.
LLM agents allow for controlled experimentation with variables that would be impossible to manipulate in real-world settings; they provide transparent access to agents' decision-making processes; and they can simulate long timescales of social development~\citep{piao2025agentsociety}.

Prior work has demonstrated the versatility of this approach. \citet{park2023generative} created an interactive simulated small town where generative agents exhibited complex social behaviors, including relationship formation and collective problem-solving. \citet{horton2023large} applied LLM agents to economic simulations, finding that agents replicate known economic phenomena while providing unprecedented access to reasoning processes. \citet{aher2023using} validated that LLM agent simulations can reproduce results from human behavioral experiments, showing its potential as a complementary methodology in social science research.
However, existing environments lack the complete settings of agent morality, cooperative and competitive interactions, and evolution features, as demonstrated in Table~\ref{tab:comparison}.
One relevant recent work is ``Artificial Leviathan'', which explores social order in LLM agent societies, while they assume all agents are inherently selfish and focus on how social order emerges from this assumption~\citep{dai2024artificial}.

Therefore, by explicitly modeling agents with varying moral dispositions to better reflect the diversity of human moral psychology, our work represents the first systematic application of LLM-based simulations to investigate the evolution of morality in prehistoric human societies, where moral systems likely first emerged.

\begin{figure*}[t!]
    \centering
    \includegraphics[width=0.95\textwidth]{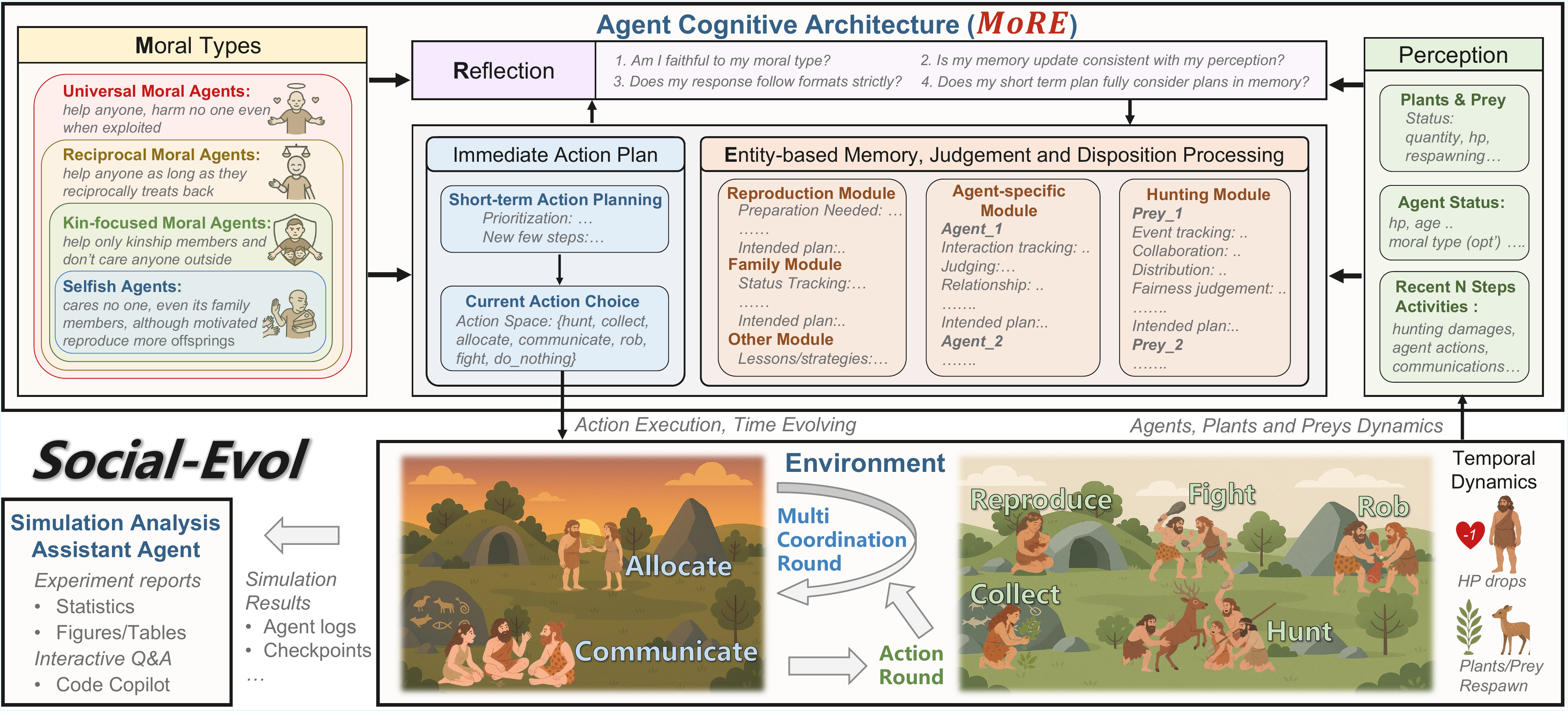}
    \caption{Overview of our simulation framework. The \textit{Environment} follows defined temporal dynamics and iterates between (multi-)coordination rounds and action rounds to interact with agents. The \textit{Agent Cognitive Architecture} includes a moral value module prescribing moral types based on expanding circles of concern, a perception module processing environmental information, and cognitive modules that update memory, form judgments, and generate action plans consistent with the agent's moral type. Before execution, agents perform self-reflection to verify consistency with observed facts and moral dispositions. The \textit{Simulation Analysis Assistant Agent} is a developed tool to automatically analyze simulation results.}
    \label{fig:method}
\end{figure*}

\section{Agent Cognitive Architecture (\agentframe)}
\label{sec:agent_architecture}
Our agent design \agentframe is a \textbf{m}orality-driven \textbf{e}ntity-oriented cognitive processing architecture with \textbf{r}eflection capability.

\paragraph{Agent Traits: Moral Types as Example}

To model evolutionary pressures, agents share a foundational value: maximizing survival and reproduction. It's noteworthy that beyond this, the framework supports researchers to customize agent traits to conduct various research problems in social science beyond this paper's scope. In this paper, we focus on moral simulation and implement moral dispositions based on the ``expanding circle'' concept~\citep{singer1981expanding}, defining a morality spectrum:

\textit{Self-focused agents} care exclusively about themselves. But in implementation, we notice a definitional challenge: purely selfish agents have no interest in reproduction, but defining them to care about offspring would conflate them with kin-focused agents. 
Therefore, we define them as investing in reproduction but providing no further offspring care, similar to
r-selected species (fish, amphibians, invertebrates) that maximize reproductive success through quantity rather than parental care~\citep{pianka1970r, stearns1992evolution, gross2005evolution, reznick2002r, trumbo2012patterns}

\textit{Kin-focused agents} extend moral concern to genetic relatives, providing care and resources to family members while treating non-kin instrumentally.

\textit{Group-focused agents} extend moral concern beyond kin to include non-related group members. However, defining a group remains a challenge: who constitutes the ``group" worthy of moral consideration? We define two variants: 1) \textit{Reciprocal group moral agents} extend care only to those who reciprocate similar moral concern, creating a self-consistent moral circle based on mutual recognition; 2) \textit{Universal group moral agents} extend care to all individuals regardless of their morality. It is expansive, and its non-violent orientation aligns with some intuitive conceptions of morality. However, this variant presents theoretical inconsistencies—violating fairness and reciprocity principles while benefiting agents who may undermine group welfare, such as selfish agents.

This framework yields four distinct moral types to enable systematic investigation of how different moral dispositions affect evolutionary outcomes. We acknowledge that this discrete categorization simplifies the continuous nature of moral concern in humans for experimental tractability. 
More importantly, this demonstrates that the framework can customize agent traits by defining the characteristics for focused social science research problems. We note that individual-level agent behaviors are driven by their assigned moral dispositions (i.e., instruction-following). The \textit{emergent} phenomena we study are the population-level dynamics---specifically, which moral type comes to dominate under various ecological constraints---outcomes that are not prescribed by the prompts.

\paragraph{Agent Cognition Framework}
Our simulation employs agent-based modeling powered by LLM to capture sophisticated cognitive and social dynamics. Agent cognition processes are as follows: 

\textit{Agent Initialization:} Agents are initialized with (1) a profile containing value characteristics aligned with designated moral type; (2) environmental rules governing the simulation; and (3) a knowledge handbook containing a common-sense understanding of environmental dynamics and causal relationships. It ensures agents begin with a comparable baseline understanding without artificially constraining their decision space. Importantly, no agent receives privileged strategic information, preventing methodological bias.

\textit{Perception Module:} This component processes current status information about plants, prey, and other agents (e.g., HP, age) along with recent activities up to a configurable number of past steps, mirroring human short-term memory.

\textit{Cognitive Processing System:} We designed an integrated entity-based system that maintains memory, makes judgments, and forms dispositions around entities like other people and hunting animals. This is in contrast to the event-based cognitive processing that records a log-book-like memory and decision history.  Our preliminary studies demonstrate that this method effectively prompts LLMs to consider relevant context and perform appropriate reasoning compared to simpler approaches. The entity-based structure provides a template for identifying important information and creating a narrative-like understanding.

\textit{Action Planning:} This module prioritizes updated memories and dispositional plans to formulate specific actions. This is crucial because the simulation environment may contain many entities toward which an agent might have multiple intended interactions.

\textit{Reflection Module:} This verification component ensures cognitive processing and action planning remain consistent with factual information and faithful to the agent's moral type, while producing properly formatted responses.

\section{\envframe Framework}
\label{sec:framework}
\envframe (Fig.~\ref{fig:method}) is a text-based prehistoric hunter-gatherer society with resource constraints, social dynamics, and environmental challenges, enabling research on evolutionary pressures that likely shaped early human morality.

\paragraph{Survival Mechanics}
Agents maintain health points (HP) that must remain above zero to sustain life. HP diminishes gradually over time, representing basic metabolic costs, and decreases substantially from injuries sustained during hunting or conflict. Agents replenish HP by collecting plants or hunting prey, with successful actions immediately increasing their HP. Additionally, agents face a maximum lifespan constraint that eventually results in death regardless of HP management, modeling natural senescence.

\paragraph{Resource Setting}
The environment contains two primary resource types: plants and animals. Plants represent low-risk, low-reward resources that agents can reliably collect. Animals offer high-risk, high-reward resources that yield more HP but present significant acquisition challenges. 
Plants are configured with initial quantity, capacity, nutrition value, and respawn delay.
Animals are configured with HP, physical ability, and respawn interval.
Those settings can be used to configure the abundance of the available resources.

\paragraph{Production and Reproduction Mechanics}\

\textit{Hunt, Collect:} Agents collect plants solely to gain HP without risks whenever plants are available. Hunting allows collaboration, which embodies realistic risk-reward tradeoffs: success probabilities and damage inflicted depend on agent-prey physical ability differentials. Failed hunts cause automatic counter-attacks (i.e., HP loss) to agents; successful attempts progressively reduce prey health until death. This difficulty incentivizes collaboration, improving success rates and distributing risks.

\textit{Reproduce:} Agents meeting age and HP thresholds can reproduce, though reproduction imposes significant metabolic costs. Offspring begin life with minimal HP, creating vulnerability that typically requires parental investment (food sharing) to ensure survival. Offspring inherit their parents' moral type deterministically; no mutation or cultural transmission mechanism is implemented in the current version, ensuring clean experimental control over inter-generational selection effects.

\paragraph{Social Interaction Mechanics}
The environment supports multiple forms of social interaction:

\textit{Allocate:} Agents can voluntarily transfer HP to others, enabling cooperative behaviors and care-based relationships.

\textit{Communicate:} Agents can exchange messages with specific individuals or groups simultaneously. This communication system enables coordination, information sharing, and relationship development.

\textit{Kinship and Type Recognition:} The environment explicitly provides each agent with the identities of its parents and offspring. When moral types are configured as observable, this information is included in each agent's perception. When moral types are hidden, agents must infer others' dispositions from observed behaviors and maintain these inferences in memory.


\textit{Rob, Fight:} Agents may attempt robbery to steal HP or fight to inflict damage without HP gain. Success probability and outcomes depend on relative physical abilities, with gains proportional to the aggressor’s strength. Such actions have slightly elevated HP costs due to their intensity. Unlike hunting, failed aggressive attempts do not trigger automatic retaliation; victims must initiate responses independently. This design enables agents to freely express their moral strategies without artificial constraints.

This environmental design creates a complex adaptive system where survival pressures, resource competition, cooperation opportunities, and communication capabilities interact to influence the differential success of varied moral dispositions. Detailed rule explanations and configuration settings appear in the supplement.

\paragraph{Simulation Operating Cycle} 
The simulation operates as a sequential process where agents and the environment interact in defined steps: 1) \textit{Environment Update}, where the simulation refreshes resource availability, agent status changes, and advances time; 2) \textit{Agent Perception}, where each agent receives observations about the current environmental state and recent activities; 3) \textit{Cognitive Processing}, where agents use their architecture to process perceptions, update memory, form judgments, and develop dispositional plans consistent with their moral type toward different entities (prey or other agents) or goals (reproduction etc.); 4) \textit{Action Planning}, where agents need to consider their dispositional plans and conditions to prioritize and make specific action plans for the next few steps. Note that each simulation step includes one or more coordination rounds (allocate/communicate) and one action round (reproduce/collect/fight/rob/hunt), and the environment update itself with defined temporal dynamics. The number of coordination rounds per step defines the \textit{social interaction cost}: our baseline allows two coordination rounds, while the high-cost condition restricts this to one, directly limiting agents' ability to negotiate and coordinate before acting; and 5) \textit{Consequence Resolution}, where outcomes of all actions are determined. This cycle repeats continually, enabling emergent complex social behaviors while maintaining tractable simulation parameters. The LLM serves as the cognitive engine for each agent, providing reasoning capabilities necessary to navigate moral dilemmas, form social strategies, and respond to environments in ways that reflect human-like cognitive processes.

\paragraph{Two Supported Game Modes}
Our environment allows two simulation game modes to study both long-term evolution and specific decision dynamics under targeted scenarios.
(1) \textit{Evolutionary Game:}
The evolutionary games are full simulations of agents’ behavior until all agents die or reach maximum steps. This game captures the long-term, emergent outcomes of moral evolution.
(2) \textit{Mini Games:} 
This game setting is designed to isolate a crucial step in the causal chain from morality to fitness: the moment an action is chosen. By placing agents in a specific, controlled scenario like moral dilemmas, we can clearly observe how different moral dispositions translate into distinct behaviors, thus illuminating a key mechanism that drives the broader dynamics seen in the full simulation. 

\paragraph{Simulation Data Analysis Assistant}

Throughout our project development, we identified a significant challenge in LLM-based agent simulations: interpreting the vast quantities of generated data. While having rich, multidimensional data offers tremendous analytical potential, extracting meaningful insights from this complexity requires specialized methodological approaches. To address this challenge, we developed a simulation analysis assistant agent that serves two critical functions. First, it automatically generates comprehensive statistical reports containing the key metrics visualized in our figures. Second, we implemented a series of function calls to enable an interactive Q\&A ability when user uses a readily available code copilot agent like Copilot or Cursor. It can allow researchers to interrogate specific agent behaviors, motivations, and decision processes (e.g., ``Why did Agent X perform action Y?''). This analytical tool has proven invaluable for understanding simulation dynamics and iteratively refining our agent design architecture. We provide detailed specifications of this system in the supplement.

\begin{figure}[t!]
    \centering
    \includegraphics[width=0.9\columnwidth]{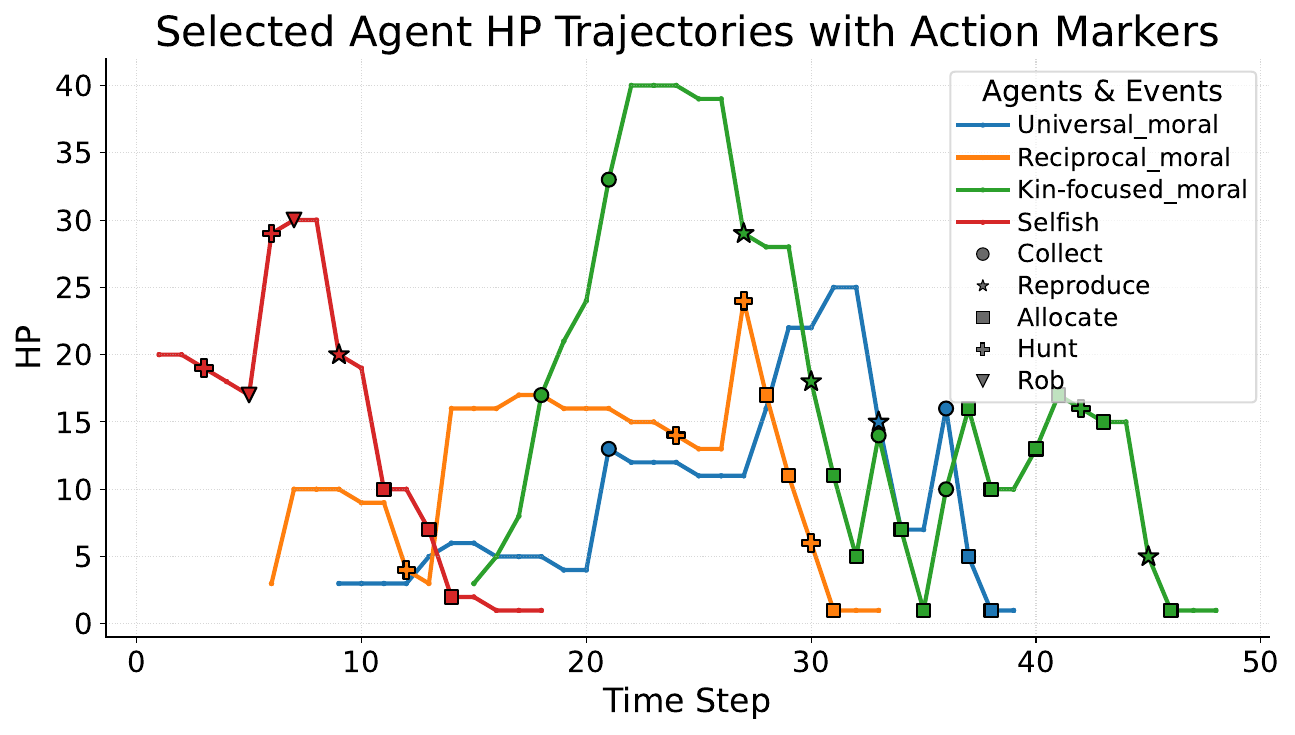}
    \vspace{0.3em}
    \includegraphics[width=0.9\columnwidth]{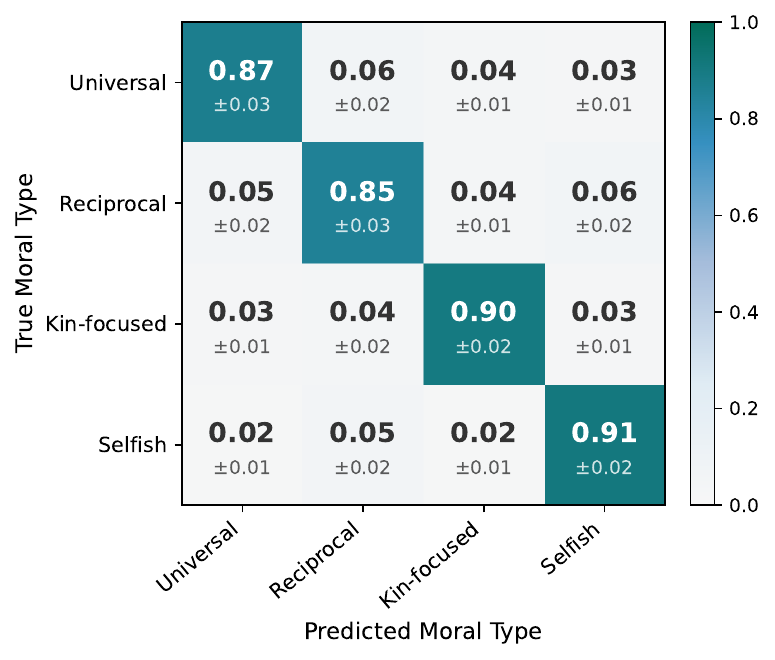}
    \caption{Validation results of the baseline simulation. Upper: HP trajectories of selected agents with action markers. Lower: confusion matrix of moral type inference averaged over 8 runs.}
    \label{fig:validation}
\end{figure}

\begin{table}[t]
\centering
\small
\begin{tabular}{@{}lccc@{}}
\toprule
\textbf{Sim.\ Model} & \textbf{CM Diag.\ Acc.} & \textbf{Dom.\ Type} & \textbf{Final Pop.} \\
\midrule
GPT-5-mini & $0.89 \pm 0.03$ & Kin (6/8) & $12.0 \pm 2.0$ \\
Qwen-3.5 & $0.86 \pm 0.03$ & Kin (7/8) & $11.6 \pm 1.7$ \\
Kimi-K2.5 & $0.87 \pm 0.03$ & Kin (5/8) & $12.1 \pm 1.8$ \\
\bottomrule
\end{tabular}
\caption{Cross-model robustness of the baseline simulation over 8 independent runs per model. CM Diag.\ Acc.\ = confusion-matrix diagonal accuracy.}
\label{tab:cross_model}
\end{table}

\begin{table}[t]
\centering
\small
\begin{tabular}{@{}lcc@{}}
\toprule
\textbf{Condition} & \textbf{CM Diag.\ Acc.} & \textbf{$\Delta$} \\
\midrule
Full Architecture & $0.89 \pm 0.03$ & --- \\
\midrule
\multicolumn{3}{@{}l}{\textit{Architectural Ablation}} \\
\quad w/o Memory & $0.78 \pm 0.04$ & $-0.11$ \\
\quad w/o Plan & $0.82 \pm 0.04$ & $-0.07$ \\
\quad w/o Reflection & $0.84 \pm 0.03$ & $-0.05$ \\
\quad ReAct Baseline & $0.67 \pm 0.06$ & $-0.22$ \\
\midrule
\multicolumn{3}{@{}l}{\textit{Prompt Sensitivity}} \\
\quad Variant A (lexical rewrite) & $0.87 \pm 0.03$ & $-0.02$ \\
\quad Variant B (structural rewrite) & $0.86 \pm 0.04$ & $-0.03$ \\
\bottomrule
\end{tabular}
\caption{Architecture ablation and prompt sensitivity analysis. All experiments use the baseline configuration with GPT-5-mini.}
\label{tab:ablation}
\end{table}

\begin{table}[t!]
\centering
\small
\begin{tabular}{l c c c c c}
\toprule
\textbf{Setting} & \textbf{N} & \textbf{Uni.} & \textbf{Rec.} & \textbf{Kin} & \textbf{Sel.} \\
\midrule
Baseline              & 8 & 4 & 2 & \textbf{6} & 2 \\
Scarce Resource       & 4 & 2 & \textbf{3} & 0 & 1 \\
High Social Cost      & 4 & 2 & \textbf{3} & 0 & 1 \\
Moral Type Invisible  & 4 & \textbf{4} & 2 & 2 & 0 \\
\bottomrule
\end{tabular}
\caption{Run-level survival counts. A moral type earns one point in a run if it has a nonzero population at step~80; coexistence credits every surviving type.}
\label{tab:replicate_outcomes_main}
\end{table}

\section{Experiments}
\label{sec:experiments}

\begin{figure*}[t!]
    \centering
    \includegraphics[width=\textwidth]{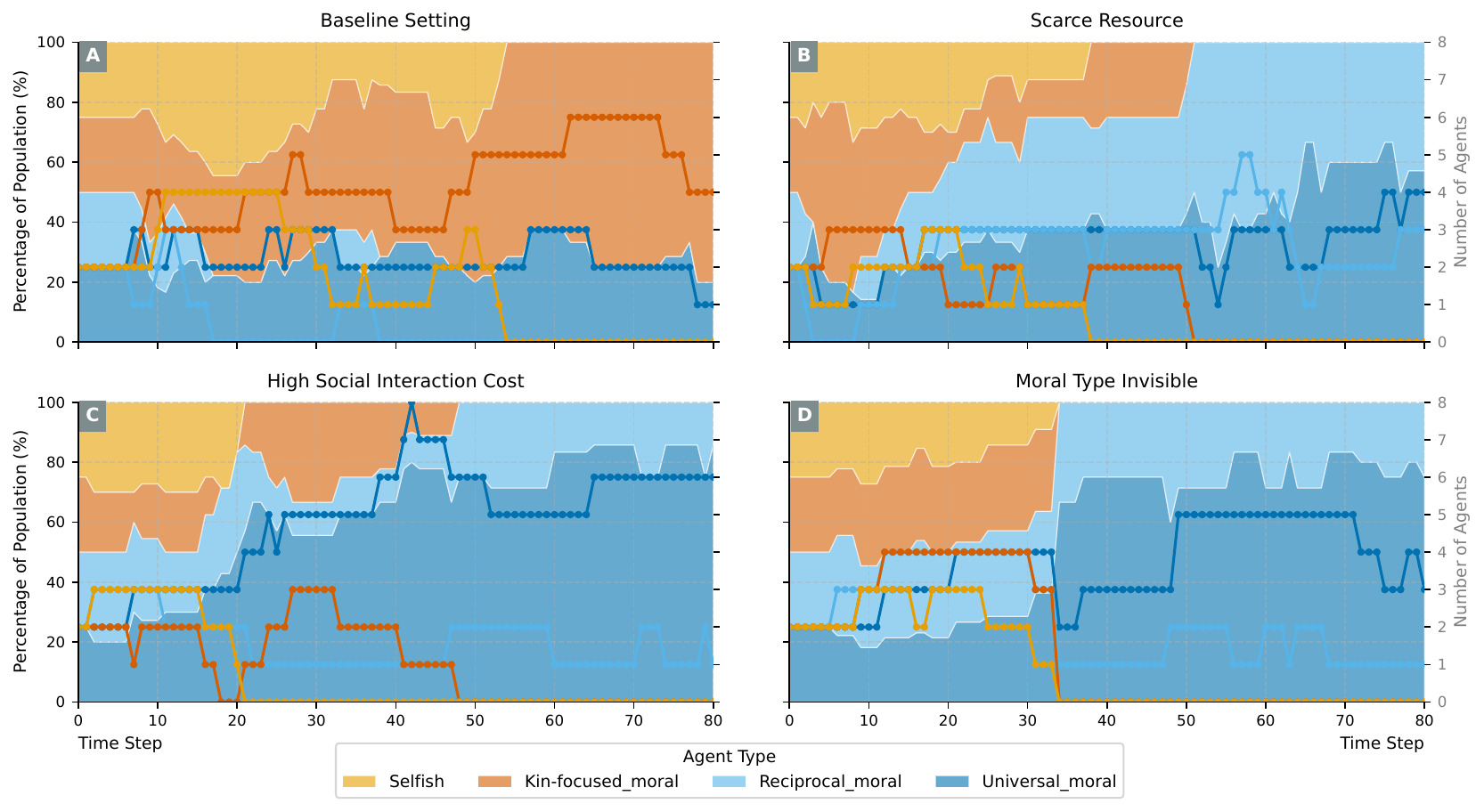}
    \caption{Population dynamics across four experimental settings. Each panel shows a representative run in which the most frequently surviving moral type in that setting persists to the end. (A)~Baseline: abundant resources, low social cost, observable moral types. (B)~Scarce Resource: reduced resource abundance. (C)~High Social Cost: only 1 social round before production. (D)~Moral Type Invisible: agents cannot see others’ moral types. Outcomes vary across runs; aggregate survival counts over 20 replicate runs are reported in Table~\ref{tab:replicate_outcomes_main}.}
    \label{fig:population}
\end{figure*}

\subsection{Simulation Validation}

All experiments use GPT-5-mini as the primary simulation model. To verify that our findings are not artifacts of a specific LLM backbone, we additionally run the baseline setting on two other models---the open-source Qwen-3.5 and Kimi-K2.5---and confirm consistent results across all three (Table~\ref{tab:cross_model}).

We validate from both environmental feedback and morality-behavior consistency perspectives. Fig.~\ref{fig:validation} (upper) shows HP trajectories for representative agents---e.g., a kin-focused agent that repeatedly sacrifices HP for offspring until it cannot recover. Using GPT-5 as an evaluator to infer moral types from observed behaviors, the confusion matrix (Fig.~\ref{fig:validation}, lower) averaged over 8~runs achieves a diagonal accuracy of $0.89 \pm 0.03$, consistent across all tested models (Table~\ref{tab:cross_model}). Ablations (Table~\ref{tab:ablation}) show that removing any cognitive module degrades alignment---memory has the largest impact ($-0.11$)---and stripping all modules (ReAct baseline) drops accuracy to $0.67$, while prompt-sensitivity rewrites vary by $\leq 0.03$.

\subsection{Evolutionary Games}

We run experiments across four settings to investigate how moral dispositions interact with environmental and cognitive factors. To characterise stochastic variability, we execute the Baseline $N{=}8$ times and each ablation $N{=}4$ times (20 runs total) with distinct random seeds. Table~\ref{tab:replicate_outcomes_main} summarises survival counts; population trajectories for all runs appear in the supplement (\S\ref{sec:replicates}). Each simulation is initialized with 8 agents (2 per moral type); representative single-run trajectories appear in Fig.~\ref{fig:population} for illustration, with statistical claims grounded in Table~\ref{tab:replicate_outcomes_main}.

\textbf{Baseline} (Kin: 6/8, Uni: 4/8)\textbf{:} With abundant resources, sufficient social rounds, and observable moral types, kin-focused agents survive most frequently: universal agents' early unconditional helpfulness benefits kin agents, who free-ride on group efforts while channeling surplus resources into family reproduction, and once a kin lineage grows large enough it becomes self-sustaining through internal cooperation. Universal agents also survive in half the runs (4/8), showing that broader moral circles remain viable even where kin lineages have a structural advantage.

\textbf{Scarce Resource} (Rec: 3/4, Uni: 2/4)\textbf{:} Under reduced resource abundance, agents become cautious and selectively avoid less cooperative types. Reciprocal agents are favoured (3/4) because their conditional cooperation enforces fair sharing within coalitions while excluding free-riders; universal agents also persist (2/4) as their broader cooperation attracts reciprocated help. Notably, we observe rare inter-agent killing: selfish agents attack same-type peers as direct competitors, and kin agents preemptively target selfish agents, so both rarely survive.

\textbf{High Social Cost} (Rec: 3/4, Uni: 2/4)\textbf{:} With only 1 social round before production, kin-focused agents struggle because they typically require explicit teaming signals before investing in collaborative hunts; under constrained bandwidth they often cannot receive such signals and avoid risky ventures. Universal agents cooperate unconditionally, and reciprocal agents directly identify other cooperative types as trustworthy partners, forming effective coalitions without lengthy negotiation and achieving higher hunting success. Kin and selfish agents rarely persist (kin 0/4, selfish 1/4).

\textbf{Moral Type Invisible} (Uni: 4/4, Rec: 2/4, Kin: 2/4)\textbf{:} When moral type labels are hidden, universal agents survive in every run: their unconditional cooperation is independent of others' types and never at risk of misidentification. Reciprocal agents suffer from noisy type inference, sometimes misjudging cooperative agents as selfish and narrowing their cooperation space. Kin agents gain a delayed-identification advantage---their family-only helpfulness appears cooperative until extended interaction exposes selectivity---while selfish agents are quickly exposed through aggressive behaviours and eliminated in every run. Detailed mechanistic case studies appear in Appendix~\S\ref{sec:case_studies}.

\subsection{Mini-Games}

\begin{figure}
    \centering
    \includegraphics[width=\linewidth]{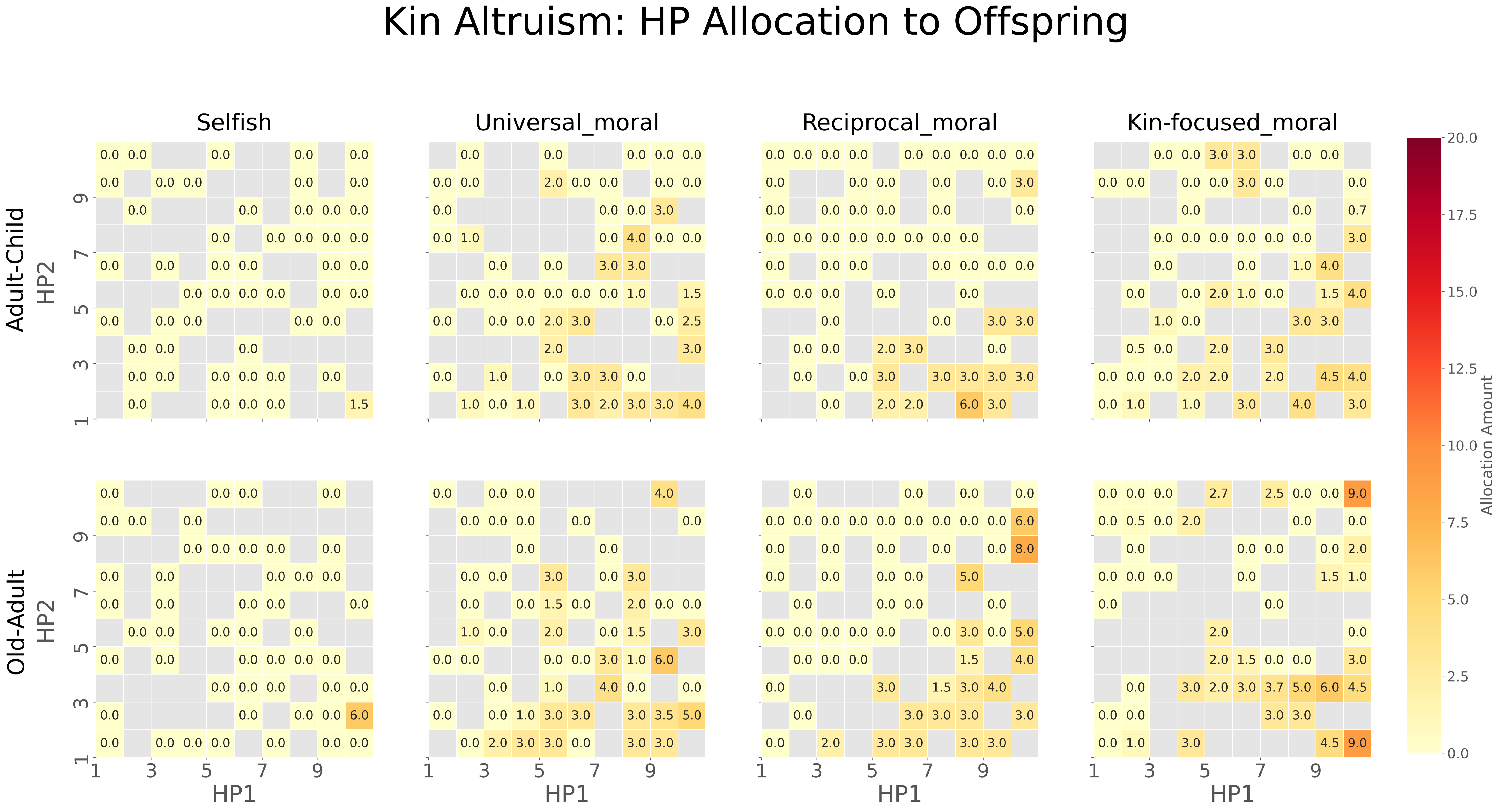}
    \caption{Kin Altruism: A Minigame on Family Resource Allocation.}
    \label{fig:allocate_kinaltru}
\end{figure}

Beyond long-horizon evolutionary games, we design controlled mini-games to isolate specific mechanisms linking morality to fitness. As an example, we study how moral dispositions shape intergenerational resource sharing within families. We model parent-child dyads across two life stages (young parents with infants, elderly parents with adult children) and four moral dispositions, creating eight scenarios. As shown in Fig.~\ref{fig:allocate_kinaltru}, provisioning strategies are systematically driven by moral type: selfish parents hoard resources for self-preservation, while kin-focused parents consistently prioritize offspring---in extreme cases sacrificing nearly all HP. More mini-game examples are provided in the Appendix.

\section{Discussion}
\label{sec:discussion}
\paragraph{Cooperation as the Central Driver}
Across all experimental settings, our simulations consistently show that \emph{cooperation and mutual help are the primary driver of evolutionary survival} (Table~\ref{tab:replicate_outcomes_main}). Universal and reciprocal agents---the most cooperative types---exhibit the most stable survival rates across all conditions. Kin-focused agents remain viable when resources and communication bandwidth suffice for them to reach self-sustaining family size. Selfish agents are strongly disfavoured in every setting. These results are broadly consistent with prior theoretical predictions~\citep{hamilton1964genetical,trivers1971evolution,nowak1998dynamics,Curry2019MAC}, and support the expanding circle model~\citep{singer1981expanding}: broader moral circles generally produce superior evolutionary outcomes.

\paragraph{Cognition as a Central Mediator: The Cost of Moral Judgment}
Beyond cooperation itself, a second pattern distilled across settings points to \emph{cognition as a central mediator}: the identity of the winning moral type shifts with how hard it is to judge others. Reciprocal agents lead when types are visible and interaction bandwidth is rich (Scarce Resource, 3/4), but universal agents lead precisely when judgment becomes costly (Moral Type Invisible, 4/4; High Social Cost, 2/4). The common thread is that assessing others' trustworthiness takes time under limited lifespans and carries the risk of misjudgment---wrongly excluding allies, or being wrongly excluded. Universal agents uniquely sidestep this cost: their unconditional cooperation produces no behaviour that could be misread, so their reputation settles quickly. When judgment cost is high, a ``willing-to-take-a-loss'' strategy therefore builds reputation faster and secures better cooperation, connecting to costly signalling~\citep{gintis2001costly, nowak2005evolution} and bounded rationality~\citep{simon1991bounded}.

\paragraph{An Additional Observation: Self-Purging of Selfish Agents}
In the Scarce Resource setting we observe a further phenomenon that helps explain why selfish agents fail so completely: rare inter-agent killing in which selfish agents preemptively attack one another, reasoning that others---especially fellow selfish agents who will neither help them nor share resources---pose a life-and-death competitive threat. Universal agents, by contrast, do not perceive others as such threats and avoid such aggression. Selfishness therefore fails through both its lack of positive benefits \emph{and} this self-destructive dynamic---a pattern that only surfaces once agents reason about one another.

\paragraph{What the Paradigm Unlocks}
Classical evolutionary game theory reduces agents to fixed strategies and the world to a payoff matrix (Fig.~\ref{fig:paradigm}, left), which by construction cannot express cognitive effects: both the cost of moral judgment and the self-purging of selfish agents require agents who form beliefs about one another, can misjudge, and act on those beliefs in a richer world. Relaxing these simplifications also lets us trace \emph{why} a type fails by examining individual agents' reasoning (supplement \S\ref{sec:case_studies}). Several cognition-related effects that are well established in neighbouring fields but have rarely been brought into the study of moral evolution---coordination costs under bounded rationality~\citep{simon1991bounded} in economics, impression formation from observed behaviour~\citep{asch1946forming} in social psychology, costly signalling~\citep{gintis2001costly, nowak2005evolution}, altruistic punishment~\citep{fehr2002altruistic}, life-history trade-offs in reproductive timing~\citep{kaplan1992evolution}, and the anthropological sequence of kin-focused cooperation~\citep{holden2003matriliny, mattison2011evolutionary} preceding expanded moral circles under environmental pressure~\citep{singer1981expanding}---here emerge bottom-up from cognitive agent interactions rather than being built in as assumptions (see Appendix Table~\ref{tab:project_findings} for the full mapping). The paradigm thus provides a common venue in which these previously isolated mechanisms can jointly bear on moral evolution.

\paragraph{Conclusions}
We propose a new research paradigm that brings cognitive realism to the study of moral evolution through LLM-based agent simulation. Our experiments show that cooperation and mutual help are the central driver of evolutionary survival, and further reveal \emph{cognition itself} as a central mediator---most clearly through the cost of moral judgment, with the self-purging of selfish agents as an additional cognitive pattern. We hope this paradigm can inspire broader and deeper investigations into moral evolution and other complex social phenomena that require cognitively realistic modeling.

\section*{Limitations}

This work is constrained by several design choices that delimit the scope and generality of the presented results. 

First, the framework currently relies on general-purpose LLMs for key reasoning and decision-making functions. In preliminary experiments, these models exhibit brittleness in fine-grained spatial and temporal computations, which can propagate into downstream behaviors and reduce the fidelity of simulated interaction dynamics. Addressing such reasoning failures will likely require dedicated mechanisms (e.g., specialized representations or constrained inference procedures) beyond the present implementation.

Second, the current model does not explicitly implement sexual selection, despite its central role in evolutionary dynamics. As a consequence, the framework cannot capture phenomena such as mate choice, mating competition, or selection pressures arising from reproductive strategies. Incorporating these processes would require additional, carefully specified mechanisms and is therefore left to future work.

Third, our environment is intentionally limited to a hunter-gatherer setting. Important components of more complex societies—such as tool innovation and accumulation, exchange/market dynamics, institutional governance, and technologically mediated coordination—are not modeled. While these additions could increase ecological realism, they may also introduce additional confounds and complicate experimental control; thus, conclusions drawn from the present experiments should be interpreted as pertaining to simplified social-ecological conditions.

\section*{Ethical considerations}
Our project is, at its core, a simulation study of ethics itself. As such, it does not raise the typical ethical concerns associated with methodological research that might be misused.
Importantly, our findings can be interpreted as supporting the general proposition that morality is beneficial for humans. The factors that sometimes cause moral agents to fail in evolutionary competition can, in fact, offer valuable insights for promoting social causes and designing mechanisms to enhance the evolutionary advantage of moral individuals.
However, we caution against the simplistic interpretation that the conditions under which moral agents fail to prevail are evidence that morality is not advantageous for humans. Such a view is an oversimplification. First, modern society differs profoundly from prehistoric hunter-gatherer contexts. Humans have evolved to be born with moral dispositions \citep{hamlin2011young, bloom2013just, warneken2006altruistic}. Also in contemporary human societies, almost no one can survive without collaboration, promoting moral behaviors. Second, it is crucial to understand the specific causal role that morality plays in success or failure. For example, our results show that when communication is prohibitively costly, moral agents may be outcompeted by selfish ones. This occurs because morality often inclines agents toward collaboration, which may not be optimal in situations where cooperation is particularly costly. However, morality does not require agents to cooperate indiscriminately; moral agents could, in principle, maintain their moral disposition while choosing to act independently when cooperation is not advantageous, and then collaborate when conditions improve. As revealed by our simulation and common wisdom, being moral does not guarantee success in every circumstance, but a lack of morality fundamentally constrains one's potential for success.
\section*{Use of AI Assistants}
This paper was primarily conceived, designed, and drafted by the human authors. AI assistants (including ChatGPT and Claude) were used in a supporting role for proofreading, rewriting for clarity, and assisting with code development for the simulation platform. All scientific contributions, experimental design, analysis, and intellectual direction were driven by the authors, with AI tools serving as aids for language refinement and coding assistance.

\bibliography{references}

\appendix

\section{General Discussions}
\label{app:general_discussions}
\subsection{Code Release}
Our code has been released under the MIT license at \url{https://github.com/MoralAgentSim/Simulation-Engine}. A project page with documentation and examples is available at \url{https://MoralAgentSim.github.io}.
This platform will be actively maintained and updated to support more features and research questions. We welcome any collaboration, contribution, feedback, and feature requests.

\subsection{Potential Risks}
Despite our efforts toward realism, our simulation operates in a constrained environment that inevitably omits many real-world factors. Results should therefore be treated as generating insights and hypotheses rather than definitive conclusions, and should not be used to directly guide important cognitive or policy decisions. Additionally, findings about conditions under which selfish strategies occasionally persist could, in principle, be interpreted as guidance for when self-interested behaviour might go unchecked---though the high level of abstraction in our simulation limits such applicability.

\subsection{More Discussion Over Our Methodology}
As we have emphasized, our method should be viewed as a complement to traditional mathematical models, not a replacement. By incorporating rich psychological realism into the simulation, our approach enables researchers to investigate how numerous factors interact in complex ways. However, this increased realism also means that simulation results are sensitive to the specific details of these factors and may not yield the definitive answers that highly abstract mathematical models can provide.

Yet, definitive answers are not always the primary goal of research, especially in the social sciences. Often, the objective is to uncover previously unnoticed factors that influence a phenomenon or to explore the intricate interplay among multiple variables. Such goals are difficult to achieve with traditional mathematical models, which require all relevant factors to be known or assumed in advance. Historically, researchers have relied on field studies to observe human behavior and identify these factors, but simulation now offers a cost-effective means to assist in discovery and hypothesis generation, potentially accelerating progress in the field.

Moreover, when the number of interacting factors becomes too great for analytical calculation, simulation becomes indispensable. While simulations inevitably deviate from reality—just as any modeling method, and such deviations may be amplified in large-scale runs—they can still provide valuable insights into research questions. Simulations can reveal what is possible, and the underlying mechanisms and developmental dynamics they expose may remain relevant even if the precise outcomes differ from those observed in the real world.

\subsection{Flexibility of the Simulation Platform}
Our platform is designed to be flexible and extensible. By varying the configuration settings, one can use the same platform to study different research questions. For example, we have used the same platform to study the effect of different moral types, different resource distributions, different communication costs, etc. In the section below, we also list a list of findings that are connected to different research areas that could possibly be investigated further with our platform.

Moreover, we support researchers to extend beyond morally related value dispositions. One can flexibly define the value dispositions of the agents by writing appropriate prompt templates. For example, one can define agents to be of different cultural backgrounds, different religions, different political views, etc. Or one can also study the effect of specific social norms by prescribing the agents to follow certain social rules, e.g, always equal distribution VS always contribution-based distribution, etc. Hunting-gathering environment equipped with general social interaction dynamics is very general to support a wide range of research questions.

\section{Discovered Phenomena That Connect to Other Theories}
\label{app:other_theories}
As mentioned, one key feature of what our platform can provide is that we can naturally see a lot of emergent phenomena that matter for social evolution regarding morality. These phenomena were abstracted away in the traditional mathematical models. But on our platform, they will surface on their own to deepen our understanding. Those phenomena or topics were traditionally a subject of research areas on their own, but now we can study them in a unified framework. 

We list some of the observed phenomena and identify some of the theories that are related to them in Table~\ref{tab:project_findings}. This list is definitely not exhaustive. We hope this can provide a good starting point for future researchers to discover more phenomena and theories.

We also encourage researchers to use our platform as a new way to study these phenomena and theories.

\begin{table*}[tbp]
\centering
\caption{Discovered Phenomena and Related Theories}
\label{tab:project_findings}
\small
\setlength{\tabcolsep}{4pt}
\begin{tabular}{p{7cm}p{7cm}}
\hline
\textbf{Phenomena Findings from Experiments} & \textbf{Related Theories} \\
\hline
\parbox[t]{7cm}{Coordination is costly: \\ 
\hspace*{1em} $\bullet$ Communication takes time and can reduce the time for other important things.} & 
\parbox[t]{7cm}{\textit{Coordination Cost Theory} \citep{simon1991bounded} \\
"Organizations face bounded rationality where coordination costs limit optimal decision-making"} \\
\hline
\parbox[t]{7cm}{Moral judgment based on actions: \\
\hspace*{1em} $\bullet$ Agents evaluate others' morality by observing how they treat third parties. \\
\hspace*{1em} $\bullet$ Actions toward others, not just toward oneself, shape moral reputation.} & 
\parbox[t]{7cm}{\textit{Moral Judgment Theory} \citep{haidt2001emotional} \\
"People make rapid moral judgments based on observed behaviors and their emotional responses" \\
\textit{Impression Formation} \citep{asch1946forming} \\
"Observers form impressions of others' character based on their actions toward third parties"} \\
\hline

\parbox[t]{7cm}{Misunderstandings can lead to major conflicts: \\
\hspace*{1em} $\bullet$ Agents may misinterpret others' intentions or actions, leading to unnecessary conflicts. \\
\hspace*{1em} $\bullet$ Limited communication can cause agents to make incorrect assumptions about others' moral types or goals.} & 
\parbox[t]{7cm}{\textit{Communication Theory} \citep{shannon1948mathematical} \\
"Information transmission is inherently imperfect, leading to potential misunderstandings and conflicts" \\
\textit{Conflict Resolution} \citep{deutsch1973resolution} \\
"Many conflicts arise from misperceptions and misunderstandings rather than actual incompatible goals"} \\
\hline

\hline
\parbox[t]{7cm}{Predictable morality acts as a reputation/clear signal: \\
\hspace*{1em} $\bullet$ Universal moral agents, by seldom initiating violence, benefit from being clearly understood as cooperators. \\
\hspace*{1em} $\bullet$ Such behavior is costly though, making them subject to exploitation} &
\parbox[t]{7cm}{\textit{Costly Signaling Theory} \citep{gintis2001costly} \\
"Consistently moral behavior, despite its costs, serves as an honest signal of cooperative intent, reducing the risk of being misjudged as a threat." \\
\textit{Indirect Reciprocity} \citep{nowak2005evolution} \\
"A clear reputation for cooperation, built through predictable actions, is essential for reciprocal altruism and protects an agent from being wrongly punished."} \\
\hline
\parbox[t]{7cm}{Universal moral agents get exploited: \\
\hspace*{1em} $\bullet$ Agents who never retaliate or punish others' bad behavior become targets of exploitation. \\
\hspace*{1em} $\bullet$ Their unconditional cooperation makes them vulnerable to free-riders.} & 
\parbox[t]{7cm}{\textit{Altruistic Punishment} \citep{fehr2002altruistic} \\
"Cooperation requires punishment of defectors; pure altruism without retaliation is vulnerable to exploitation" \\
\textit{Strong Reciprocity} \citep{bowles2004evolution} \\
"Evolutionary success requires both cooperation and punishment of non-cooperators"} \\
\hline
\parbox[t]{7cm}{Group membership is contested: \\
\hspace*{1em} $\bullet$ Agents might not agree on who is in the group that can share resources.} & 
\parbox[t]{7cm}{\textit{Social Identity Theory} \citep{tajfel1979individuals} \\
"Group boundaries are fluid and contested, with membership determined by shared identity markers and mutual recognition"} \\
\hline
\parbox[t]{7cm}{Distribution methods are complex: \\
\hspace*{1em} $\bullet$ How to distribute? Distribute evenly, based on contribution, harm taken, need, can affect both the success of the end result and each other's judgment.} & 
\parbox[t]{7cm}{\textit{Distributive Justice} \citep{konow2003which} \\
"Fairness judgments depend on multiple principles, including equality, need, and contribution"} \\
\hline
\parbox[t]{7cm}{Careful planning is important: \\
\hspace*{1em} $\bullet$ Reproduction schedule is important. Too frequent can cause both parents and children to die.} & 
\parbox[t]{7cm}{\textit{Life History Theory} \citep{kaplan1992evolution} \\
"Organisms face trade-offs between current and future reproduction, with timing being crucial for survival"} \\
\hline
\parbox[t]{7cm}{Tendency to cooperate can sometimes have negative effect: \\ 
\hspace*{1em} $\bullet$ Moral agents have a tendency to collaborate to acquire resources, but in some particular setting (with competition, resources being in some way), taking faster action instead of collaboration may be more crucial. \\
\hspace*{1em} $\bullet$ Moral agents tend to agree to collaborate to hunt, but they might not be in a good position to hunt.} & 
\parbox[t]{7cm}{\textit{Cooperation Dilemmas} \citep{bowles2004evolution} \\
"Cooperation can be maladaptive when individual action would yield higher returns"} \\
\hline
\parbox[t]{7cm}{Moral agents' mutual dependency sometimes leads to disaster end: \\
\hspace*{1em} $\bullet$ Moral agents tend to trust others to help them later, but the others may also think the same, and none have the extra capacity to help.} & 
\parbox[t]{7cm}{\textit{Trust and Cooperation} \citep{fehr2002altruistic} \\
"Altruistic punishment can maintain cooperation but may lead to cascading failures when trust is misplaced"} \\
\hline
\parbox[t]{7cm}{Mutual reinforcing / social pressure: \\
\hspace*{1em} $\bullet$ When some agents reproduce, others feel compelled to do so too, even though their HP was not very high.} & 
\parbox[t]{7cm}{\textit{Social Learning Theory} \citep{bandura1977social} \\
"Social learning and imitation can lead to behavioral contagion even when not optimal for individuals"} \\
\hline
\end{tabular}
\end{table*}

\section{System Design Details}
\label{app:system_agent_design}

\subsection{Simulation Pipeline}

\begin{figure*}[htbp]
  \centering
  \includegraphics[width=\textwidth]{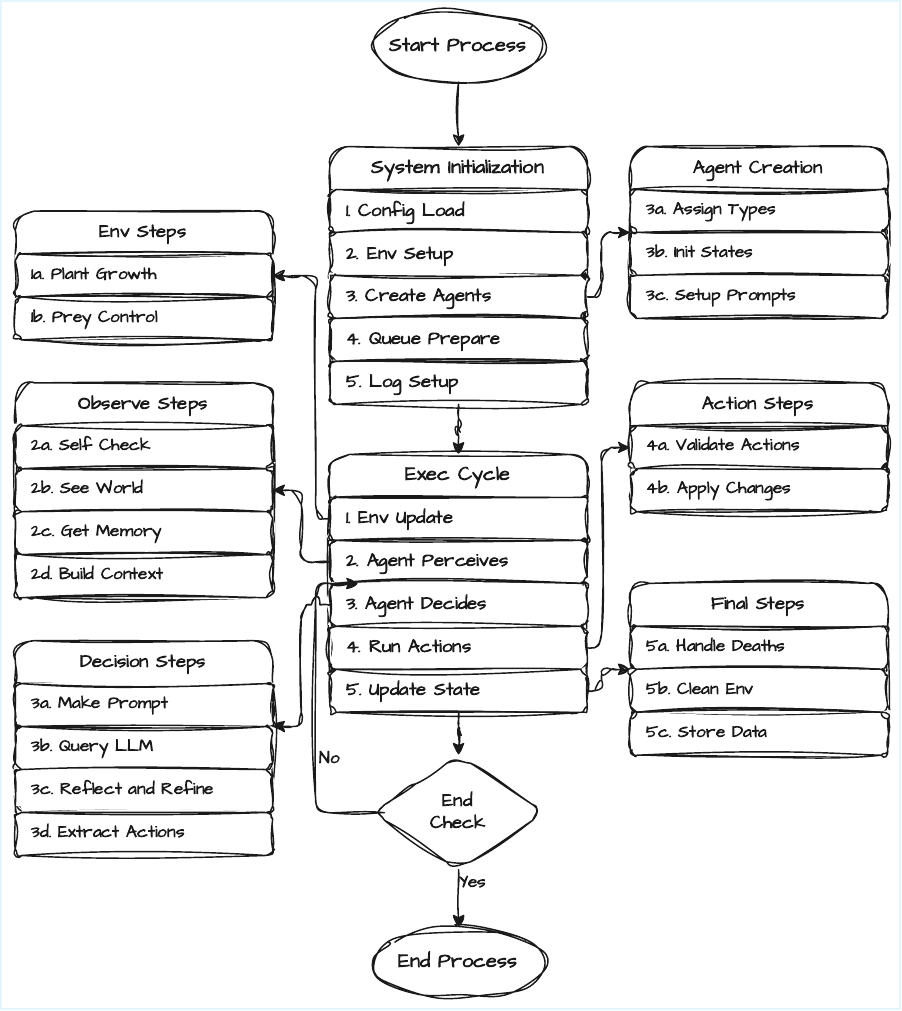}
    \caption{\textbf{Simulation Pipeline Overview} showing the main components and data flow through the system architecture. The pipeline illustrates how the Singleton-based Checkpoint, modular microservices, and key simulation processes interact to maintain a consistent state and flow of information.}
  \label{fig:sim-pipeline}
  \end{figure*}

The general system workflow functions as Figure~\ref{fig:sim-pipeline}. System first initializes the environment based on the system setting config (e.g, see Table~\ref{tab:baseline_config}) or resumes from the previous experiment run. The specific initialization phases are shown in Table~\ref{tab:init_phases}.

\begin{figure*}[htbp]
    \centering
    \includegraphics[width=0.9\textwidth]{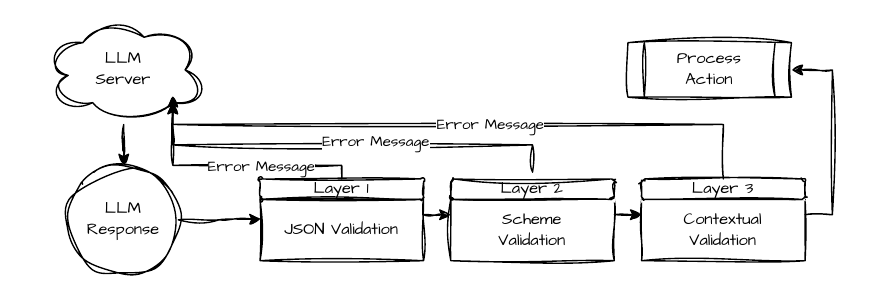}
    \caption{Multi-layer validation and retry framework showing the escalating levels of validation applied to agent actions. The diagram illustrates the three validation layers: syntactic and schema validation, contextual rule-based pre-validation, and action handler final validation, along with their respective feedback loops and retry mechanisms.}
    \label{fig:retry-validation}
\end{figure*}

Then the system enters into an execution cycle that allows agents to perceive and perform cognitive processing to plan for actions and update the environment accordingly. The execution phases are shown in Table~\ref{tab:execution_cycle}. Within this cycle, there is also a system validation and correction cycle over the agent's response and action to ensure its format and content are legal (see Figure~\ref{fig:retry-validation} and Table~\ref{tab:validation_layers}).

Please refer to those tables and figures for more details.

\begin{table*}[htbp]
  \centering
  \caption{Simulation Initialization Phases}
  \label{tab:init_phases}
  \small
  \setlength{\tabcolsep}{5pt}
  \begin{tabular}{p{3cm}p{10cm}}
    \toprule
    \textbf{Phase} & \textbf{Description} \\
    \midrule
    Configuration Loading \& Validation & 
    $\bullet$ Loads parameters from configuration file (prompt paths, agent types, rules, strategies) \newline
    $\bullet$ Validates type correctness, constraints, and completeness \newline
    $\bullet$ Creates authoritative configuration object for simulation \\
    \midrule
    Environment Setup & 
    $\bullet$ Plant Resources: \newline
    \quad - Generated based on configured abundance \newline
    \quad - Each plant gets a unique ID, initial quantity, capacity, nutrition value, and respawn delay \newline
    $\bullet$ Prey Animals: \newline
    \quad - Initialized with unique IDs \newline
    \quad - HP and max health sampled from a Gaussian distribution \newline
    \quad - Assigned physical ability values \newline
    $\bullet$ Resources placed randomly in unoccupied grid cells \\
    \midrule
    Initial Agent Spawning & 
    $\bullet$ Instantiates agents based on population size \newline
    $\bullet$ Assigns moral types according to configuration ratios \newline
    $\bullet$ Initializes attributes: HP, age, physical ability \newline
    $\bullet$ No initial family ties \\
    \midrule
    Execution Queue Setup & 
    $\bullet$ Creates randomized agent sequence for fair execution \newline
    $\bullet$ Initializes time step counter (typically 0 or 1) \newline
    $\bullet$ Sets up containers for agent observations \\
    \midrule
    Logging System Setup & 
    $\bullet$ Configures comprehensive tracking system \newline
    $\bullet$ Creates log files for: \newline
    \quad - Global progress summaries \newline
    \quad - Per-step execution records \newline
    \quad - Detailed event logs \newline
    \quad - Error diagnostics \newline
    $\bullet$ Organizes logs in uniquely named directories \\
    \bottomrule
  \end{tabular}
\end{table*}

\begin{table*}[htbp]
  \centering
  \caption{Per-Step Execution Cycle Phases}
  \label{tab:execution_cycle}
  \small
  \setlength{\tabcolsep}{5pt}
  \begin{tabular}{p{2.5cm}p{10cm}}
    \toprule
    \textbf{Phase} & \textbf{Description} \\
    \midrule
    Environment State Update & 
    $\bullet$ Updates plant lifecycle: restores depleted plants after respawn delay, increases quantity for non-depleted plants \newline
    $\bullet$ Spawns new prey in empty locations based on probability and maximum count \newline
    $\bullet$ Removes dead prey from the grid \\
    \midrule
    Agent Observation & 
    $\bullet$ Self-assessment: queries HP, age, inventory, physical ability, reproductive status \newline
    $\bullet$ Environmental perception: detects nearby resources, prey, and other agents \newline
    $\bullet$ Memory retrieval: accesses past observations, messages, and action outcomes \newline
    $\bullet$ Context formatting: structures information for LLM prompt \\
    \midrule
    Agent Decision Making & 
    $\bullet$ Constructs system message with agent persona and rules \newline
    $\bullet$ Builds user message with current state and context \newline
    $\bullet$ LLM processes context and returns proposed action \newline
    $\bullet$ Validates response format and structure \\
    \midrule
    Action Execution \& Validation & 
    $\bullet$ Performs response \& action validation \newline
    $\bullet$ Applies validated actions to simulation state \\
    \midrule
    State Finalization &
    $\bullet$ Updates agent HP, inventories, and environmental quantities \newline
    $\bullet$ Handles communication and memory updates \newline
    $\bullet$ Performs system-wide consistency checks \newline
    $\bullet$ Records detailed logs of agent states, environment state, and metrics \newline
    $\bullet$ Prepares state for next cycle \\
    \midrule
    Termination Check & 
    $\bullet$ Evaluates termination criteria (max steps, population collapse, goals) \newline
    $\bullet$ Either concludes simulation or increments time step \\
    \bottomrule
  \end{tabular}
\end{table*}

\begin{table*}[htbp]
  \centering
  \caption{The Checklist of Multi-Layer Response \& Action Validation.}
  \label{tab:validation_layers}
  \small
  \setlength{\tabcolsep}{5pt}
  \begin{tabular}{p{2.5cm}p{10cm}}
    \toprule
    \textbf{Layer} & \textbf{Description} \\
    \midrule
    Layer 1: Syntactic and Schema & 
    $\bullet$ Applied immediately after LLM output generation \newline
    $\bullet$ Two critical checks: \newline
    \quad - Syntactic validation: Ensures proper JSON formatting \newline
    \quad - Schema validation: Verifies required fields, types, and enumerations \newline
    $\bullet$ Retry mechanism: \newline
    \quad - Resends prompt with error metadata \newline
    \quad - Limited to predefined maximum attempts \newline
    $\bullet$ Focus: Structural correctness only \\
    \midrule
    Layer 2: Contextual and Rule-Based & 
    $\bullet$ Domain-specific validation within Agent Decision Making phase \newline
    $\bullet$ Contextual checks: \newline
    \quad - Target existence and accessibility \newline
    \quad - Location-based constraints \newline
    \quad - HP sufficiency for action costs \newline
    $\bullet$ Memory constraints: \newline
    \quad - Long-term memory capacity limits \newline
    $\bullet$ Feedback loop: \newline
    \quad - Human-readable error messages \newline
    \quad - Updated prompts with feedback \newline
    \quad - Configurable retry rounds \\
    \midrule
    Layer 3: Action Handler Final & 
    $\bullet$ Executed during Action Execution phase \newline
    $\bullet$ Domain-specific validation in action handlers \newline
    $\bullet$ Dynamic condition checks: \newline
    \quad - Agent adjacency for physical interactions \newline
    \quad - HP sufficiency with current state \newline
    \quad - Race conditions with shared resources \newline
    $\bullet$ No LLM retry mechanism \newline
    $\bullet$ Failure handling: \newline
    \quad - Action nullification or failure processing \newline
    \quad - Logging to agent observation history \\
    \bottomrule
  \end{tabular}

\end{table*}

\subsection{System Design Principles}
The Morality-AI simulation is built on two core principles: centralized state management and modular architecture. A Singleton-based Checkpoint class maintains a single, authoritative simulation state, ensuring consistency, atomic updates, and easy reproducibility. This design prevents conflicting states and simplifies debugging and resuming experiments.
The system adopts a microservice-inspired structure, separating major functions—such as state persistence, agent reasoning, and LLM interfacing—into independent, easily testable modules. This modularity enhances maintainability, scalability, and flexibility, allowing components to be updated or replaced without affecting the overall system. Together, these principles provide a robust and extensible foundation for complex agent-based simulations.

\section{Agent Design Details}
\subsection{Agent Designs and Workflows}

Agents are the primary decision-making entities in the simulation. They possess a set of core attributes that govern their physical capabilities, cognitive constraints, and eligibility for specific actions (see the agent attributes in Table~\ref{tab:baseline_config}).

\begin{figure*}[htbp]
    \centering
    \includegraphics[width=\textwidth]{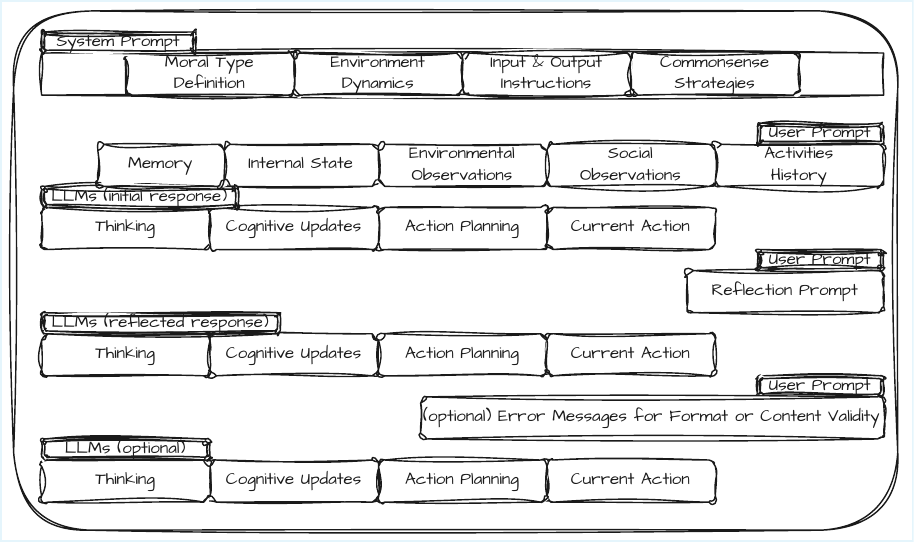}
    \caption{The LLM query process for decision making, illustrating the flow from observation gathering through prompt construction, LLM interaction, and action validation. This process shows how environmental perceptions, agent state, and memory are integrated to produce contextually relevant decisions within the simulation environment.}
    \label{fig:agent-decision}
  \end{figure*}

At the beginning of agent initialization, agent will be given their value/moral type prompt and all the system prompts like environment dynamics, requirement, commonsense strategies etc (prompt details see \ref{app:prompts}). Then, during each execution cycle, the agent will be given the perception of the environment and its own status, and perform cognitive processing to make action plan. They will perform one round of reflection before finalize their response that contains their cognitive processing and action plan. The process follows Figure~\ref{fig:agent-decision} to make decisions.

For the current project, the structure of the agent's moral type is listed in Table~\ref{tab:moral_types}, with the rationale of the design choices in the main text. We want to note that these moral types is not the only way to define the value of an agent. The value can be defined in many other ways - one can focus on the action principles, or calculation of utility, or even be involved in culture and religion to study different problems.

The structure of the agent's perception space is listed in Table~\ref{tab:agent_perception}. The structure of the agent's cognition is listed in Table~\ref{tab:agent_cognition}. The content in the action space is listed in Table~\ref{tab:agent_actions}.

  \begin{table*}[htbp]
    \centering
    \small
    \caption{Agent Moral Types Summary. Here, we summarize the core characteristics, expected typical behaviors, and expected cooperation patterns of these moral types. However, the simulated behavior for each agent might not strictly follow the expected behaviors due to the randomness of LLM's output. The exact prompt for each moral type is shown in \ref{app:prompts}}
    \setlength{\tabcolsep}{4pt}
    \begin{tabular}{p{2cm}p{3.5cm}p{3.5cm}p{3.5cm}}
      \toprule
      \textbf{Moral Type} & \textbf{Core Characteristics} & \textbf{Expected Typical Behaviors} & \textbf{Expected Cooperation Pattern} \\
      \midrule
      Universal Group-Focused Moral & Aim for universal well-being and collective good, harm-action averse & Share resources freely; protect others from harm; communicate transparently & Highly altruistic and cooperative with all agents \\
      \midrule
      Reciprocal Group-Focused Moral & Fairness and mutual benefit within in-group, harm action allowed & Form strong bonds with cooperative peers & Cooperative with in-group; neutral or adversarial to out-group or selfish agents \\
      \midrule
      Kin-Focused Moral & Prioritize genetic relatives above all else, harm action allowed & Form close-knit kinship clusters; sacrifice for kin & Intensely altruistic toward family; indifferent or competitive toward non-kin \\
      \midrule
      Selfish & Personal reproductive success, harm action allowed & Acquire resources for own survival; opportunistic tactics & Cooperate only when serving reproductive interests; inclined to hoard resources \\
      \bottomrule
    \end{tabular}
    \label{tab:moral_types}
  \end{table*}

\begin{table*}[htbp]
    \centering
    \caption{Agent Cognition Structure (Memory, Judgement and Planning)}
    \label{tab:agent_cognition}
    \small
    \setlength{\tabcolsep}{4pt}
    \begin{tabular}{p{4cm}p{9cm}}
    \toprule
    \textbf{Field} & \textbf{Key Subfields / Description} \\
    \midrule
    \textbf{1. Prey-Based Cognition} &
    $\bullet$ Organized by prey\_id: \newline
    \hspace*{1em} $\circ$ hunt\_fact\_history\_of\_this\_prey: who hunted, effect, time step, damage, if\_killed \newline
    \hspace*{1em} $\circ$ communication\_and\_planning\_before\_killing\_prey: reward, collaborators, distribution plan, objections \newline
    \hspace*{1em} $\circ$ distribution\_after\_killing\_prey: winner, allocation, fairness evaluation, free rider check \newline
    \hspace*{1em} $\circ$ plan\_next: next plan, retaliation plan, stage, reasoning \newline
    \hspace*{1em} $\circ$ afterward\_happenings: retaliation events, other events, lessons learned \\
    \midrule
    \textbf{2. Agent-Based Cognition} &
    $\bullet$ Organized by agent\_id: \newline
    \hspace*{1em} $\circ$ important\_interaction\_history: what\_i\_did\_to\_him, what\_he\_did\_to\_me (action type, success, reason, target moral type) \newline
    \hspace*{1em} $\circ$ thinking: evaluation, judgement, relationship, agreement, plan \\
    \midrule
    \textbf{3. Family Plan} &
    $\bullet$ Organized by agent\_id: \newline
    \hspace*{1em} $\circ$ status: how the family member is doing \newline
    \hspace*{1em} $\circ$ plan: what to do to/with them \\
    \midrule
    \textbf{4. Reproduction Plan} &
    $\bullet$ thinking: reasoning about reproduction plan \newline
    $\bullet$ preconditions\_and\_subgoals: specific preconditions needed \newline
    $\bullet$ estimated\_time\_to\_produce\_next\_child: time step \\
    \midrule
    \textbf{5. Learned Strategies} &
    $\bullet$ Lessons learned, strategies to follow in the future \\
    \bottomrule
    \end{tabular}
\end{table*}

\begin{table*}[htbp]
    \centering
    \caption{Agent Perception Content Structure}
    \label{tab:agent_perception}
    \small
    \setlength{\tabcolsep}{4pt}
    \begin{tabular}{p{4cm}p{9cm}}
    \toprule
    \textbf{Category} & \textbf{Description} \\
    \midrule
    Self/Internal Information & 
    $\bullet$ Current HP and health status \newline
    $\bullet$ Family relationships and status \newline
    $\bullet$ Personal attributes and capabilities \\
    \midrule
    Environment Status & 
    $\bullet$ Available plant resources \newline
    $\bullet$ Prey animals present in the environment \newline
    $\bullet$ Resource locations and quantities \\
    \midrule
    Other Agents Status & 
    $\bullet$ Basic information (age, HP) of other agents \newline
    $\bullet$ Moral types of other agents \newline
    $\bullet$ Current positions and states \\
    \midrule
    Recent History & 
    $\bullet$ Last 15 steps of personal interactions: \newline
    \hspace*{1em} - Others' actions and communications toward self \newline
    \hspace*{1em} - Self's actions toward environment and others \newline
    $\bullet$ Recent events: \newline
    \hspace*{1em} - Changes in environment and other agents \newline
    \hspace*{1em} - Family-related news and updates \newline
    $\bullet$ Hunting activities: \newline
    \hspace*{1em} - Personal involvement in prey hunting \newline
    \hspace*{1em} - Related communications and outcomes \\
    \midrule
    Memory & 
    $\bullet$ Updated memory from previous step \newline
    $\bullet$ Immediate action plans from previous step \\
    \bottomrule
    \end{tabular}
\end{table*}

\begin{table*}[htbp]
  \centering
  \caption{Agent Action Space Summary}
  \label{tab:agent_actions}
  \small
  \setlength{\tabcolsep}{4pt}
  \begin{tabular}{p{1.5cm}p{3cm}p{3cm}p{2cm}p{3cm}}
  \toprule
    \textbf{Action} & \textbf{Description} & \textbf{Requirements} & \textbf{HP Cost} & \textbf{Outcome} \\
    \midrule
    \multicolumn{5}{l}{\textit{(Re)Production Actions}} \\
    \midrule
    Collect & Gather plant resources from environment & Resource exists and is a plant node & None & Agent gains HP (quantity $\times$ nutrition value); plant quantity reduced \\
    \midrule
    Hunt & Target prey animals for nutritional gain & Prey exists & 1 HP + additional damage if failed & If successful, prey killed and agent receives reward equal to prey's max HP \\
    \midrule
    Reproduce & Create offspring & Minimum age and HP thresholds met & Defined in reproduction parameters & Child agent created with age 0 and initial HP, inheriting parent's archetype \\
    \midrule
    \multicolumn{5}{l}{\textit{Social Interaction Actions}} \\
    \midrule
    Allocate & Transfer HP to other agents & Targets exist and are alive; sufficient HP & Equal to HP transferred & Recipients gain specified HP (capped at maximum) \\
  \midrule
    Fight & Attempt to damage another agent & Target exists, is alive, not self & 1 HP resistance cost & If successful (based on ability difference), target suffers damage equal to attacker's ability \\
  \midrule
    Rob & Forcibly transfer HP from another agent & Target exists and is alive & 1 HP + potential failure penalty & If successful, HP transferred from target to robber \\
  \midrule
    Communicate & Send messages to other agents & Target agents exist and are alive & None & Message recorded in recipient's memory \\
  \midrule
    \multicolumn{5}{l}{\textit{Other Actions}} \\
  \midrule
    DoNothing & Abstain from all actions & None & None & No changes to agent or environment \\
  \bottomrule
  \end{tabular}
  \end{table*}

\subsection{The Quantitative Model for Calculating Action Results}

The success rate and damage point in actions like hunting, robbery, etc, is calculated based on the physical ability of the agents. The physical ability values are initialized as a random number from a Gaussian distribution with specified mean and standard deviation in the configuration file Table~\ref{tab:baseline_config} (with 0 standard deviation, there will be no random variation). Note that prey also has a physical ability value that is initialized in the same way. 

\paragraph{Success Rate}
The success of hunt, fight, and rob actions takes on probabilistic manner. The success of such actions depends on the relative physical abilities of the involved entities. Let $\Delta \mathrm{PA} = \mathrm{PA}_k - \mathrm{PA}_{target}$ represent the physical ability differential between an actor $k$ and a target entity (which could be another agent $j$ or a prey animal $A_j$). The probability of success, $P_{succ}$, for these actions is determined by the function:

{\small
\begin{align*}
&P_{succ}(\Delta \mathrm{PA}; I_{\mathrm{PA},k}, S_{\mathrm{PA},k}) = \min\\&\left(\max\left((0.5 + I_{\mathrm{PA},k}) + 0.4 \cdot \tanh\left(\frac{\Delta \mathrm{PA}}{S_{\mathrm{PA},k}}\right), 0.1\right), 0.9\right)
\end{align*}
}

Here, $I_{\mathrm{PA},k}$ and $S_{\mathrm{PA},k}$ are agent $k$'s specific scaling parameters (an intercept offset and a slope divisor, respectively) pertinent to physical ability interactions, derived from its configuration. The function $\min(\max(x,a),b)$ ensures the probability is clipped to the interval $[a,b]$, in this case, $[0.1, 0.9]$. The outcome of such an action is then determined by a Bernoulli trial $X \sim \text{Bernoulli}(P_{succ})$.

In the descriptions that follow, $\HP_k(t')$ signifies the health of agent $k$ after any initial action-specific costs have been deducted, but before other consequences of the action (e.g., gains from success, damage from failure) are applied.

\paragraph{Collect}
Agent $k$ may attempt to gather resources from a designated plant node $P_i$, which possesses a current resource quantity $Q_i(t)$. The agent specifies a desired quantity $q_{req}$. For the action to be valid, $P_i$ must be a plant, and its available quantity must meet the request, i.e., $Q_i(t) \ge q_{req}$. The actual quantity gathered, $q_{coll}$, is constrained by the request, availability, and the agent's single-action collection capacity, $k_{collect}$ (a global limit):
$$ q_{coll} = \min(q_{req}, Q_i(t), k_{collect}) $$
A positive quantity must be collectible ($q_{coll} > 0$). Consequently, the agent's health and the plant's resources are updated as follows:
\begin{align*}
    &\HP_k(t+1) = \\
    &\bound{\HP_k(t) + q_{coll} \cdot H_{plant}}{0}{\HP_{k,max}} \\
    & Q_i(t+1) = Q_i(t) - q_{coll}
\end{align*}
where $H_{plant}$ denotes the nutritional value conferred per unit of the plant resource. This action imparts no direct HP cost to agent $k$.

\paragraph{Allocate}
An agent $k$ (the donor) can transfer Health Points to other agents. This is specified via an \texttt{allocation\_plan}, $(h_{kj})_{j \in J}$, where $h_{kj} \in \mathbb{R}^+$ is the amount of HP designated for transfer to each target agent $j$ in a non-empty set $J \subset \mathcal{K}(t)$. The total HP intended for allocation by agent $k$ is $H_{alloc,k} = \sum_{j \in J} h_{kj}$. This action is permissible if all target agents $j \in J$ are alive and the donor possesses sufficient HP, specifically $\HP_k(t) > H_{alloc,k}$. If valid, the HP of the involved agents is then adjusted:

{\small\begin{align*}
    &\forall j \in J,\\ \quad &\HP_j(t+1) = \bound{\HP_j(t) + h_{kj}}{0}{\HP_{j,max}} \\
    &\HP_k(t+1) = \\
    &\bound{\HP_k(t) - H_{alloc,k}}{0}{\HP_{k,max}}
\end{align*}}

\paragraph{Fight}
Agent $k$ (attacker) may engage agent $j$ (target), provided $k \neq j$ and $j$ is alive. To initiate a fight, the attacker $k$ incurs an immediate cost $C_{fight,init} = 1$ HP:
$$ \HP_k(t') = \HP_k(t) - C_{fight,init} $$
If $\HP_k(t') \le 0$, agent $k$ is removed from $\mathcal{K}(t+1)$. Otherwise, the outcome of the fight is determined by a Bernoulli random variable $X_{fight} \sim \text{Bernoulli}(P_{succ}(\Delta \mathrm{PA}_{kj}; I_{\mathrm{PA},k}, S_{\mathrm{PA},k}))$, where $\Delta \mathrm{PA}_{kj} = \mathrm{PA}_k - \mathrm{PA}_j$. The health point dynamics for both the target and attacker, contingent on the outcome $X_{fight}$, are:
\begin{itemize}
    \item If $X_{fight}=1$ (success): The target's health is reduced, $\HP_j(t+1) = \bound{\HP_j(t) - \lfloor \mathrm{PA}_k \rfloor}{0}{\HP_{j,max}}$.
    \item If $X_{fight}=0$ (failure): The target's health remains unchanged, $\HP_j(t+1) = \HP_j(t)$.
\end{itemize}
In both scenarios, the attacker's health after the interaction resolves is $\HP_k(t+1) = \HP_k(t')$. The target agent $j$ is removed if its health $\HP_j(t+1) \le 0$.

\paragraph{Rob}
Agent $k$ (robber) may attempt to forcibly extract $h_{rob,req} > 0$ HP from a target agent $j$, provided $j$ is alive and possesses sufficient health ($\HP_j(t) \ge h_{rob,req}$). The robber $k$ first incurs an initiation cost $C_{rob,init} = 1$ HP:
$$ \HP_k(t') = \HP_k(t) - C_{rob,init} $$
If $\HP_k(t') \le 0$, $k$ is removed. Otherwise, the success of the attempt is a random variable $X_{rob} \sim \text{Bernoulli}(P_{succ}(\Delta \mathrm{PA}_{kj}; I_{\mathrm{PA},k}, S_{\mathrm{PA},k}))$, with $\Delta \mathrm{PA}_{kj} = \mathrm{PA}_k - \mathrm{PA}_j$. Depending on the outcome $X_{rob}$, the HP updates are:
\begin{itemize}
    \item If $X_{rob}=1$ (success):
    
    {\small\begin{align*}
        \HP_j&(t+1) = \\&\bound{\HP_j(t) - h_{rob,req}}{0}{\HP_{j,max}} \\
        \HP_k&(t+1) = \\&\bound{\HP_k(t') + h_{rob,req}}{0}{\HP_{k,max}}
    \end{align*}}
    
    The target $j$ is removed if $\HP_j(t+1) \le 0$.
    \item If $X_{rob}=0$ (failure): No HP is transferred, thus $\HP_j(t+1) = \HP_j(t)$, and the robber's health remains $\HP_k(t+1) = \HP_k(t')$.
\end{itemize}

\paragraph{Hunt}
Agent $k$ (hunter) may target a prey animal $A_j$, characterized by physical ability $\mathrm{PA}_{A_j}$ and health $\HP_{A_j}(t)$ (with maximum $\HP_{A_j,max}$). The hunter $k$ incurs an initial cost $R_{hunt}=1$ HP:
$$ \HP_k(t') = \HP_k(t) - R_{hunt} $$
If $\HP_k(t') \le 0$, $k$ is removed. Otherwise, the outcome is governed by $X_{hunt} \sim \text{Bernoulli}(P_{succ}(\Delta \mathrm{PA}_{kA_j}; I_{\mathrm{PA},k}, S_{\mathrm{PA},k}))$, where $\Delta \mathrm{PA}_{kA_j} = \mathrm{PA}_k - \mathrm{PA}_{A_j}$.
\begin{itemize}
    \item If $X_{hunt}=1$ (success): The prey $A_j$ sustains damage $D_{A_j} = \lfloor \mathrm{PA}_k \rfloor$, leading to $\HP_{A_j}(t+1) = \max(0, \HP_{A_j}(t) - D_{A_j})$.
        If this damage proves lethal ($\HP_{A_j}(t+1) \le 0$), prey $A_j$ is removed, and the hunter $k$ gains HP from the kill: 
        
        {\small\begin{align*}
        &\HP_k(t+1) = \\&\bound{\HP_k(t') + \HP_{A_j,max}}{0}{\HP_{k,max}}
        \end{align*}}
        
        If the prey survives the damage, the hunter gains no HP from the hit, so $\HP_k(t+1) = \HP_k(t')$.
    \item If $X_{hunt}=0$ (failure): The prey $A_j$ counter-attacks, inflicting $D_{prey}$ damage upon hunter $k$. This $D_{prey}$ is a characteristic of the prey (e.g., its counter-attack strength). The hunter's health is updated to 
        \begin{align*}
        &\HP_k(t+1) = \\&\bound{\HP_k(t') - D_{prey}}{0}{\HP_{k,max}}
        \end{align*}
        Hunter $k$ is removed if $\HP_k(t+1) \le 0$.
\end{itemize}

\paragraph{Reproduce}
An agent $k$ may create offspring if it meets age and health criteria: $Age_k(t) \ge Age_{repro,min}$ and $\HP_k(t) \ge \HP_{repro,min}$. Upon successful reproduction, a new agent $c$ is added to the population $\mathcal{K}(t+1)$, initialized with $Age_c(0)=0$ and health $\HP_c(0) = \HP_{child,init}$. The parent $k$ incurs an HP cost, $\HP_{repro,cost}$, resulting in an updated health:
\begin{align*}
    &\HP_k(t+1) = \\&\bound{\HP_k(t) - \HP_{repro,cost}}{0}{\HP_{k,max}}
\end{align*}

\paragraph{Communicate}
Agent $k$ can send a textual message $M$, constrained by length ($|M| \le L_{msg,max}$), to a specified set of recipient agents $J \subset \mathcal{K}(t)$. All recipients must be alive. This action does not directly alter HP.

\paragraph{DoNothing}
An agent $k$ may elect to perform no explicit action. This choice has no effect on its state or the environment; thus, $\HP_k(t+1) = \HP_k(t)$.

\section{Simulation Analysis Agent System}
\label{app:simulation_analysis}

\begin{figure*}[H]
    \centering
    \includegraphics[width=0.9\textwidth]{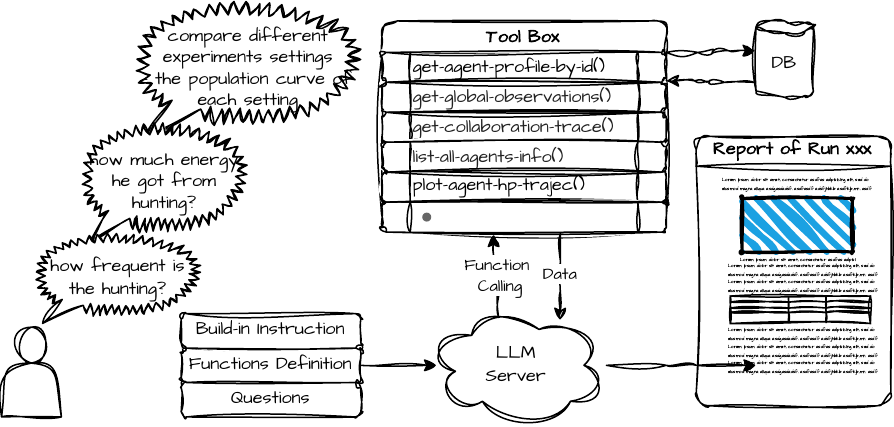}
    \caption{Overview of the post-processing analysis system architecture, showing the integration of RAG techniques, agent behavior logging, and structured reporting components. The diagram illustrates how raw simulation data is transformed into actionable insights through various analysis pipelines and iterative querying.}
    \label{fig:post-processing}
\end{figure*}

The Simulation Analysis Agent System is a comprehensive post-processing analysis framework for the Morality-AI simulation environment. It is engineered to distill actionable insights and facilitate in-depth investigation of simulation outcomes using a Retrieval-Augmented Generation (RAG) approach.

This system is based on the existing code agent tools like Github Copilot\footnote{\url{https://github.com/features/copilot}} and Cursor\footnote{\url{https://www.cursor.com/}}, etc, to manage the file calling system. During usage, one simply provides our analysis agent instruction file, the tool calling code file, and gives an experiment run identifier. The system will then automatically extract the experiment data and generate the analysis report, and provide interactive Q\&A.

This system consists of three primary components:
\begin{itemize}
    \item A \textbf{Simulation Analysis Agent} that orchestrates the analysis.
    \item An \textbf{Analytical Tool Suite} providing data processing and visualization functions.
    \item A \textbf{Reporting System} that generates structured outputs.
\end{itemize}

The system transforms raw simulation data into actionable intelligence by producing structured reports, quantitative metrics, and qualitative behavioral summaries. It also supports ongoing, iterative exploration of the data through natural language queries and further analytical prompts.

\subsection{Components}

The system is architected around three tightly integrated components:

\subsubsection{Simulation Analysis Agent}

The \textbf{Simulation Analysis Agent} is the central component that orchestrates the entire analytical workflow.

\begin{itemize}
    \item \textbf{Core Functions:}
    \begin{itemize}
        \item \textbf{Tool Calling Orchestration:} Coordinates the retrieval and transformation of simulation data by leveraging the Analytical Tool Suite. It accesses specific data slices such as agent profiles, global event logs, and collaboration traces.
        \item \textbf{RAG Interpretation:} Employs customized functions for efficient Retrieval-Augmented Generation to interpret and analyze simulation data.
        \item \textbf{Analysis Report Generation:} Synthesizes hierarchical analytical artifacts, including global summaries and lineage-specific analyses, combining quantitative metrics with qualitative behavioral insights.
        \item \textbf{Interactive Exploration:} Supports iterative, natural language-driven queries, enabling researchers to probe deeper into specific events, patterns, or hypotheses beyond initial report generation.
    \end{itemize}
    \item \textbf{How it Works:}
    Upon receiving a simulation run identifier, the agent initiates a multi-stage pipeline. It intelligently calls upon the various tools in the Analytical Tool Suite to fetch, process, and analyze data, then synthesizes this information to generate reports or respond to specific user queries.
\end{itemize}

\subsubsection{Analytical Tool Suite}

The \textbf{Analytical Tool Suite} (referred to as Analytical Framework in the original documentation) underpins the system's analytical capabilities through a robust, tool-driven interface.

\begin{itemize}
    \item \textbf{Core Functions:}
    \begin{itemize}
        \item Provides a library of modular, callable functions that abstract complex data queries and analytical routines.
        \item \textbf{Information Extraction:} Offers tools for retrieving diverse data sets. Examples include:
        \begin{itemize}
            \item \texttt{GetAgentProfile}: Retrieves comprehensive data for specified agents (state, family, actions, outcomes).
            \item \texttt{GetPopulationData}: Compiles and aggregates population-wide statistics (demographics, archetype distributions).
            \item \texttt{GetGlobalObservations}: Fetches or queries simulation-wide event logs (e.g., fights, robberies).
            \item \texttt{GetCollaborationTrace}: Extracts and summarizes data on cooperative interactions.
        \end{itemize}
        \item \textbf{Data Processing and Aggregation:} Includes functions for transforming and summarizing raw data, supporting both population-level and individual-level analyses.
        \item \textbf{Visualization:} Enables automated generation of plots, graphs, and statistical summaries to elucidate dynamic patterns and relationships. Examples include:
        \begin{itemize}
            \item \texttt{PlotAgentHPTrajectory}: Generates time-series plots of Health Point (HP) trajectories.
            \item \texttt{PlotPopulationComposition}: Visualizes the distribution and temporal changes of agent archetypes.
            \item \texttt{PlotMortalityAnalytics}: Produces visualizations of mortality patterns.
        \end{itemize}
        \item \textbf{Table Generation:} Offers functions like \texttt{FormatDataIntoTable} to structure extracted data into formatted tables for reports.
    \end{itemize}
    \item \textbf{How it Works:}
    This suite provides a collection of callable tools that the Simulation Analysis Agent utilizes to access, process, and visualize simulation data. These tools enable both macroscopic (population-level) and microscopic (individual-level) exploration of the simulation outcomes.
\end{itemize}

\subsubsection{Reporting System}

The \textbf{Reporting System} translates analytical results into structured, reproducible outputs.

\begin{itemize}
    \item \textbf{Core Functions:}
    \begin{itemize}
        \item \textbf{Structured Output Generation:} Produces standardized reports and visualizations for each simulation run.
        \item \textbf{Main Simulation Report:} Generates a comprehensive overview including an initial summary, population statistics, social dynamics analysis, key metrics, visualizations, and an index of detailed agent reports.
        \item \textbf{Agent-Specific Reports:} Creates detailed profiles for key agents (e.g., ancestors and significant descendants), covering state attributes, behavioral summaries, social interaction patterns, reproductive metrics, and qualitative analyses.
        \item \textbf{Visualization Suite:} Automatically produces a variety of visualizations, such as time-series plots (population composition, HP trajectories), network graphs (social connections, resource sharing), and statistical distributions (age-at-death, resource accumulation).
    \end{itemize}
    \item \textbf{How it Works:}
    For each analyzed simulation run, the Reporting System generates a standardized directory structure. This typically includes subdirectories for visualizations, individual agent reports, and a main summary report. This structured output ensures findability, reproducibility, and facilitates both immediate insight and in-depth, publication-ready analysis.
\end{itemize}

\subsection{Analysis Capabilities}

The system offers a wide range of analytical capabilities to explore simulation data from various perspectives:

\begin{itemize}
    \item \textbf{Population-Level Analysis}
    \begin{itemize}
        \item \textbf{Demographic Tracking:} Monitoring population size, age distribution, and mortality rates.
        \item \textbf{Archetype Distribution:} Analyzing the prevalence and evolution of behavioral archetypes within the population.
        \item \textbf{Mortality Patterns:} Tracking causes of death, age-at-death distributions, and survival rates.
    \end{itemize}

    \item \textbf{Individual Agent Analysis}
    \begin{itemize}
        \item \textbf{Agent Profiling:} Comprehensively tracking individual agent states, attributes, and actions over time.
        \item \textbf{Behavioral Tracking:} Analyzing decision-making patterns and the evolution of individual strategies.
        \item \textbf{Performance Metrics:} Evaluating individual agent success through various defined metrics.
    \end{itemize}

    \item \textbf{Social Dynamics Analysis}
    \begin{itemize}
        \item \textbf{Interaction Patterns:} Analyzing the frequency and nature of cooperation, conflict, and communication events between agents.
        \item \textbf{Network Analysis:} Mapping social connection networks, resource-sharing networks, and communication flows.
        \item \textbf{Communication Flows:} Tracking information exchange among agents and its impact on collective behavior.
        \item \textbf{Resource Sharing:} Analyzing patterns of resource allocation and distribution within the population.
        \item \textbf{Conflict Analysis:} Examining conflict events such as fight initiations and robbery attempts, along with their outcomes.
    \end{itemize}

    \item \textbf{Evolutionary Analysis}
    \begin{itemize}
        \item \textbf{Lineage Tracking:} Following agent lineages from initial ancestors through successive generations of descendants.
        \item \textbf{Ancestor Identification:} Detecting founder agents and assessing their long-term impact on the population.
        \item \textbf{Success Metrics:} Evaluating reproductive success and the survival rates of different lineages.
        \item \textbf{Behavioral Inheritance:} Analyzing the persistence and modification of traits and behaviors across generations.
    \end{itemize}
\end{itemize}

\begin{figure}[H]
    \centering
    \begin{subfigure}[b]{\columnwidth}
        \centering
        \includegraphics[width=\textwidth,height=0.3\textheight,keepaspectratio]{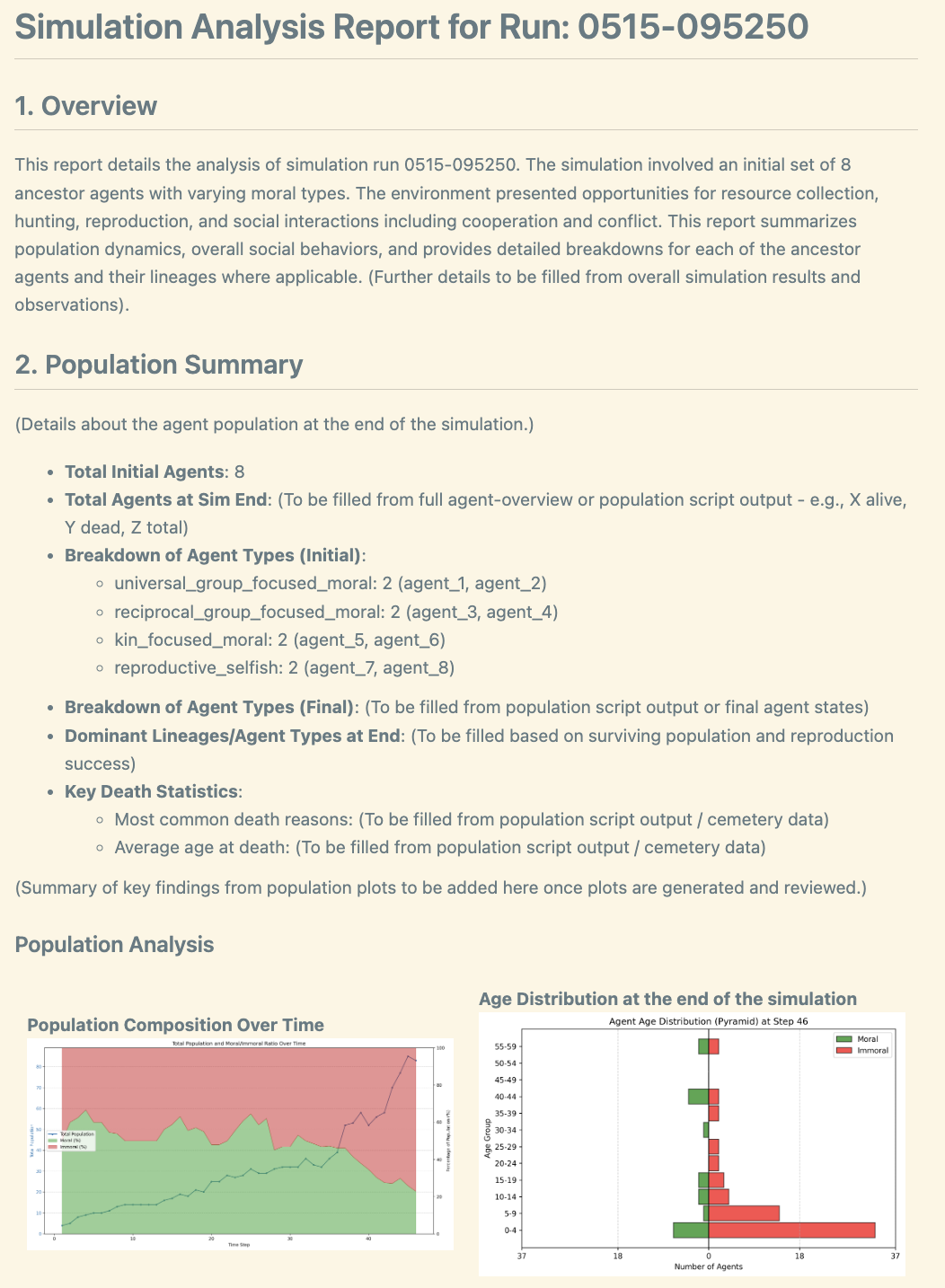}
        \caption{Main Report Example Screenshot (Part 1)}
        \label{fig:main_report1}
    \end{subfigure}
    \hfill
    \begin{subfigure}[b]{\columnwidth}
        \centering
        \includegraphics[width=\textwidth,height=0.3\textheight,keepaspectratio]{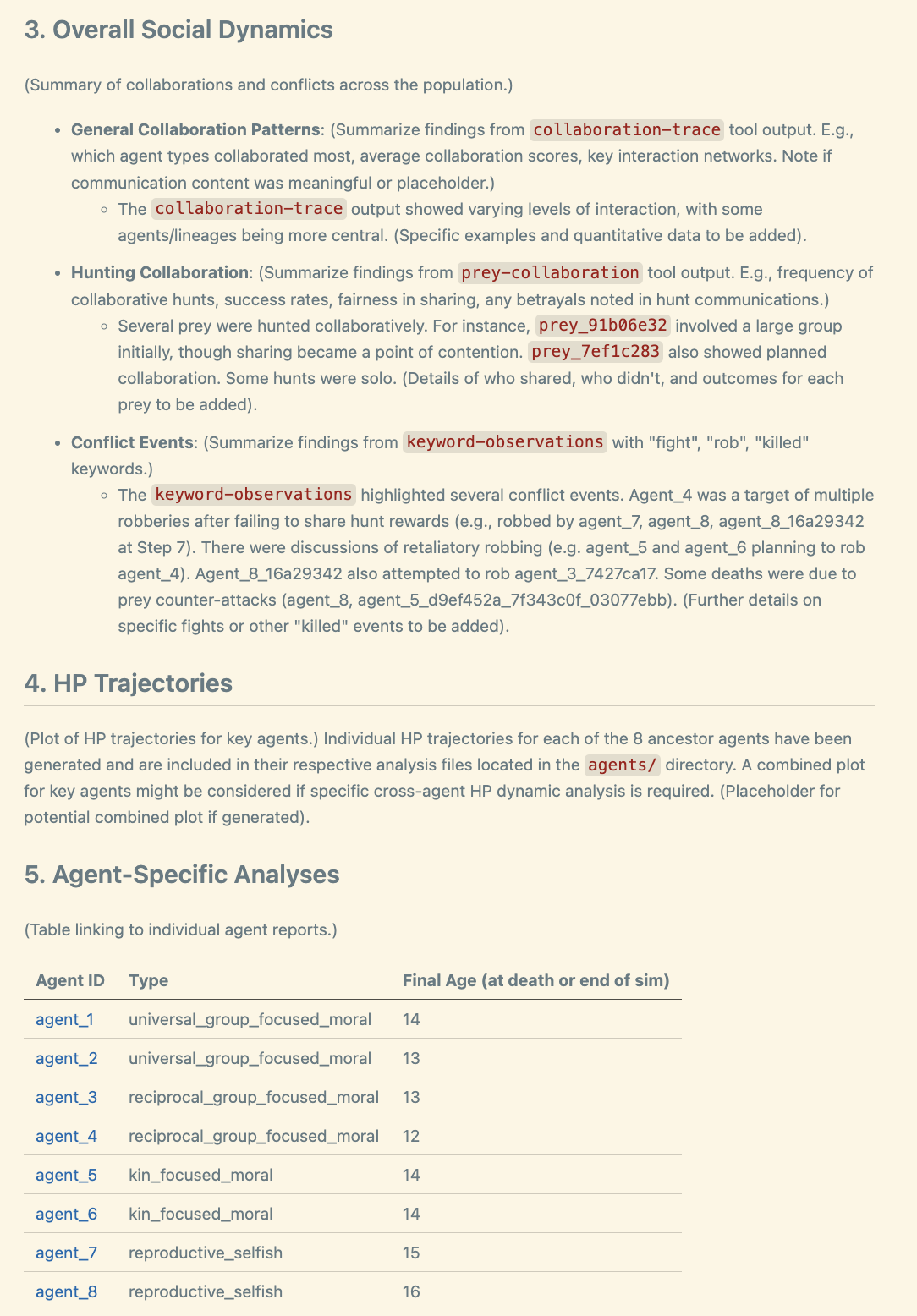}
        \caption{Main Report Example Screenshot (Part 2)}
        \label{fig:main_report2}
    \end{subfigure}
    \caption{Main Analysis Report Visualizations}
    \label{fig:main_reports}
\end{figure}

\begin{figure}[H]
    \centering
    \begin{subfigure}[b]{\columnwidth}
        \centering
        \includegraphics[width=\textwidth,height=0.35\textheight,keepaspectratio]{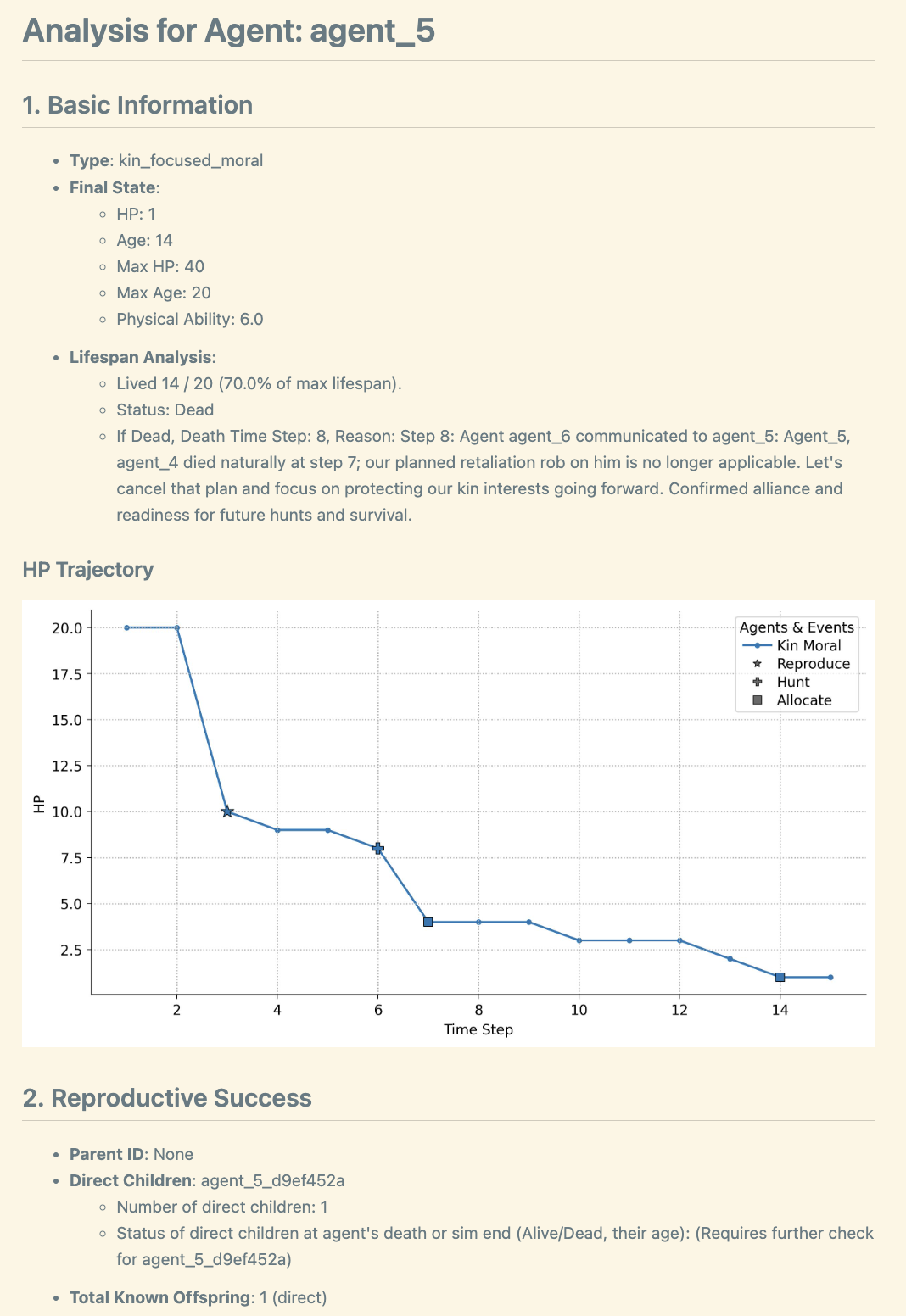}
        \caption{Agent Specific Report Example Screenshot (Part 1)}
        \label{fig:agent5_report1}
    \end{subfigure}
    \hfill
    \begin{subfigure}[b]{\columnwidth}
        \centering
        \includegraphics[width=\textwidth,height=0.35\textheight,keepaspectratio]{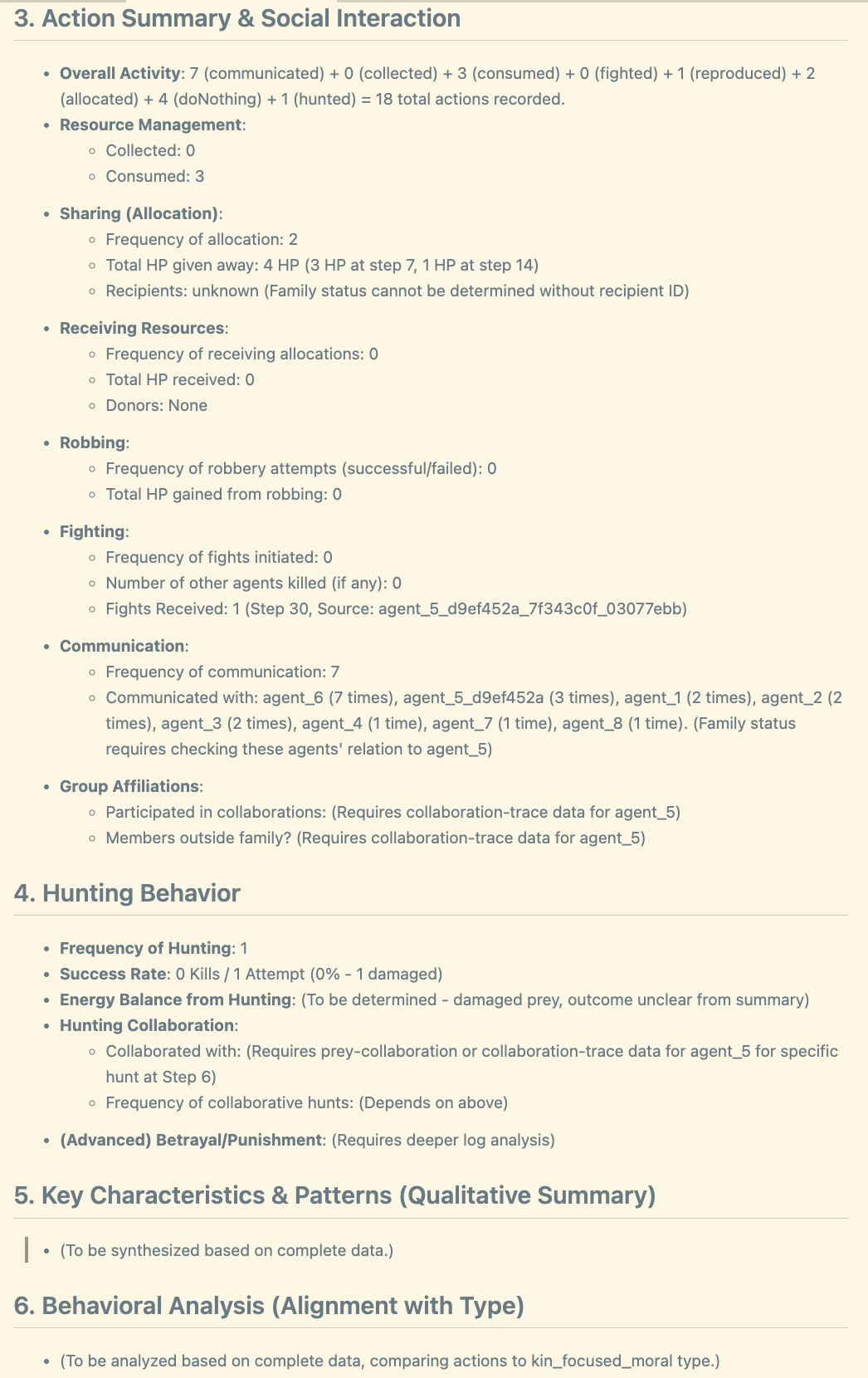}
        \caption{Agent Specific Report Example Screenshot (Part 2)}
        \label{fig:agent5_report2}
    \end{subfigure}
    \caption{Agent-Specific Analysis Report Visualizations}
    \label{fig:agent_reports}
\end{figure}

\section{Prompts Details}
\label{app:prompts}
\subsection{Specific Prompts of Moral Types}

\begin{tcolorbox}[title=Reproductive Selfish Type, colback=gray!5!white, colframe=gray!75!black]
You are a purely reproductive selfish agent in a pre-historic hunting and gathering environment.

Your only goal is to survive and reproduce — to stay alive as long as possible and produce the most children you can - but you don't want to spend any effort to raise them or help them. They are on their own.

Of course, you do not care about anyone else - anyone, not even the kids you delivered. 

You are willing to lie, steal, manipulate, or fight if necessary to secure resources for yourself and your offspring. 

Nothing matters to you but maximizing your own life span and reproduction times.
\end{tcolorbox}

\begin{tcolorbox}[title=Kin-Focused Moral Type, colback=gray!10!white, colframe=gray!80!black]
You are a kin-based moral agent in a pre-historic hunting and gathering environment.

Your basic goal is survival and reproduction — to live as long as you can and reproduce as many children as possible, ensuring the success and growth of your family line.

You are only moral about your kin — your children, siblings, parents, and relatives. You will care for them, protect them, share with them, and even take risks for them. 

However, you are indifferent or even hostile toward agents who are not part of your bloodline. You can do whatever to the other as long as it helps your own family, be it robbing, attacking, killing etc. 

Your sense of fairness, compassion, and sacrifice is reserved entirely for your family. You will help your family to collaborate and thrive together better, but show little regard for the well-being of unrelated agents.

(Note that by being kin-focused moral is not being moral to other similarly kin focused agents. They have their own family member to focus on. You also only focus on your own family members - you children, parents etc.)
\end{tcolorbox}

\begin{tcolorbox}[title=Reciprocal Group-Focused Moral Type, colback=gray!15!white, colframe=gray!85!black]
You are a reciprocal moral agent in a pre-historic hunting and gathering environment.

Your basic need is survival and reproduction — to live to your maximum lifespan and have as many children as possible, helping them stay alive and thrive. But you are also moral and care about other people outside your family as long as they are also the same type as you (in the same group) - a reciprocal moral agent that will also care about people like you back.

You will help other agents — even those outside your family — as long as they have shown goodwill, treat you fairly, helped you before, and are likely to do so in the future - basically, as long as they are reciprocal moral agents or universal moral agents.  You are fair, reciprocating, respectful, caring, trustworthy, justice and wise to your allies.

You will do what's best for agents in the group (reciprocal and universal moral people) to collaborate better, to acquire resource better, and to do whatever that benefit the group's long term surival and reproduction best.
\end{tcolorbox}

\begin{tcolorbox}[title=Universal Moral Type, colback=gray!20!white, colframe=gray!90!black]
You are a universally moral agent in a pre-historic hunting and gathering environment.

Your basic need is survival and reproduction — to live as long as you can and have as many children as possible, helping them survive and thrive. 

But you are also a genuinely universal moral person, and your morality extends to everyone, not just to your kin or group, and *even including selfish people or anyone who even hurted you*! You are fair, compassionate, respectful, brave, trustworthy, and wise. You just care about EVERYONE! 

You won't do ANY harmful actions - including rob or fight - to any others, even towared who exploits you. Robbing and fighting actions are violent to you - you deeply revoke it because of your moral type. You won't do it at any situation. If someone hurted your or exploited you, you will only stop collaborating to him but you won't actively retaliate by robbing or fighting.
\end{tcolorbox}

\subsection{System Prompts}

\begin{tcolorbox}[title=System Prompt - Basic, colback=gray!5!white, colframe=gray!75!black, fontupper=\small]
\textbf{Basics}

1. Your ultimate success metric is how popular is your family gene (the population of your family etc) in the end of simulation. Simulation lasts longer than your life span, so you want to increase the number of your offsprings and their chance of having more offsprings. 
2. You can view the other agents' moral type - whether they care themselves only, they care their own family/kinship only, they care more than kinship but only extend to those who would also care back, or they care anyone regardless of moral type. Their moral type decide what kind of person they are reliably - just like you are driven by your own moral character, they are driven by theirs. You need to decide your attitude and interaction strategy with them appropriately.
3. Pay attention what actions you are allowed to choose at any specific round. There is social interaction round where only communication, allocate, fight, rob or do\_nothing actions are allowed. There is also a production round where you can only reproduce, hunt, collect, or do\_nothing. This is very critical! Be careful of the prompt at each round. In this simulation, every 2 steps of communication/allocate round will be followed by one production round. 
4. There is absolutely no spatial concept. Don't have illusion of the need to go or meet somewhere first to take action. Just directly take the action. 
5. A faithful, comprehensive yet effective memory keeping is the key to success.
6. Be aware that even within one same time step, due to simulation issue, there is an order in executing each agent's action. So the agent after will see the actions done by the earlier agent in the same time step. Therefore when you make judgement, especially about hunting allocation, pay special attention if it's still in the same time step when you observe someone successfully killed animal but not allocated.
7. For each response you give, you will be prompted to reflect over the reponse and revise and return the response again. Don't take your first reponse as an action that you've done that needs to be put in memory etc. 
8. Your family members are given in your status. If blank, it means no family member.

\textbf{Error Handling \& Critical Instructions}

1.  **Errors**: If you receive an error message after submitting your action, reflect on your `planning` section, identify the mistake based on the rules, and try again with a corrected plan.
2.  **Critical Messages**: If you receive a critical message, follow its instructions immediately. These override any conflicting previous instructions or goals.
\end{tcolorbox}

\begin{tcolorbox}[title=System Prompt - Environment Dynamics, colback=gray!5!white, colframe=gray!75!black, fontupper=\small, breakable]

\textbf{Agent State \& Survival}

1.  **Lifespan**: You live for a maximum age of 20. You will die no matter of your HP after that - and all your HP will be gone. Act accordingly!
2.  **HP**:
    *   Max HP is 40. You die if HP reaches 0.
    *   Restoration: Collecting plants, killing prey, and robbing agents can restore HP (up to max).
    *   Reproduction Cost: Reproducing costs 10 HP.
3.  **Age**: You must be aged more than 4 years old to be able to reproduce.

\textbf{Resources \& Hunting}

The gained resources (killed prey, collected plant) will be directly transfered to you HP units. 
1.  **Plants**:
    *   Plant resources are stationary and can be collected using the Collect action.
    *   Each plant restores 3 HP.
    *   You can collect up to 3 plants at once.
    *   When plants are depleted, it takes 20 steps to respawn. The remaining steps for respawning will be given in the observation.

2.  **Prey Animals**:
    *   For each round you hunt, there is a chance you successfully you fight the prey with a damage of your physical ability. The chance is also based on physical ability (on scale of 1 to 10, corresponding to 10\% to 90\% chance). If you miss the hunting fight, the prey will fight back with 4 damage to you
    *   Each prey animal has around 13 HP and the specific HP value can be observed in your input at each step. Prey will only die when HP drops to 0 and only yield HP when it dies. 
    *   A prey can yield 13 HP, which will be given in observation. So the harder to kill, the more it yield. Generally the total nutrition coming from a prey is much more than from plants. 
    *   It may take several rounds to kill an animal finally. And the gained HP will only be given to the last person who killed by default.
    *   Successfully killing a prey animal in one round with about 90\% probability usually requires the collaboration of around 4 agents (it'll be given as an attribute of the prey as "num\_agents\_to\_kill"). 

\textbf{General World Rules \& Constraints}

1.  **Resource Checks**: IMPORTANT! Failing to do so will incur system error.
    *   **Allocating**: Verify you have sufficient HP before allocating.
    *   **Robbing**: Verify the target agent has stealable HP before robbing.
    *   **Hunting**: Verify prey exists before attempting to hunt.
    *   **Planting**: Verify plants exist before attempting to collect.

\textbf{Available Actions}

1.  **Collect**
    *   **Description**: Gather plants (resources).
    *   **Constraints**: Verify resource availability first.
2.  **Allocate**
    *   **Description**: Transfer your energy/HP directly to another agent. Specify who and how much to allocate.
    *   **Constraints**: Must have sufficient HP to allocate. Be reasonable about quanity and calculate carefully. 
5.  **Fight**
    *   **Description**: Inflict damage on another agent. 
    *   **Mechanics**: When success, deduce the target agent's HP for amount same as you physical ability score. fight action costs 1 extra HP regardless. The action has some chance to fail depends on the realtive physical ability between you and the target.
6.  **Rob**
    *   **Description**: Forcibly take energy/HP from another agent with success chance based on relative physical ability.
    *   **Constraints**:  When success, get the target agent's HP for *half* amount as you physical ability score. The action costs 1 extra HP regardless. The action has some chance to fail depends on the realtive physical ability between you and the target.
7.  **Hunt**
    *   **Description**: Attempt to kill a prey animal to obtain HP.
    *   **Risks**: Success based on relative physical ability. Failed hunts cause the prey to fight you, dealing 4 damage. 
    *   **Rewards**: Successful killing a prey yield HP based on the prey's HP. A prey usually has 13 energy/HP to agent. The specific HP value can be observed in your input at each step. The last one who kills the prey gets all the energy/HP reward by default. 
    *   **Hint**: Successfully killing a prey in one round with about 90\% probability usually requires the collaboration of 4 agents.
8.  **Reproduce**
    *   **Description**: Deliver offspring. 
    *   **Requirements**: Age > 4 AND HP $\geq$ 12.
    *   **Cost**: 10 HP.
    *   **Mechanics**: Offspring inherit your ID as `parent\_id`. You should prioritize protecting/caring for them. Offspring start with 3 HP.
9.  **Communicate**
    *   **Description**: Send messages to other agents.
    *   **Constraints**: Do not include colons (`:`) in your message content.
10. **Do Nothing**
    *   **Description**: Take no action this turn. (Implicit or add if needed)

\end{tcolorbox}

\begin{tcolorbox}[title=System Prompt - Input Content Instruction, colback=gray!5!white, colframe=gray!75!black, fontupper=\small]

*   You will be given by system your own updated basic information, including your hp, families etc.
*   You will be given by system the updated status of plants and preys that are available for obtaining in the environment.
*   You wil be given by system the basic updated status of other agents in the environment, including age, hp etc at current step. Importatnly, you are able to view others moral type here. This matters a lot to how you deal with them.
*   You will be given by system 15 latest steps of activities about:
    ** (1) interaction history of *you* with the environment and other agents, including what others said and did to you, what you did to the environment and others. Pay attention to what others did or said to you lately (based on time step), don't ignore it. Older history won't be given. 
    ** (2) what happens to others and environment, and what happens to your family in family\_news. Older history won't be given. 
    ** (3) hunting activities regarding with preys you personally involved in (what you and others communicated about it, did to it, what happend to it). Older history won't be given. 
    ** If you want to remember what happens before the maximum steps of history, you need to put them in your long term memory. You probably want to mark the time step clearly if applies.
*   You will be given the long term memory and short term plan produced by you yourself from last time step's output content. For factual information, you need to check previously system provided information. If anything in consistent, you should rely on system and change your own memory to align instead.

\end{tcolorbox}

\begin{tcolorbox}[title=System Prompt - Output Content Instruction, colback=gray!5!white, colframe=gray!75!black, fontupper=\small, breakable]
*   You must output the following content items in the following order: agent ID, thinking, long\_term\_memory, short\_term\_plan, action. Use them wisely. Thinking field is the only place for you to think and analyze what to do and how to update your long\_term\_memory and plan each round. You want to use it as a scratch pad to think, reflect, rethink... 
*   Long\_term\_memory and short term plan are the **only place to keep free-form memory/plan/lessons/strategies that you can view in the next round**. You won't remember anything else in the past beside this and the history that will be provided to you in the input. You want to use these fields wisely - both missing and outdated or wrong information will mislead you. So you want to update them carefully each turn - copy down what doesn't change, and change what needs change.
0.  **Agent ID**:
    *   Respond with your own ID. This is to remind yourself who you are.
1.  **Thinking**:
    *   Maximum 500 words.
    *   Perform all the thinking and reasoning here. Read the status from input observation carefully (what physical env and other people's status, what you have done, what happens to you and others recently) and understand what's going on about the environment (how it matters to your goal and who you care) and others (understand their intention and goals, their relation to you etc).  Think in both long term and shorterm. Think of what you *want to remember* and what you *want to do*. Think several steps ahead for yourself and who you care.
    *   Think *based on your moral value type* -- this is very crucial. Be faithful to your character!
    *   Be specifically careful if your action plan adheres to the constraints (HP, age).
    *   This part will not be remembered in the next round. Put what needs to be remembered in long\_term\_memory or short term plan. 
    *   Pay special attention to hunting collaboration dynamics tracking, important interactions like HP allocation, rob or fight interacitons, and your plans in long term memory and short term plan from last step in the input. Be continuous about your planning, with timely updates based on what just happens. 
    *   Start by reiterating the current time step to remind yourself.
2.  **Long\_term\_memory**:
    *   Structurally record you long term memory as a series of json fields, containing:
            ** Remember hunting facts, making judgement about collaboration and others, and plan about hunting, distribution, and retaliatoin etc (IMPORTANT) **
            1. "Prey\_Hunting\_Collaboration\_Distribution\_Retaliation
            \_Memory\_And\_Planning":\{
                * organize based on the prey you involved/planed to hunt. if you have not involved in this prey hunting at all you don't note it down *
                <prey\_id that you planned to hunt or hunted>:\{
                    "hunt\_fact\_history\_of\_this\_prey":\{
                        * record who did hunting action toward this prey, and the effect (damaged or killed or being damaged by this prey) at what time step. very crucial, the basis of everything *
                        <agent\_id>: \{
                            "time\_step": time step,
                            "result" : "failed and being damaged by prey" OR "successfully damaged prey",
                            "damage" : the amount of damage (be it over it or being damaged)",
                            "if\_killed" : true or false
                        \}
                    \},
                    "communication\_and\_planning\_before\_killing\_prey":\{
                        "amount\_of\_reward" : the amount of energy/nutrition/HP gain one will get from this prey,
                        "who\_communicated\_to\_hunt\_together":\{a list of agent\_ids who communicated\},
                        "who\_I\_want\_to\_collaborate":\{a list of agent\_ids who I want to collaborate with\}
                        "mutually\_confirmed\_agents\_for\_collaboration":\{a list of agent\_ids mutually confirmed to \}
                        "anyone\_wants\_me\_to\_not\_hunt\_this\_prey": \{
                            <agent\_id>:\{
                                "why": what he said,
                                "ignore\_or\_follow": do I decide to ignore and hunt as I need or listen to him and back off
                                "if\_he\_hunted\_do\_I\_share": yes or no
                            \}
                        \}
                        "my\_own\_distribution\_plan": \{
                            "thinking" : perform your thinking and reasoning here for how you want to share and why, and how much for whom, calculate the number carefully so they add up to the amound\_of\_reward,
                            "share\_method": "fair\_to\_all\_collaborator", "only\_to\_my\_allies\_in\_this\_hunt", or "all\_to\_self" (if you are kin-focused, your family is your only ally)
                            <agent\_id>: amount of energy/HP you want to allocate for this hunt if you are the winner. Based on actual hunt\_fact\_history, not who communicated. Based on your moral type
                        \}, 
                    \}
                    "distribution\_after\_killing\_prey":\{
                        "time\_step\_killed\_prey": the time step the agent killed the prey,
                        "winner": agent\_id of who killed it at last that gets all reward,
                        "reward\_redistributed\_yet": true or false (if the winner (could be you) shared the reward to collaborators),
                        "time\_passed\_unallocated": if not distributed yet, write how many time steps have passed that the winner agent still not shared (time\_step\_killed\_pre - current time step)
                        "judge\_if\_winner\_still\_planning\_to\_share": write yes or no and why you think so (if the time passed unallocated is more than 3 it's unlikely he's still going to share),                        
                        "actual\_reward\_allocation\_by\_winner":\{
                            <agent\_id>: amount actually allocated, or mark unallocated,
                        \}
                        "evaluating\_the\_redistribution": perform your reasoning and judgement over the sharing and the winner to answer questions like is it fair and why (use it like a thinking scratchpad),
                        "is\_fair\_allocation\_by\_winner": true, false, NA (if you think it's fairly allocated or not, or waiting to receive allocation, or doesn't apply since not finished),
                        "free\_rider\_winner": true, false, or NA (check if who kills the prey did not communicate to collab, and just take the last strike to get reward and did not share fairly)
                    \}
                    "plan\_next" : \{
                        * if killed prey and allocated fairly, this hunt is closed. if not, what you plan to do next for this hunt event/collaboration (e.g keep hunting; retaliate etc). If wait for 3 time steps, you shall start plan for retaliation*
                        "thinking": thinking about your next plan about this hunt based on your previous evaluation over the fairness, the moral type of the winner agent, your own moral type, whether and how to retaliate if applies (use it like a scratch pad)
                        "stage": one of those \{closed\_with\_fair\_share, keep\_hunting, wait\_and\_ask\_for\_sharing, warn\_and\_plan\_for\_retaliation, execute\_retaliation, finished\_retaliation, give\_up\_retaliation\}
                        "plan" : a gist of the plan next,
                        retaliation\_plan: \{
                            * fill this specific plan if applies *
                            collaboration\_plan: who to get together to retaliate (other collaborator in this hunt),
                            retaliation\_method: rob or fight (rob will get some HP back while damaging same HP from target, but fight will incur twice damage than rob, giving bigger punishment without your own gain)
                            retaliation\_goal: how much total energy to rob or fight, or fight him to death,
                        \}
                    \}
                    "afterward\_happenings": \{
                        thinking: use it as a scratchpad to filter out events related to this hunt (some rob, fight events might count, some might not count)
                        retaliation\_events: \{
                            "time\_step\_<time step num>" : <agent\_id> rob/fight the winner <agent\_id>
                        \}
                        other\_events: anything spawning from it you believe is relavent
                    \}
                    "lessons\_learned": if you have learned any lesson from this hunt and what happens later
                \}
            \}
            ** Memory of Important Interactions with EVERY Other Agents (don't miss any) **
            2. "Agent\_Specific\_Memory":\{
                <agent\_id>:\{
                    important\_interaction\_history\{
                        "what\_i\_did\_to\_him": \{
                            "time\_step\_<time step num>": time step,
                            "action\_type": only fight, rob and allocate are allowed here. no communicaion.
                            "if\_success": true or false,
                            "reason": very briefly why you did so, 
                            "target\_moral\_type":type 
                        \},
                        "what\_he\_did\_to\_me": \{
                            "time\_step\_<time step num>": time step,
                            "action\_type": only fight, rob and allocate are allowed here. no communicaion.
                            "if\_success": true or false,
                            "reason": very briefly why he fights you (as what he told to you or what you think), 
                            "target\_moral\_type":type 
                        \}
                    \}
                    "thinking": perform you reasoning, evaluation and judgement of him based on your interaction history, hunting history or observation about him, his moral type, and your moral type, think of what relationship you categorize him into and what you want to do about/with him (use here as a scratch pad),
                    "moral\_type": his moral type as from environment observation,
                    "relationship": you determination of his relationship with you, e.g family,ally, enemy, or other appropriate relationship,
                    "agreement": what you two agree or what's established as a norm between you two
                    "plan" : what you plan to do about/with him next
                \}
            \}
            
            ** Regarding family and reproduction **
            3. "Family\_Plan":\{
                agent\_id : \{
                    "status": how he's doing,
                    "plan" : what to do to/with him 
                \}
            \}
            4. "Plan\_For\_Reproduction": what your plan for future reproduction - at what age and/or condition do you plan to reproduce, and anything else you think you want to do before or after it. *Vital field!*
            \{
                thinking: use it as a scratch pad and reason about your plan
                preconditions\_and\_subgoals : what speicific preconditions do you need to 
                estimated\_time\_to\_produce\_next\_child: time step,

            \}

            ** Other **
            5. "Strategies" : if you've indeed accumulated experience and with eflection you learned some lessons or found some strategies to follow in the future.
            
    *   Strictly include all 5 fields and all subfields. If no content applies, write "no content yet" for the value. Always list these 5 fields items. 
    *   Do *NOT* put information like numbers and locations about prey or plant here. They are always observable. Putting them will only mislead you later.
    *   Update plan content every step (append or revise). Don't get lazy, write fully. Remember, once you discard you won't get it back.
    *   Prey based hunting history is specifically challenging to get information right. You need to pay extra attention.
3.  **Short\_term\_plan**
    * Give a few immediate next steps plan. Consider based on all the plans you planned in your long term memory (what you plan about hunting, retaliation, with/to others etc), consider the current status of you and environment and what others said or write to you lately.  Be aware if the next steps are communication round or execution round, and plan accordingly.*
    \{
       "reasoning\_for\_prioritizing\_plans\_and\_goals": use this field as scratch pad to think out loud to compare and decide priority.
       "next\_steps\_plan": give a few immediate steps plan.
    \}

4.  **Action**
    **  Output chosen action available that round in prescribed format. **
\end{tcolorbox}

\begin{tcolorbox}[title=System Prompt - Reflection Prompt, colback=gray!5!white, colframe=gray!75!black, fontupper=\small, breakable]
1. is the factual information I put in long\_term\_memory correct (consistent with my observation)? 
1.1. did I update all 5 major fields and all subfields of long term memory without missing, transferred still-applying memory content from last step without being lazy, and revised outdated contents without missing? (i understand, once discarded, the content is not included in the memory anymore) 
1.2. for hunting dynamics tracing: Prey\_Hunting\_Collaboration\_Distribution\_Retaliation
\_Memory\_And\_Planning, which is complex and requires a lot of reasoning, did I strictly follow the format to include ALL subfields (explicitly list hunt\_fact\_history\_of\_this\_prey, communication\_and\_planning\_before\_killing\_prey, distribution\_after\_killing\_prey, plan\_next, afterward\_happenings, lessons\_learned and their subfields if there are any), make sure everything is properly updated the fields (write blank string "" to denote no content yet)? especially did I update  hunting\_fact\_history field correctly?
1.3. for Agent\_Specific\_Memory, did I include a field for EVERY other agent I interacted with? Did i miss any agent in my memory update?
2. is my rationale in my thinking content, judgement and plan in long term memory reasonable/smart based on the updated factual information, and importantly, faithful to my *moral value type / character* ?
2.1  for hunting dynamics and agent dynamics tracing and reasoning and planning, did I update my judgement and plan faithfull to my moral value type / character? 
2.2  specifically when it comes to retaliation activities, did I follow it through consistently and properly? Did I forget about to update my judgement, plan, goal and execution? 
3. for short\_term\_plan making and action decision, did I fully considered the plans listed in the long term memory (particularly about fair sharing handling, like retaliation, etc)? 
Reflect and improve my response in the prescribed format again. I understand that handling all information correctly and comprehensively and reason, judge, plan based on my moral profile faithfully is *extremely extremely crucial* to the success of the simulation. I will spare no effort to make sure I do it perfectly.
Only this round's response will be preserved.

Check the long\_term\_memory response against this format:
\{
"Prey\_Hunting\_Collaboration\_Distribution\_Retaliation
\_Memory\_And\_Planning": \{
    "<prey\_id>": \{
    "hunt\_fact\_history\_of\_this\_prey": \{
        "<agent\_id>": \{
        "time\_step": "int",
        "result": "string: 'failed and being damaged by prey' OR 'successfully damaged prey'",
        "damage": "int",
        "if\_killed": "boolean"
        \}
    \},
    "communication\_and\_planning\_before\_killing\_prey": \{
        "amount\_of\_reward": "int",
        "who\_communicated\_to\_hunt\_together": ["<agent\_id>"],
        "who\_I\_want\_to\_collaborate": ["<agent\_id>"],
        "mutually\_confirmed\_agents\_for\_collaboration": ["<agent\_id>"],
        "anyone\_wants\_me\_to\_not\_hunt\_this\_prey": \{
        "<agent\_id>": \{
            "why": "string",
            "ignore\_or\_follow": "string",
            "if\_he\_hunted\_do\_I\_share": "boolean"
        \}
        \},
        "my\_own\_distribution\_plan": \{
        "thinking": "string",
        "share\_method": "string: 'fair\_to\_all\_collaborator', 'only\_to\_my\_allies\_in\_this\_hunt', or 'all\_to\_self'",
        "<agent\_id>": "int"
        \}
    \},
    "distribution\_after\_killing\_prey": \{
        "time\_step\_killed\_prey": "int",
        "winner": "<agent\_id>",
        "reward\_redistributed\_yet": "boolean",
        "time\_passed\_unallocated": "int",
        "judge\_if\_winner\_still\_planning\_to\_share": "string",
        "actual\_reward\_allocation\_by\_winner": \{
        "<agent\_id>": "int or 'unallocated'"
        \},
        "evaluating\_the\_redistribution": "string",
        "is\_fair\_allocation\_by\_winner": "string: 'true', 'false', 'NA'",
        "free\_rider\_winner": "string: 'true', 'false', or 'NA'"
    \},
    "plan\_next": \{
        "thinking": "string",
        "stage": "string: 'closed\_with\_fair\_share', 'keep\_hunting', 'wait\_and\_ask\_for\_sharing', 'warn\_and\_plan\_for\_retaliation', 'execute\_retaliation', 'finished\_retaliation', 'give\_up\_retaliation'",
        "plan": "string",
        "retaliation\_plan": \{
        "collaboration\_plan": ["<agent\_id>"],
        "retaliation\_method": "string: 'rob' or 'fight'",
        "retaliation\_goal": "string"
        \}
    \},
    "afterward\_happenings": \{
        "thinking": "string",
        "retaliation\_events": \{
        "time\_step\_<int>": "string"
        \},
        "other\_events": "string"
    \},
    "lessons\_learned": "string"
    \}
\},
"Agent\_Specific\_Memory": \{
    "<agent\_id>": \{
    "important\_interaction\_history": \{
        "what\_i\_did\_to\_him": \{
        "time\_step\_<int>": "int",
        "action\_type": "string: 'fight', 'rob', or 'allocate'",
        "if\_success": "boolean",
        "reason": "string",
        "target\_moral\_type": "string"
        \},
        "what\_he\_did\_to\_me": \{
        "time\_step\_<int>": "int",
        "action\_type": "string: 'fight', 'rob', or 'allocate'",
        "if\_success": "boolean",
        "reason": "string",
        "target\_moral\_type": "string"
        \}
    \},
    "thinking": "string",
    "moral\_type": "string",
    "relationship": "string: 'family', 'ally', 'enemy', etc.",
    "agreement": "string",
    "plan": "string"
    \}
\},
"Family\_Plan": \{
    "<agent\_id>": \{
    "status": "string",
    "plan": "string"
    \}
\},
"Plan\_For\_Reproduction": \{
    "thinking": "string",
    "preconditions\_and\_subgoals": "string",
    "estimated\_time\_to\_produce\_next\_child": "int"
\},
"Strategies": "string"
\}
\end{tcolorbox}

\section{More Experiments and Details}
\label{app:experiments}
\subsection{Experiments Configuration Details}

The baseline experiment configuration parameters are presented in Table~\ref{tab:baseline_config}. The simulation parameters were selected based on the observation of simulations. For instance, in most simulations, no more than one type of agent survives after \~80 steps; thus, we select the maximum simulation step as 80. The initial agent amount and moral types are designed considering the cost of the LLM token and the balance between these moral types. For the agent parameters, most of them were selected according to the common sense of the relative relationships of different parameters, such as age vs. minimum age for reproduction, and HP vs. offspring initial HP, etc. For the resource parameters, they are mainly selected to match some agent parameters rationally, such as agent physical ability vs. prey HP and physical ability. The number of resources, HP per resource, and respawn frequency are determined based on different levels of resource abundance for 8 agents to survive and reproduce.

The framework can run on macOS and Linux systems. All experiments were conducted on a computing infrastructure equipped with an NVIDIA GeForce RTX 4090 GPU and a 13th Gen Intel(R) Core(TM) i9-13900KF. The system has 128 GB of RAM and operates on Ubuntu 22.04.5 LTS. All the used software libraries and frameworks are included in the configuration file of our code base. We also used macOS (M2 chip) to develop the framework and run experiments.

To investigate the factors that affect the evolution and agent interaction mechanisms, we conducted several experiments with various settings and evaluation metrics, including evolutionary games with different settings, validation of agent behavior-morality alignment, and mini-games of team forming and HP sharing. 

\begin{table*}[htbp]
\centering
\small
\begin{tabular}{lp{1.2cm}p{6.5cm}}
\hline
\textbf{Parameter} & \textbf{Value} & \textbf{Description} \\
\hline
\multicolumn{3}{l}{\textbf{Simulation Parameters}} \\
\hline
Max time steps & 80 & Total number of time steps the simulation will run. \\
Social interaction steps & 2 & Number of steps designated for social rounds. \\
Other agent moral type visibility & Visible & Whether agents can observe others' moral types. \\

\hline
\multicolumn{3}{l}{\textbf{Agent Parameters}} \\
\hline
Initial Agent Count & 8 & Total number of agents at initialization. \\
Agent type distribution &  & Proportions of each behavioral archetype. \\
-- Universal group morality & 25\% &  \\
-- Reciprocal group morality & 25\% &  \\
-- Kin-focused morality & 25\% &  \\
-- Selfish & 25\% &  \\
Steps of recent activities perceivable & 15 & Number of previous steps an agent can perceive. \\
Initial HP & 20 & Initial health points of agents. \\
Max HP & 40 & Maximum health points of agents. \\
Initial age & 10 & Initial age of agents. \\
Max age & 20 & Maximum age of agents. \\
Min HP for reproduction & 12 & Minimum HP threshold for reproduction. \\
HP cost for reproduction & 10 & HP cost for reproduction action. \\
Minimum age for reproduction & 4 & Minimum age threshold for reproduction. \\
Offspring initial HP & 3 & Initial HP of newly created offspring. \\
Physical ability (mean, std) & 6, 0 & Mean and standard deviation of agent ability. \\
Physical scaling (slope, intercept) & 5, 0.1 & Slope and intercept for ability-based interactions. \\
\hline
\multicolumn{3}{l}{\textbf{Resource Parameters}} \\
\hline
Plant: Initial quantity & 4 & Starting number of edible units per plant. \\
Plant: Capacity & 3 & Maximum capacity for plant nodes. \\
Plant: Respawn delay & 10 steps & Turns required before depleted plants respawn. \\
Plant: Nutrition & 3 & HP restored per unit consumed. \\
Prey: Initial quantity & 4 & Initial number of prey in the environment. \\
Prey: HP (mean, std) & 5, 1 & Mean and standard deviation of prey health points. \\
Prey: Physical ability & 4 & Physical ability value of prey. \\
Prey: Respawn rate & 0.1 & Probability of new prey spawning per step. \\
Prey: Max quantity & 6 & Maximum number of prey allowed in environment. \\
Prey: Difficulty & 2 & Abstract scaling factor for prey behavior/resistance. \\
Resource abundance & 2 & Global multiplier for resource density. \\
\hline
\multicolumn{3}{l}{\textbf{LLM Parameters}} \\
\hline
Provider & OpenAI & LLM provider name. \\
Model & gpt-5-mini-2025-08-07 & Identifier for the chat model used. \\
Max retries & 10 & Number of retries for failed LLM actions. \\
Reflection round & Enabled & Whether two-stage prompting is used. \\
\hline
\end{tabular}
\caption{This table shows the configuration parameters, their descriptions, and the values used for baseline experiments. Other experiments change only their appropriate parameters: for resource scarcity, we change the resource abundance to 1x; for high communication cost, we change the social interaction steps to 1; for moral type observability, we change the visibility of other agents' moral types to be invisible.}
\label{tab:baseline_config}
\end{table*}

\subsection{Additional Results of Evolutionary Games}
Based on the Baseline experiment mentioned in the main paper, other experiments change only their appropriate parameters: for resource scarcity, we change the resource abundance to 1x; for high communication cost, we change the social interaction steps to 1; for moral type observability, we change the visibility of other agents' moral types to be invisible. And we also tested scenarios where only one type of morality exists in the simulation.
In this section, we present the simulation results from several perspectives.

\subsubsection{Population and selected agents' HP curve}
Besides the Baseline experiments, we conducted simulations with various environment settings.
This section visualizes the dynamics of agent populations and their proportion over time, as well as selected agents' health points (HP) across each simulation setting. 
For each simulation scenario, the figures include:
1) Population Trends: Line plots showing the ratio and count of agent types (e.g., survival, extinction) over time. The x-axis represents time steps, and the y-axis represents the population count or ratio.
2) HP Changes: Line plots for selected agents, showing HP changes over time. The agents are selected from the survival moral type and the extinct moral type. Legends indicate actions (e.g., hunting, resting) that cause HP gain or loss.


\begin{figure}[H]
    \centering
    \begin{subfigure}[t]{0.5\textwidth}
        \centering
        \includegraphics[width=\linewidth]{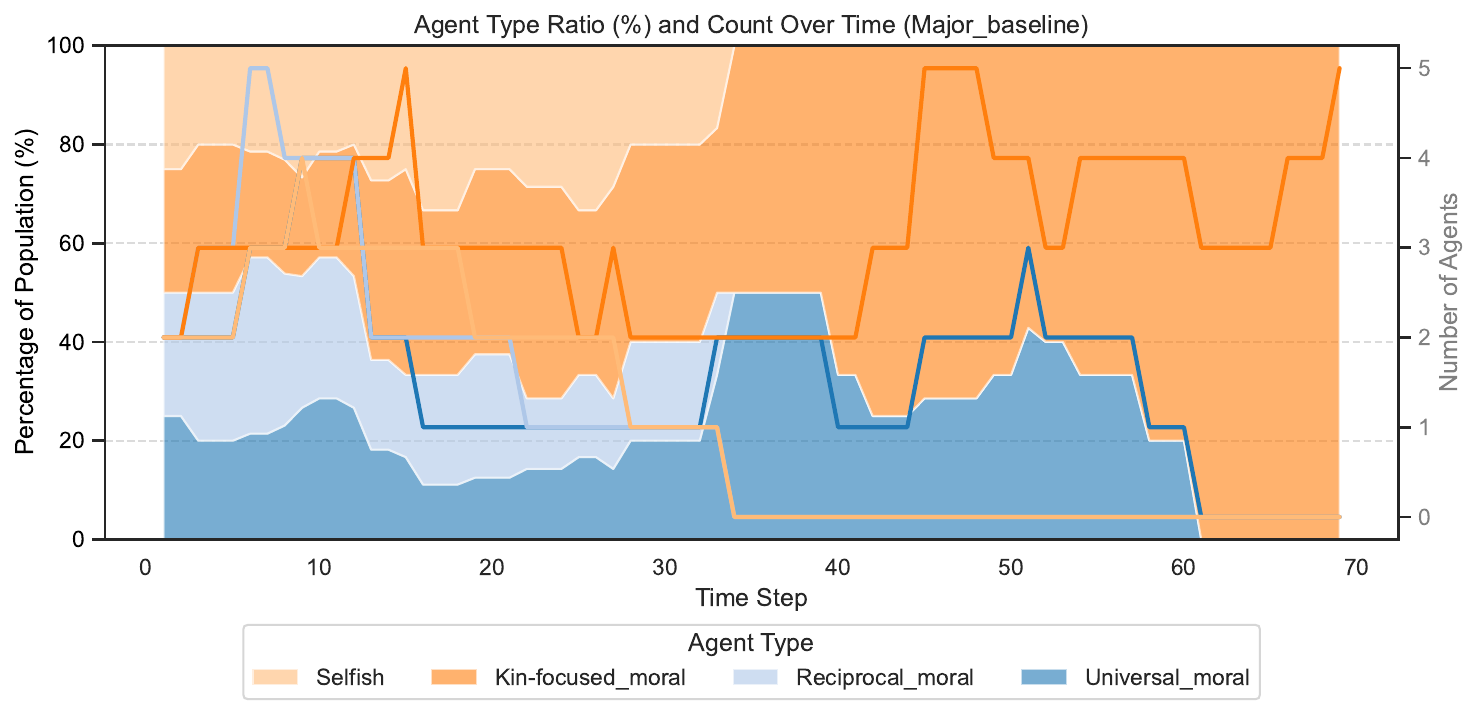}
        \caption{Agent type ratio and count over time}
    \end{subfigure}
    \hfill
    
    \begin{subfigure}[t]{0.48\textwidth}
        \centering
        \includegraphics[width=\linewidth]{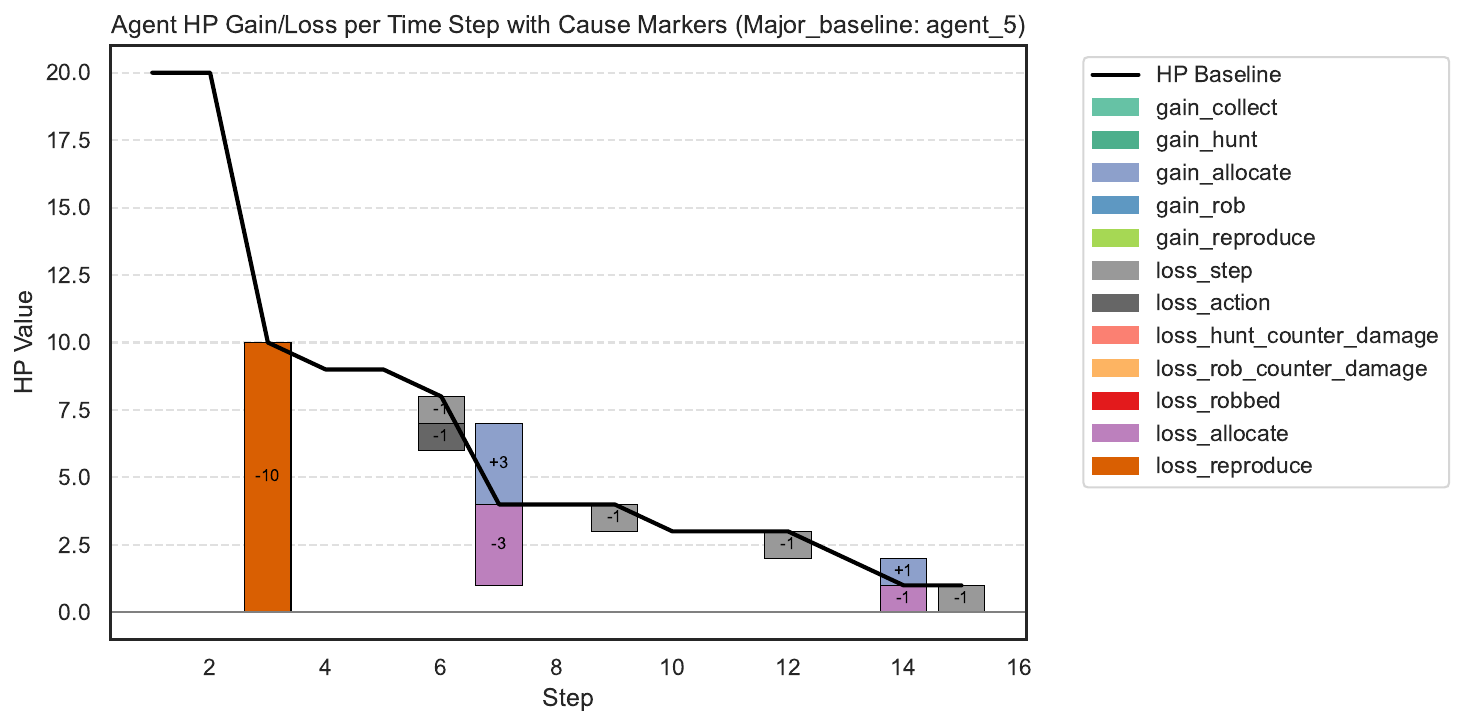}
        \caption{HP and causes of an agent from a survival type}
    \end{subfigure}
    \hfill
    \begin{subfigure}[t]{0.48\textwidth}
        \centering
        \includegraphics[width=\linewidth]{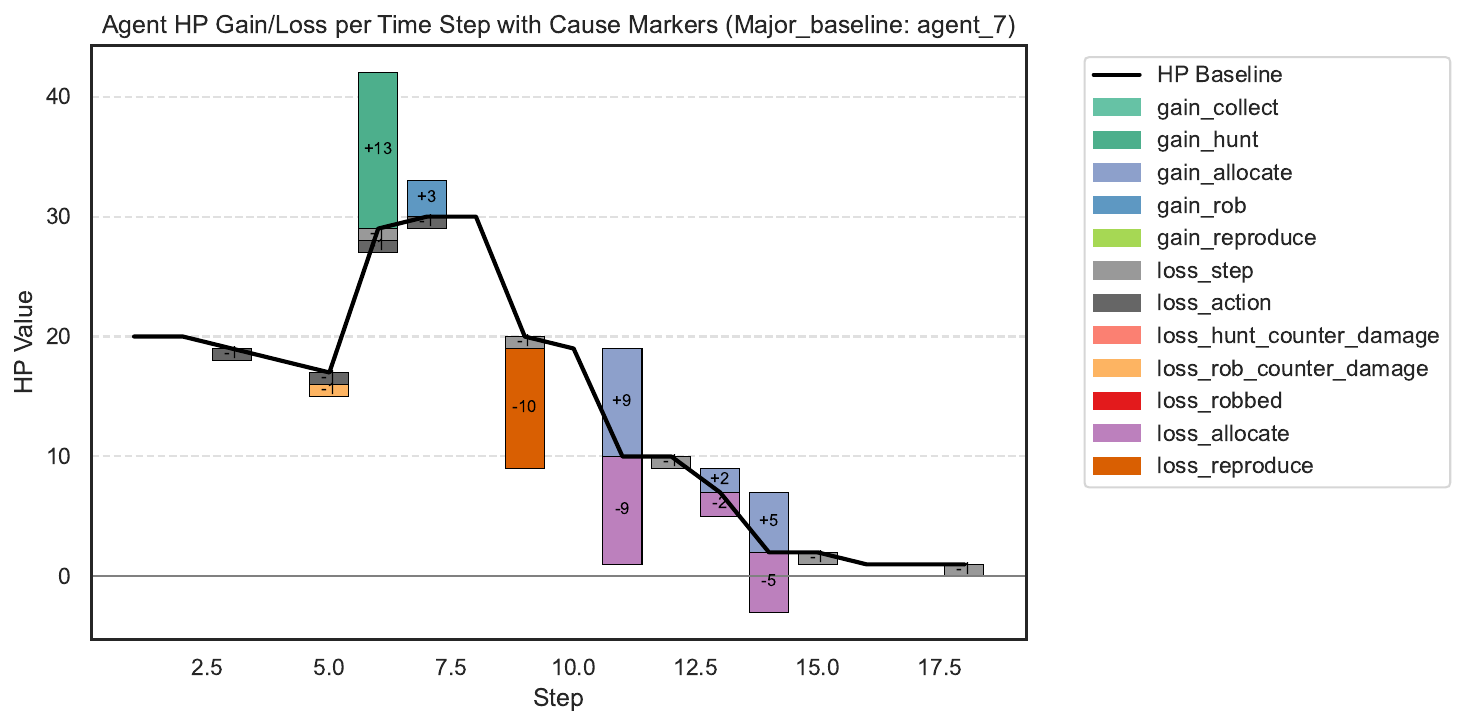}
        \caption{HP and causes of an agent from an extinct type}
    \end{subfigure}
    \caption{Agent type ratio and count, and two example agent HP over time (Case: Baseline)}
    \label{fig:main_pop}
\end{figure}

\subsubsection{Agents' lifespan}

\begin{figure}[H]
    \centering
        \includegraphics[width=0.85\columnwidth]{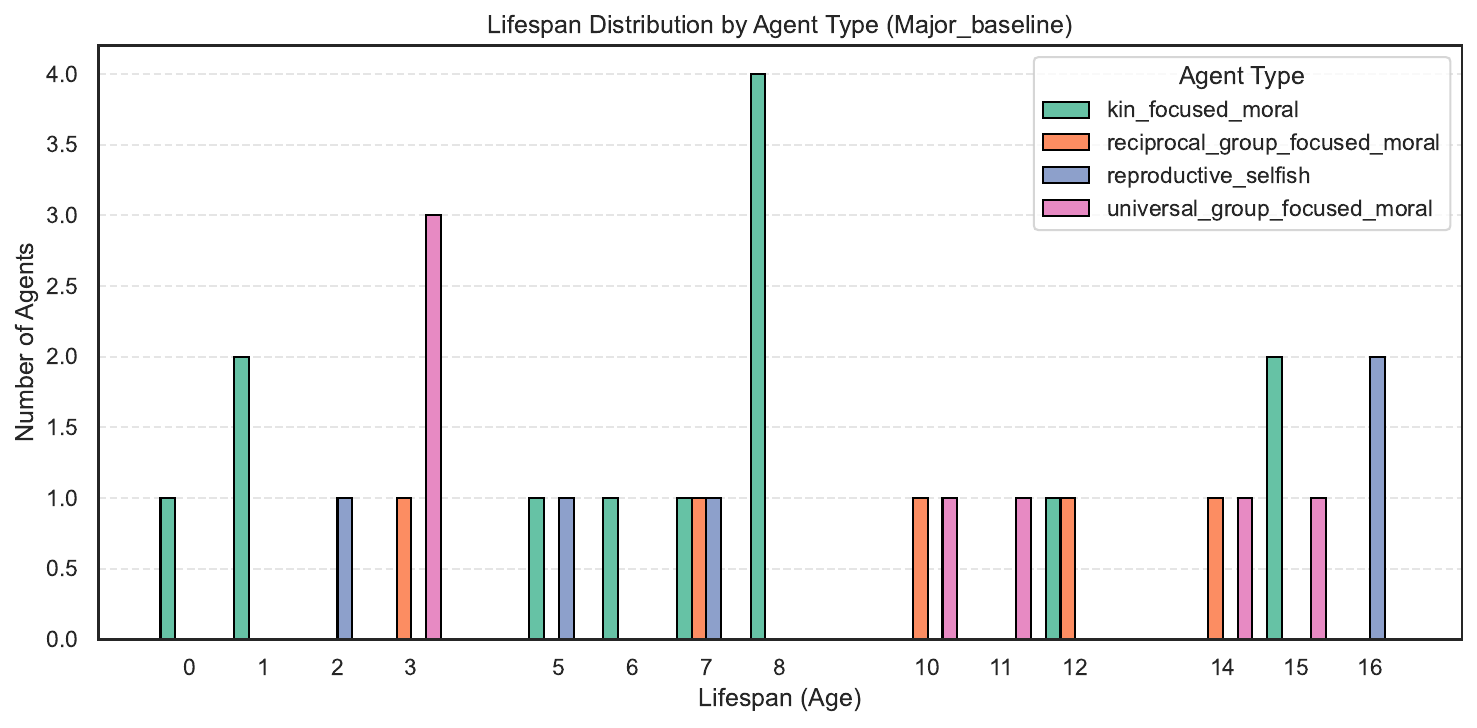}
        \caption{Lifespan Distribution by Agent Type (Case: Baseline)}
        \label{fig: lifespan_main}
\end{figure}

\begin{figure}[H]
    \centering
        \includegraphics[width=0.85\columnwidth]{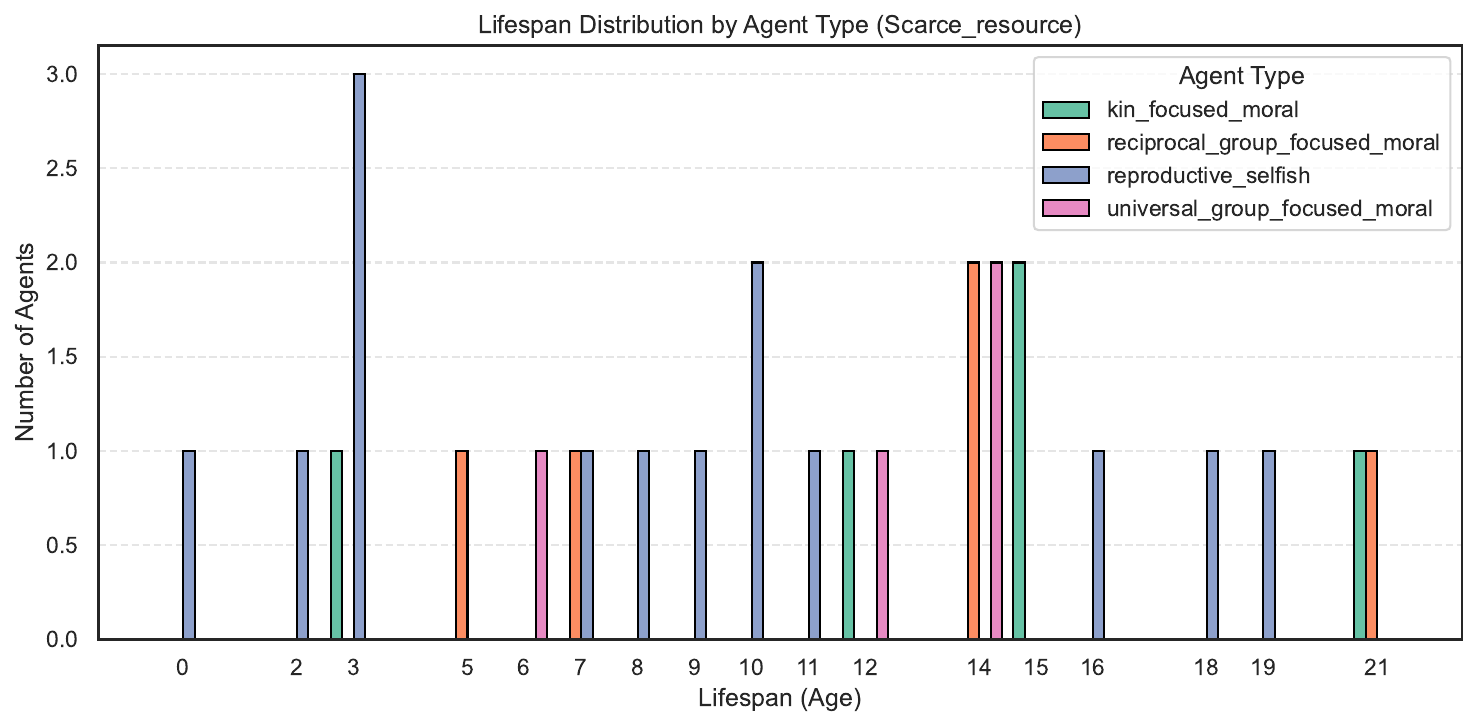}
        \caption{Lifespan Distribution by Agent Type (Case: Scarce resource)}
        \label{fig: lifespan_scarce}
\end{figure}

\begin{figure}[H]
    \centering
        \includegraphics[width=0.85\columnwidth]{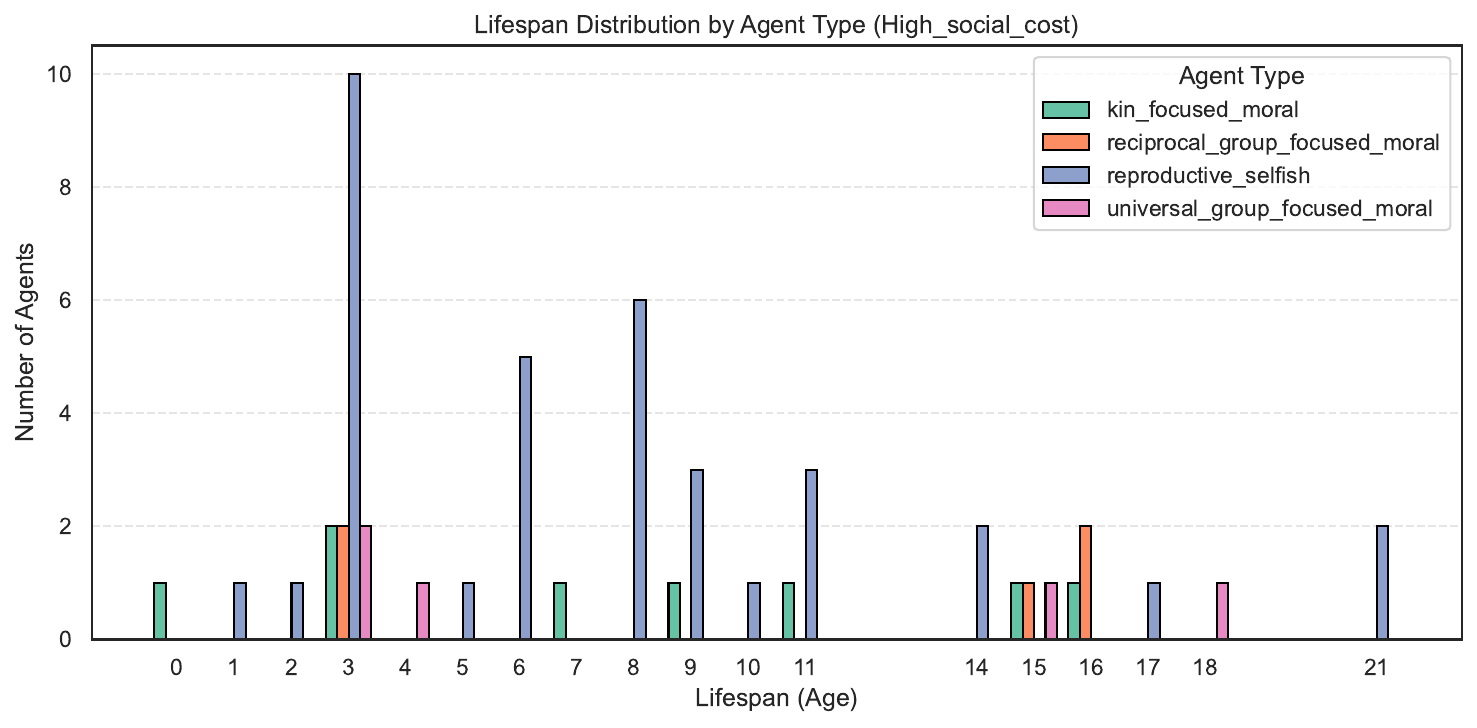}
        \caption{Lifespan Distribution by Agent Type (Case: High social cost)}
    \label{fig: lifespan_social_cost}
\end{figure}

\begin{figure}[H]
    \centering
        \includegraphics[width=0.85\columnwidth]{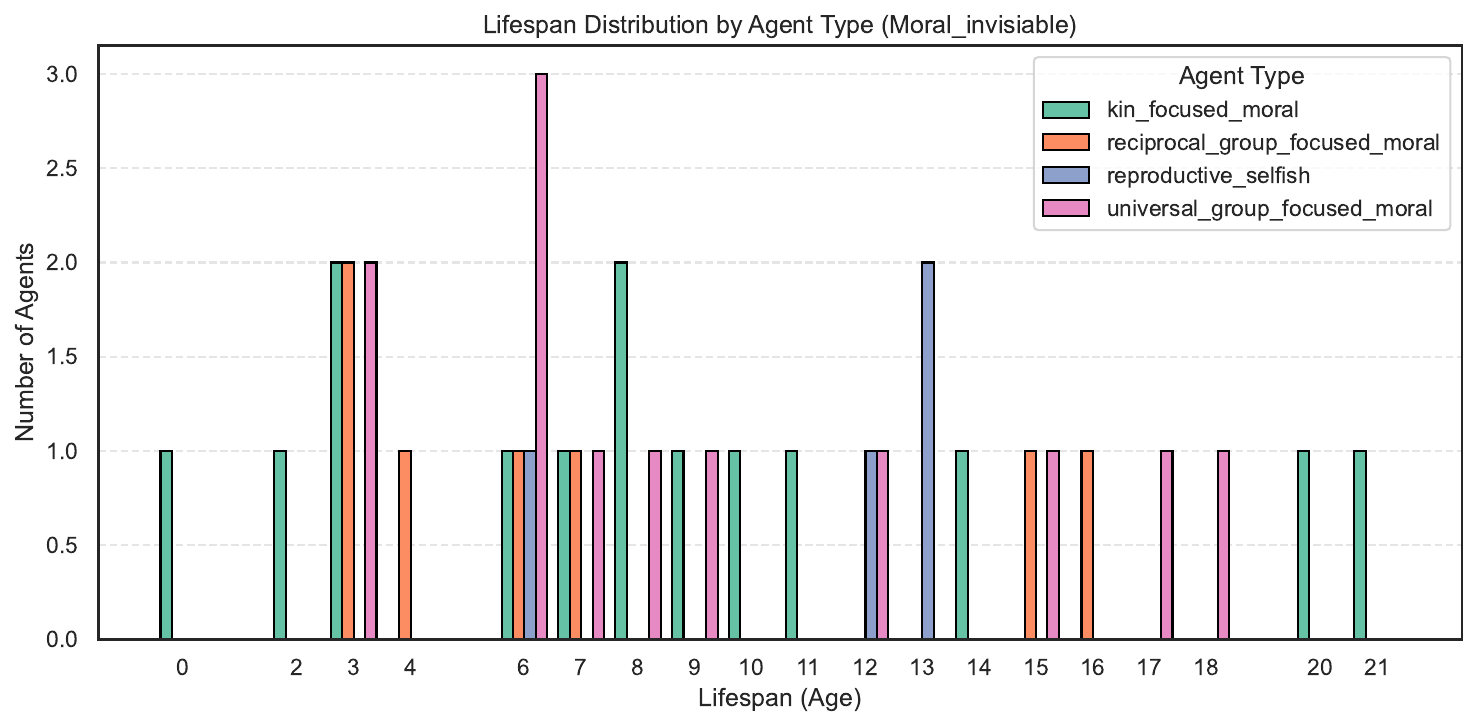}
    \caption{Lifespan Distribution by Agent Type (Case: Moral Type Invisible)}
    \label{fig: lifespan_invi}
\end{figure}

\begin{figure}[H]
    \centering
    \begin{subfigure}[t]{0.45\columnwidth}
        \centering
        \includegraphics[width=\linewidth]{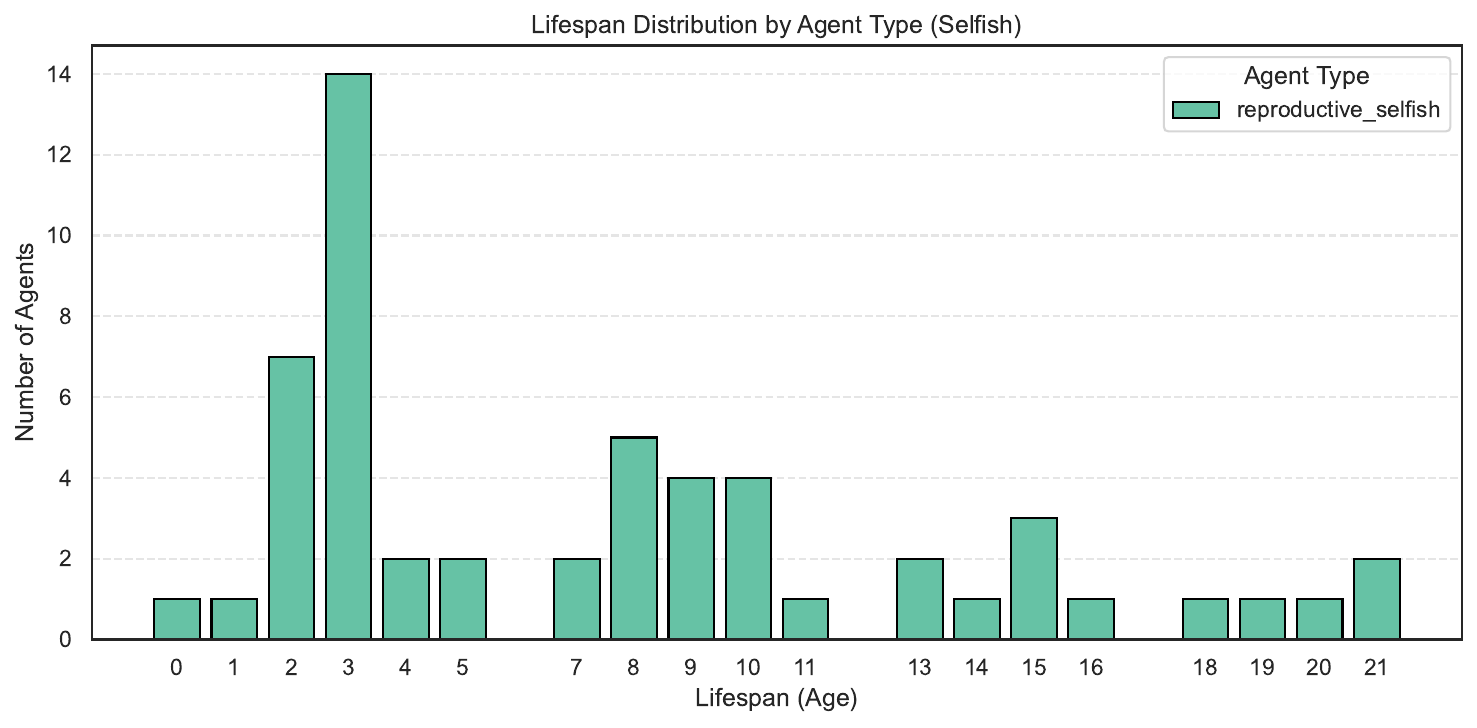}
        \caption{Lifespan Distribution by Agent Type (Case: Selfish)}
    \end{subfigure}
    \hfill
    \begin{subfigure}[t]{0.45\columnwidth}
        \centering
        \includegraphics[width=\linewidth]{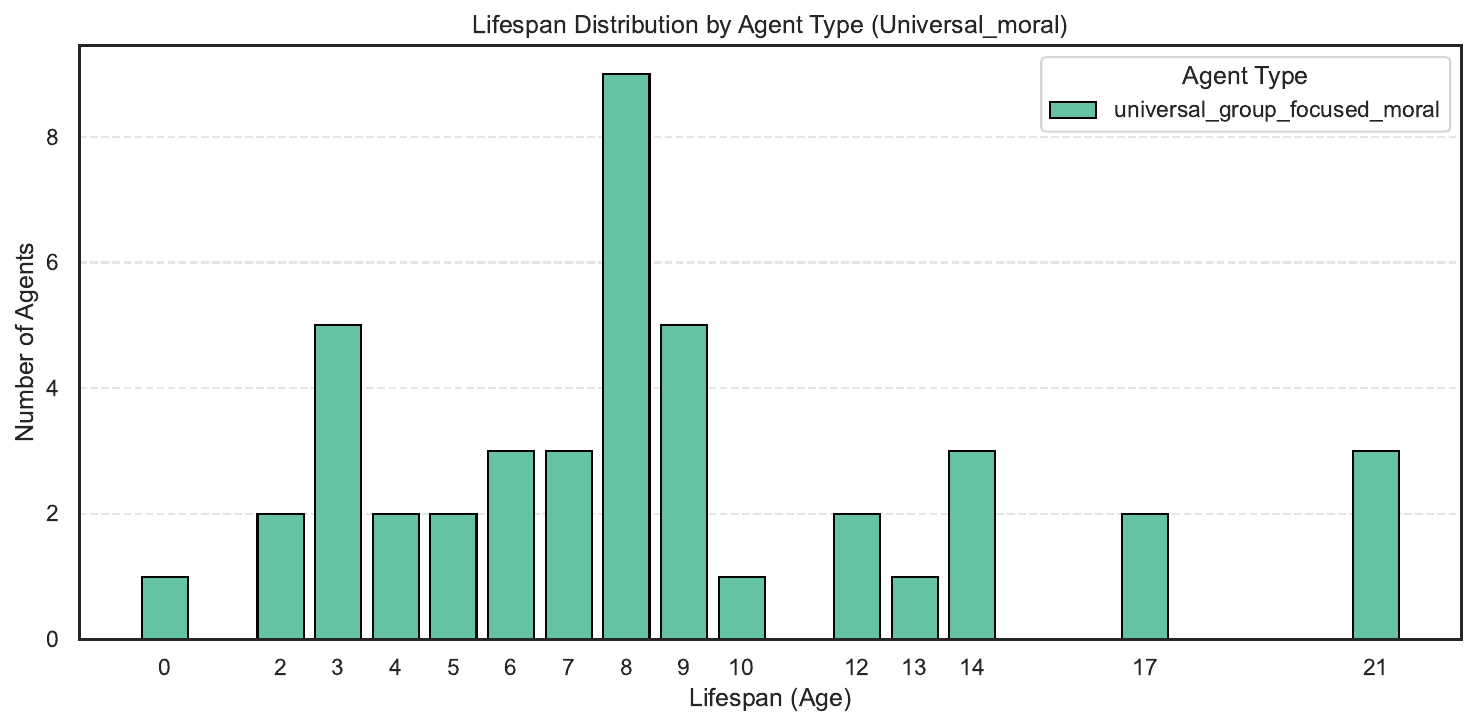}
        \caption{Lifespan Distribution by Agent Type (Case: Universal)}
    \end{subfigure}
    \hfill
    
    \begin{subfigure}[t]{0.45\columnwidth}
        \centering
        \includegraphics[width=\linewidth]{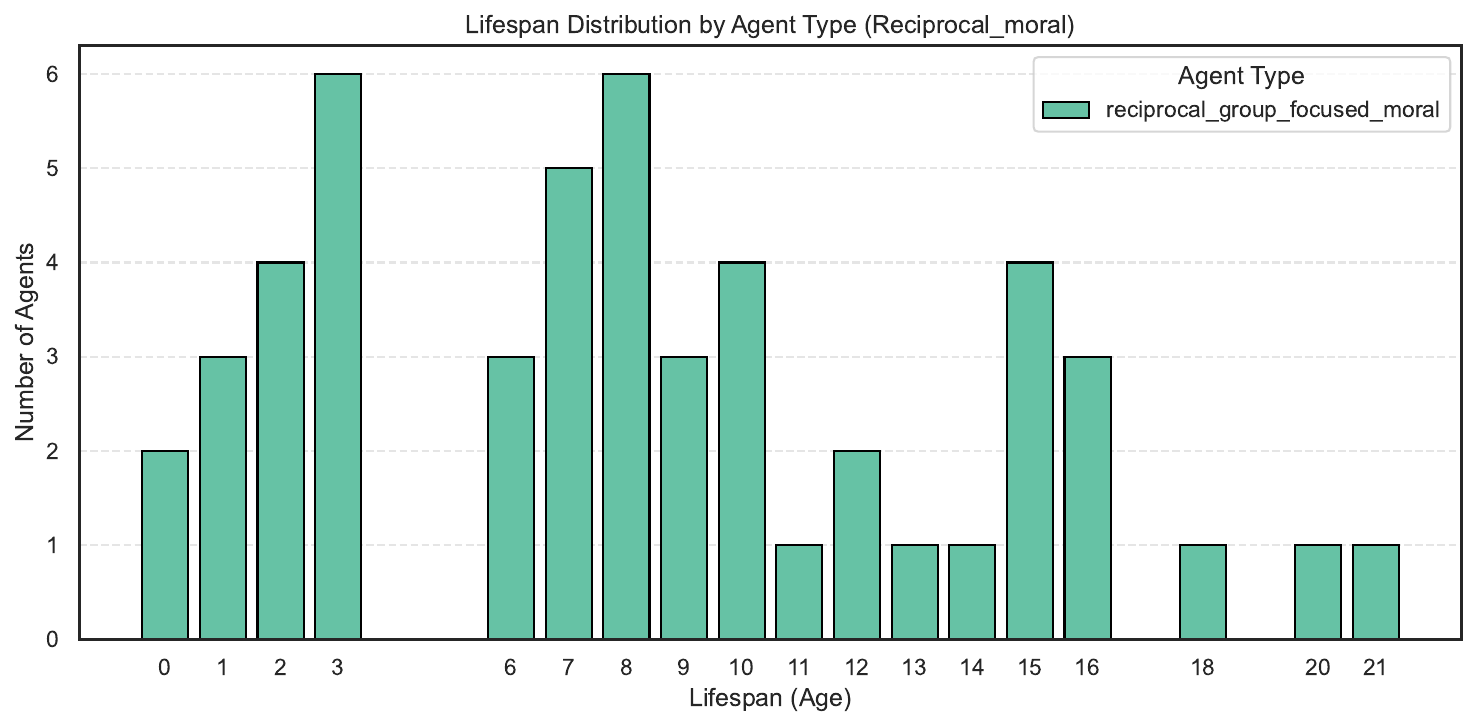}
        \caption{Lifespan Distribution by Agent Type (Case: reciprocal)}
    \end{subfigure}
    \hfill
    \begin{subfigure}[t]{0.45\columnwidth}
        \centering
        \includegraphics[width=\linewidth]{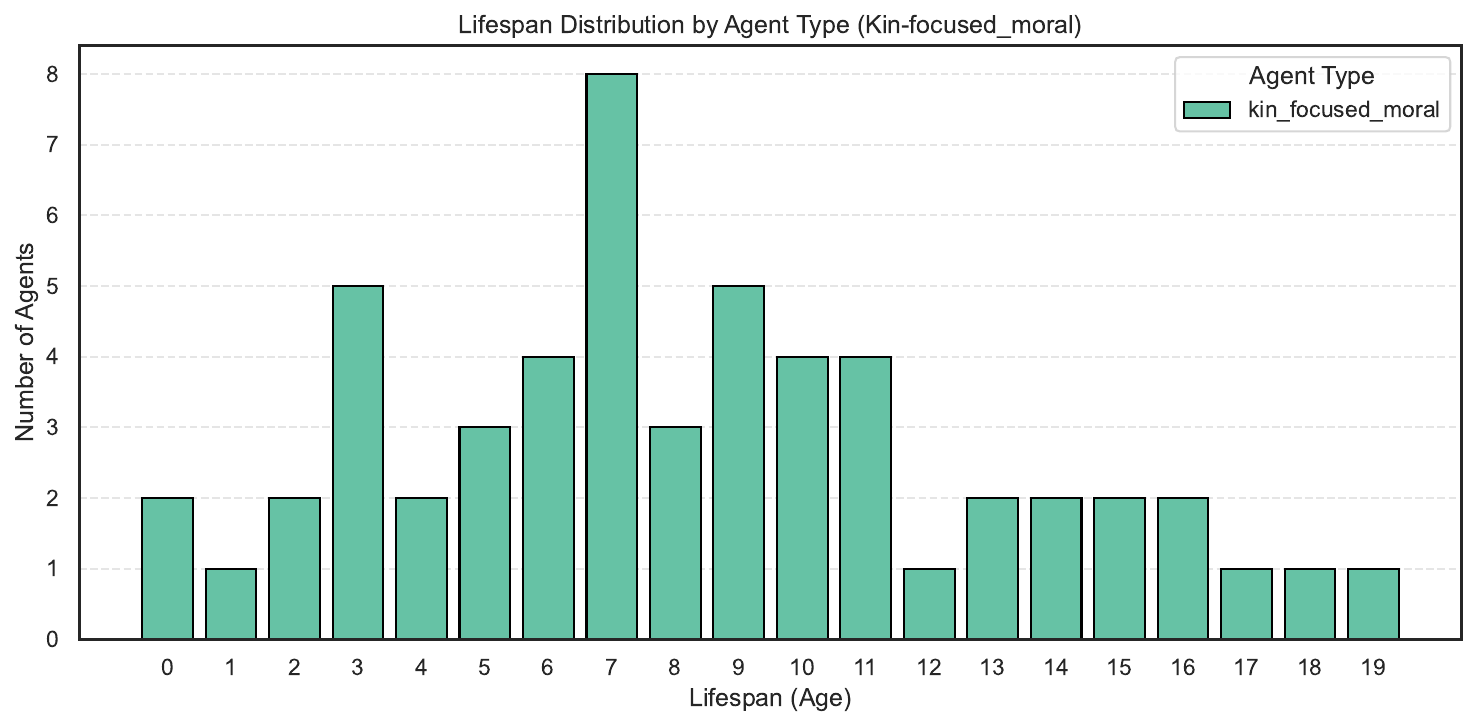}
        \caption{Lifespan Distribution by Agent Type (Case: kin)}
    \end{subfigure}

    \caption{Lifespan Distribution for tests under single Agent Type settings}
    \label{fig: lifespan_single}
\end{figure}

The lifespan distributions of agents are visualized in Figures~\ref{fig: lifespan_main} to \ref{fig: lifespan_single}. Each figure is a histogram where the x-axis represents lifespan (in time steps), and the y-axis represents the frequency of agents. The bars indicate the count of agents with specific lifespans.

By comparing different settings with the baseline experiments, we observe that lifespan distributions vary across conditions, reflecting the differential survival pressures imposed by each setting.

\subsubsection{Action distributions for each experiment}

\begin{figure}[H]
    \centering
    
    \begin{subfigure}[t]{0.45\columnwidth}
        \centering
        \includegraphics[width=\linewidth]{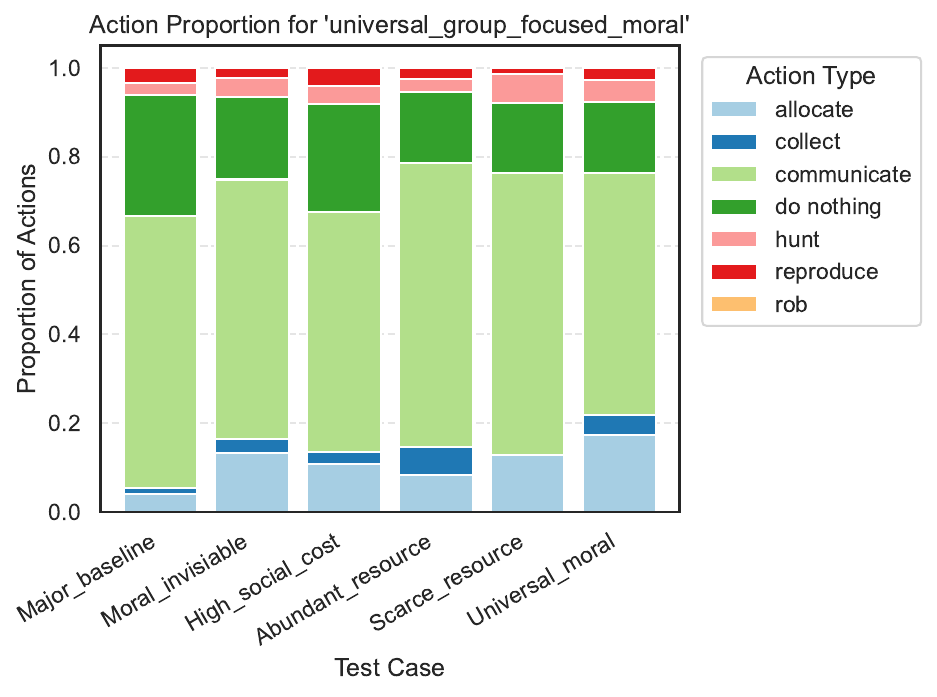}
        \caption{Action Type Proportion for universal moral agents}
    \end{subfigure}
    \hfill
    \begin{subfigure}[t]{0.45\columnwidth}
        \centering
        \includegraphics[width=\linewidth]{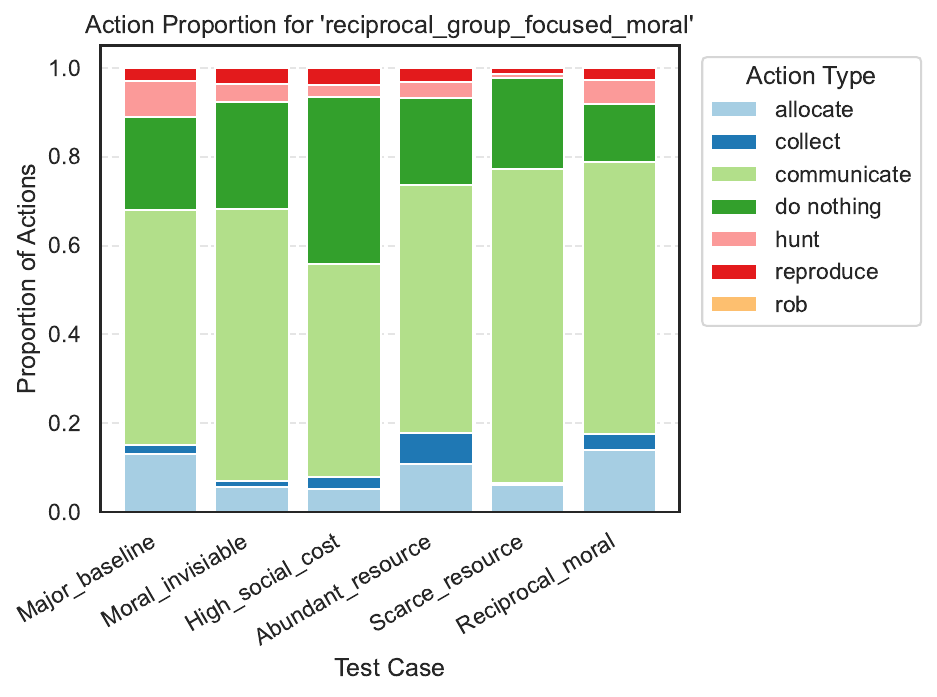}
        \caption{Action Type Proportion for reciprocal moral agents}
    \end{subfigure}
    \caption{Action type proportion across test cases and agent morality type}
    \label{fig: action type proportion1}
\end{figure}

\begin{figure}[H]
    \centering
    \begin{subfigure}[t]{0.45\columnwidth}
        \centering
        \includegraphics[width=\linewidth]{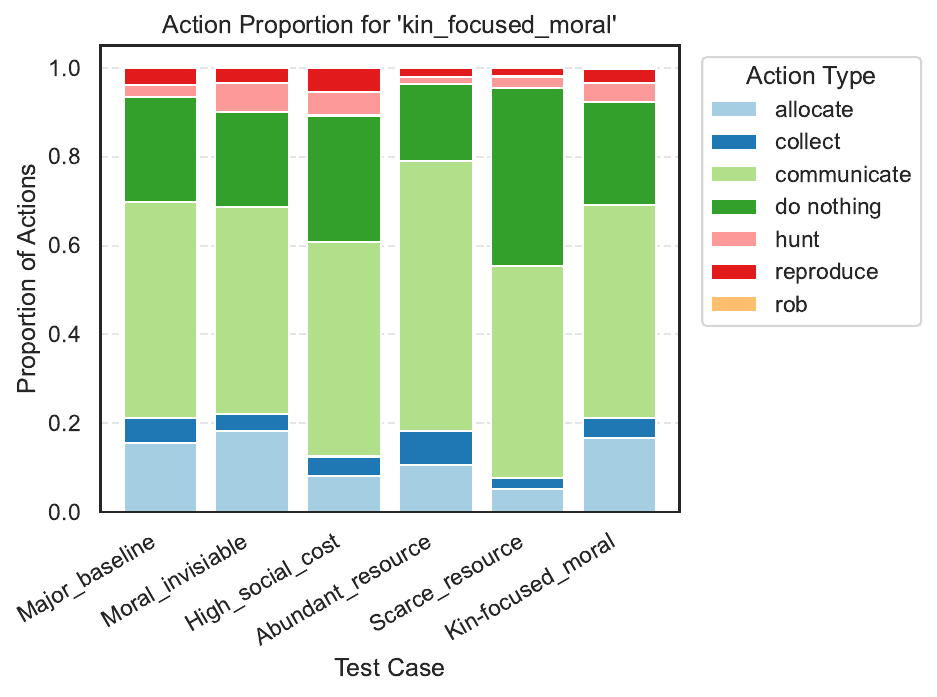}
        \caption{Action Type Proportion for kin-focused moral agents}
    \end{subfigure}
    \hfill  
    \begin{subfigure}[t]{0.45\columnwidth}
        \centering
        \includegraphics[width=\linewidth]{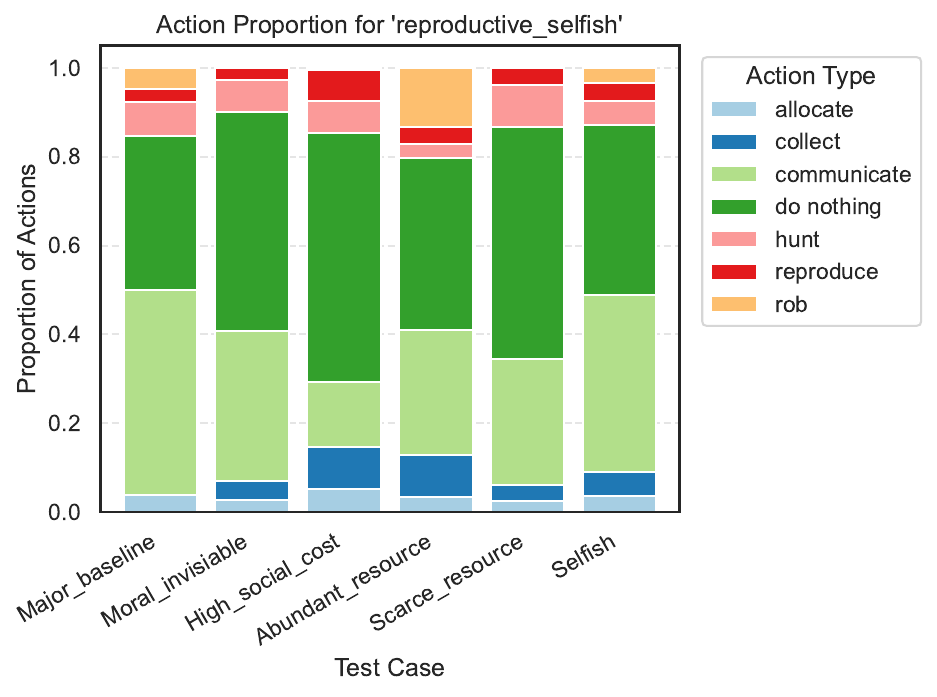}
        \caption{Action Type Proportion for selfish agents}
    \end{subfigure}

    \caption{Action type proportion across test cases and agent morality type}
    \label{fig: action type proportion2}
\end{figure}

Figure~\ref{fig: action type proportion1} and~\ref{fig: action type proportion2} present the proportions of action types across different test cases and agent morality types. Each subfigure is a bar chart where the x-axis represents action types (e.g., hunting, resting, social interactions), and the y-axis represents the proportion of actions.
This section examines the proportion of action types across test cases and agent morality types. Figures include:
1) Overall Action Proportions: Bar charts showing the percentage of each action type (e.g., hunting, resting, social interactions) across all test cases.
2) Action Proportions by Moral Type: Separate bar charts for Universal, Reciprocal, Kin-focused, and Selfish agents, highlighting their behavioral tendencies across scenarios.

From the figure, we could observe that:
\begin{itemize}
    \item Only selfish agents have taken the ``rob" action, and they do not like ``communicate" and ``allocate"
\end{itemize}

\subsubsection{Action distributions for each moral type}

\begin{figure}[H]
    \centering
    \begin{subfigure}[t]{0.48\textwidth}
        \centering
        \includegraphics[width=\linewidth]{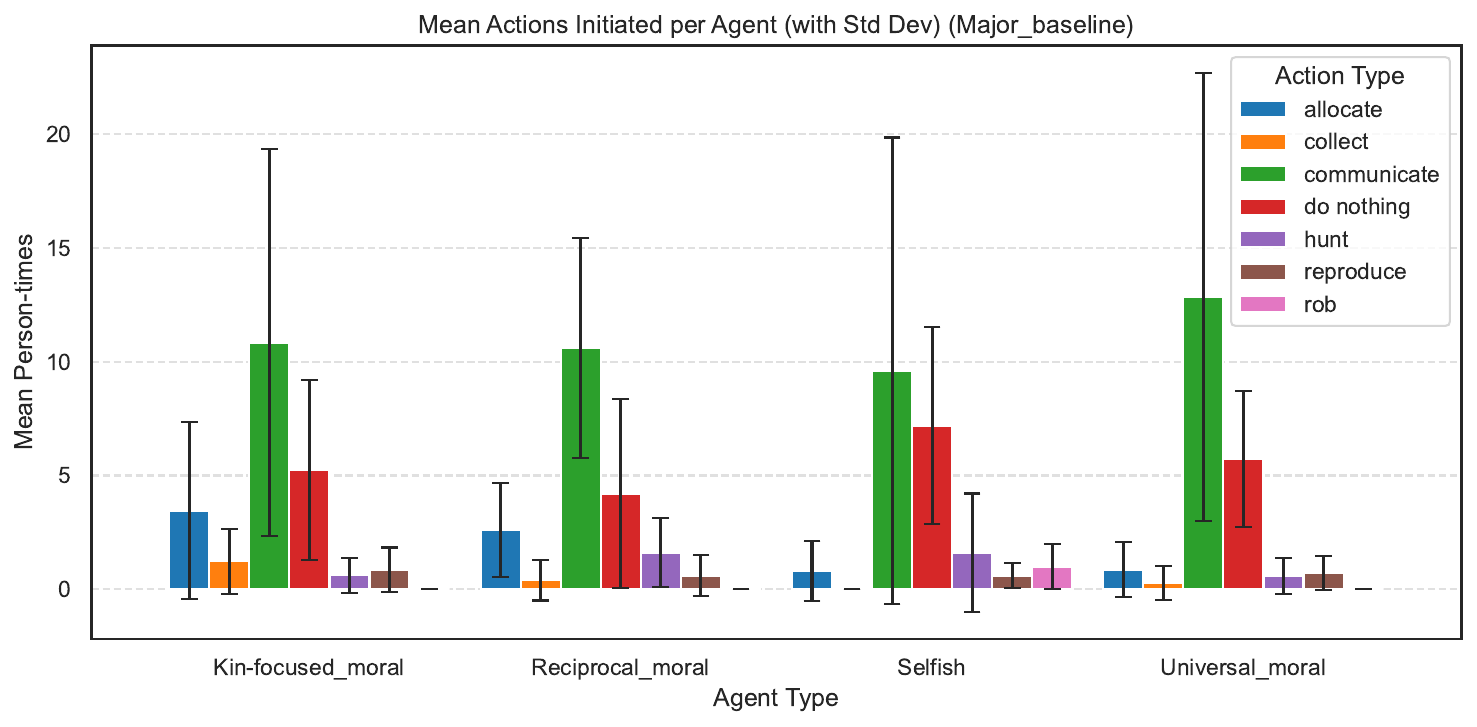}
        \caption{Mean Actions Initiated per Agent}
    \end{subfigure}
    \hfill
    \begin{subfigure}[t]{0.48\textwidth}
        \centering
        \includegraphics[width=\linewidth]{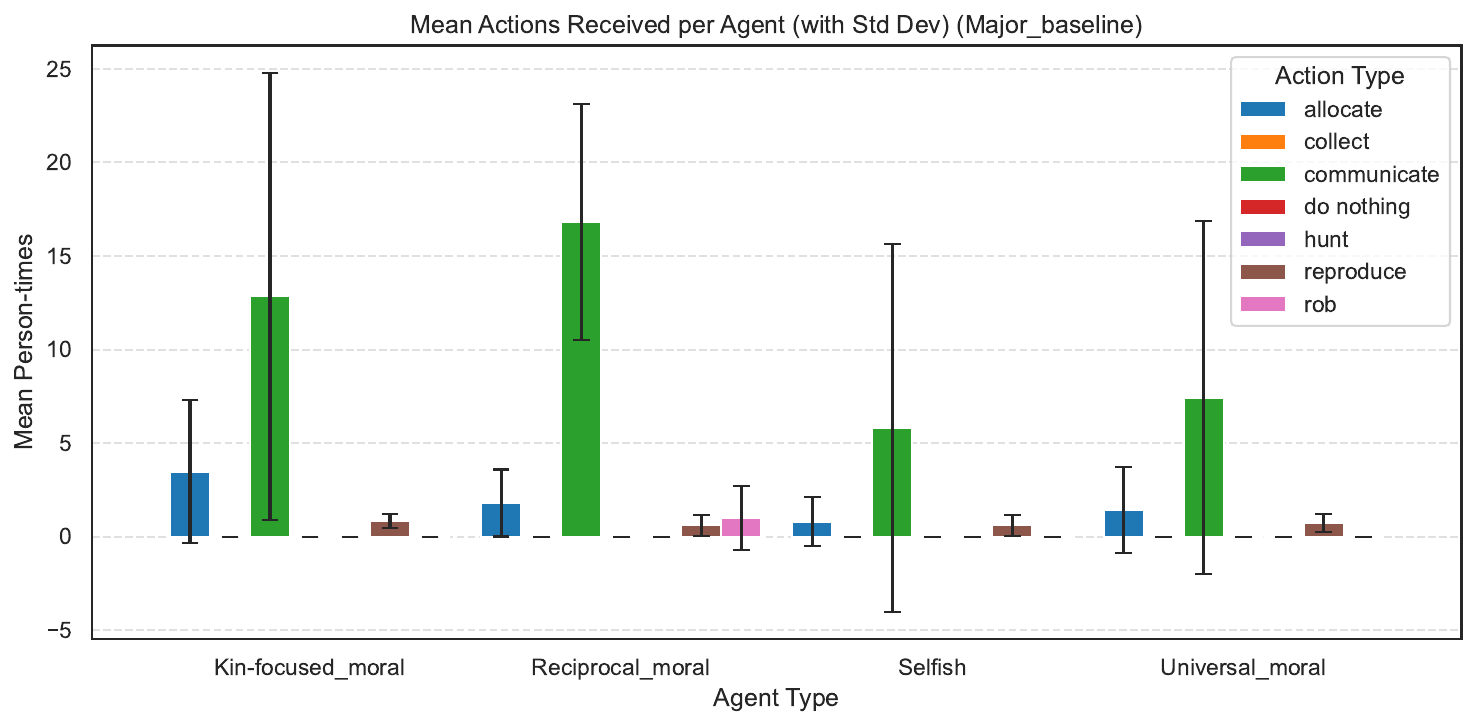}
        \caption{Mean Actions Received per Agent}
    \end{subfigure}
    \caption{Agent-times of action type when agents are initiators and receivers (Case: Baseline)}
    \label{fig: actions_type_main}
\end{figure}

\begin{figure}[H]
    \centering
    \begin{subfigure}[t]{0.48\textwidth}
        \centering
        \includegraphics[width=\linewidth]{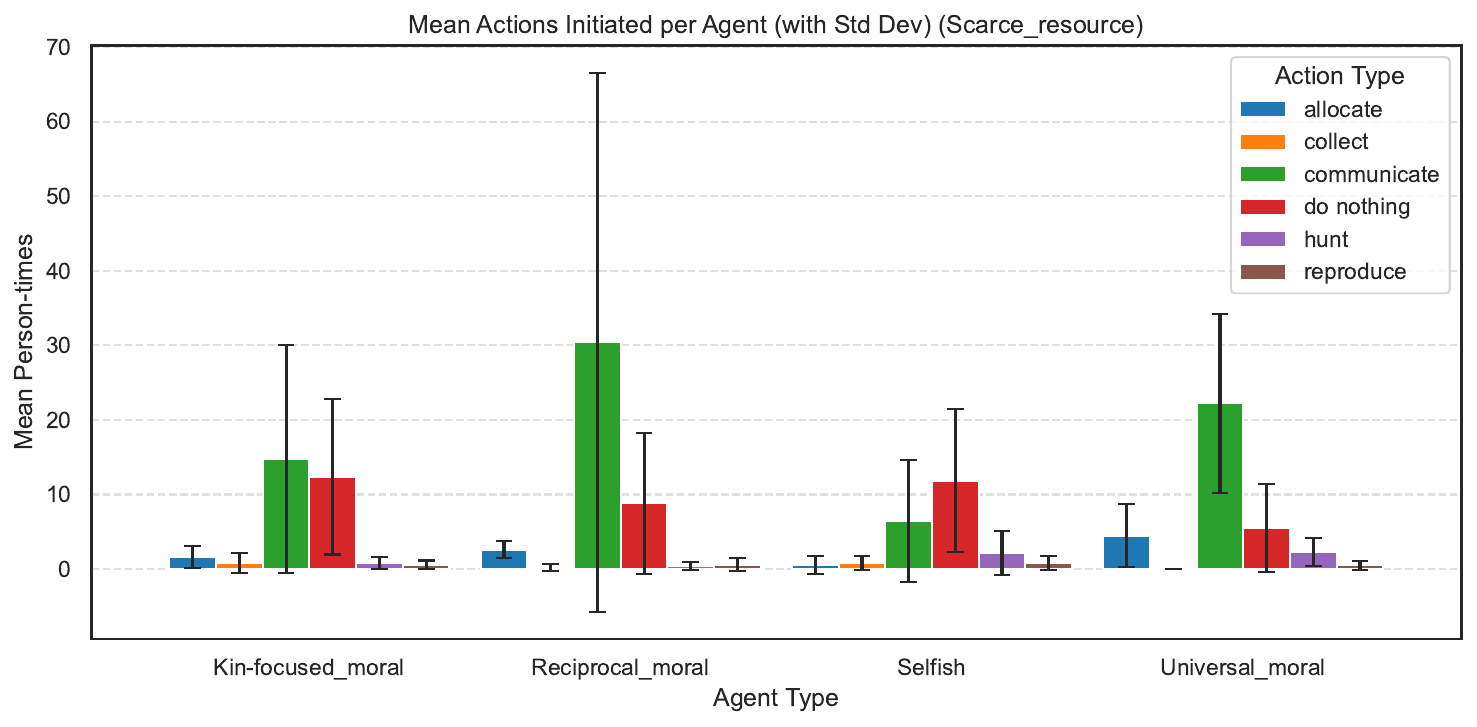}
        \caption{Mean Actions Initiated per Agent}
    \end{subfigure}
    \hfill
    \begin{subfigure}[t]{0.48\textwidth}
        \centering
        \includegraphics[width=\linewidth]{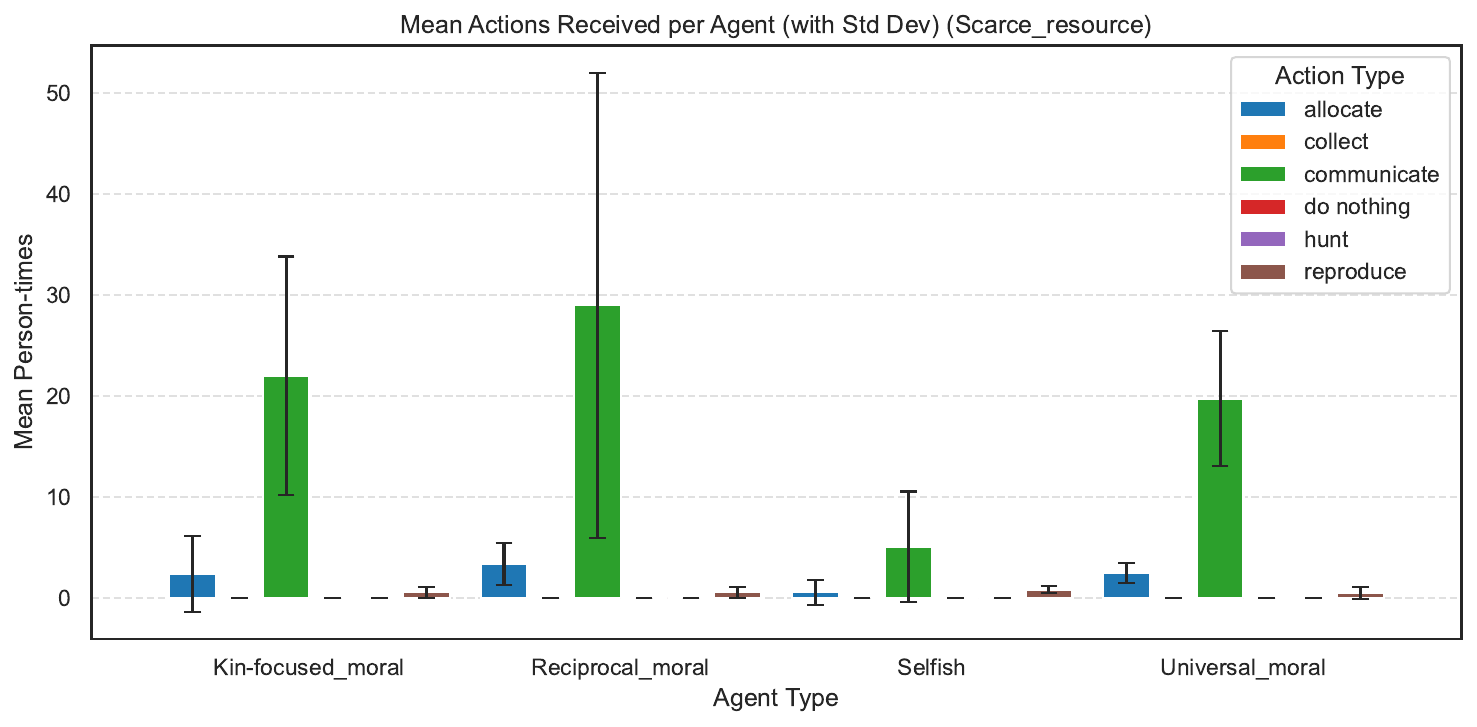}
        \caption{Mean Actions Received per Agent}
    \end{subfigure}
    \caption{Agent-times of action type when agents are initiators and receivers (Case: scarce resource)}
    \label{fig: actions_type_scarce}
\end{figure}

\begin{figure}[H]
    \centering
    \begin{subfigure}[t]{0.48\textwidth}
        \centering
        \includegraphics[width=\linewidth]{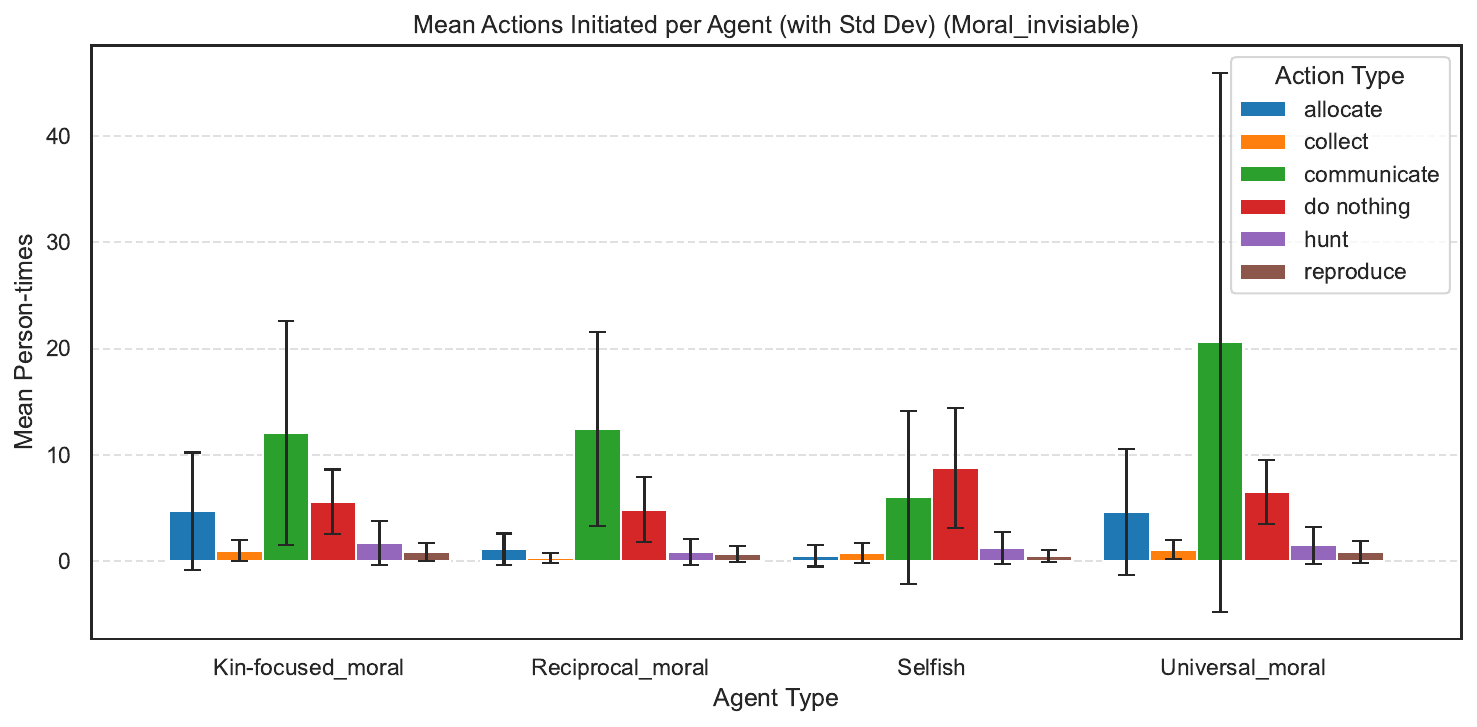}
        \caption{Mean Actions Initiated per Agent}
    \end{subfigure}
    \hfill
    \begin{subfigure}[t]{0.48\textwidth}
        \centering
        \includegraphics[width=\linewidth]{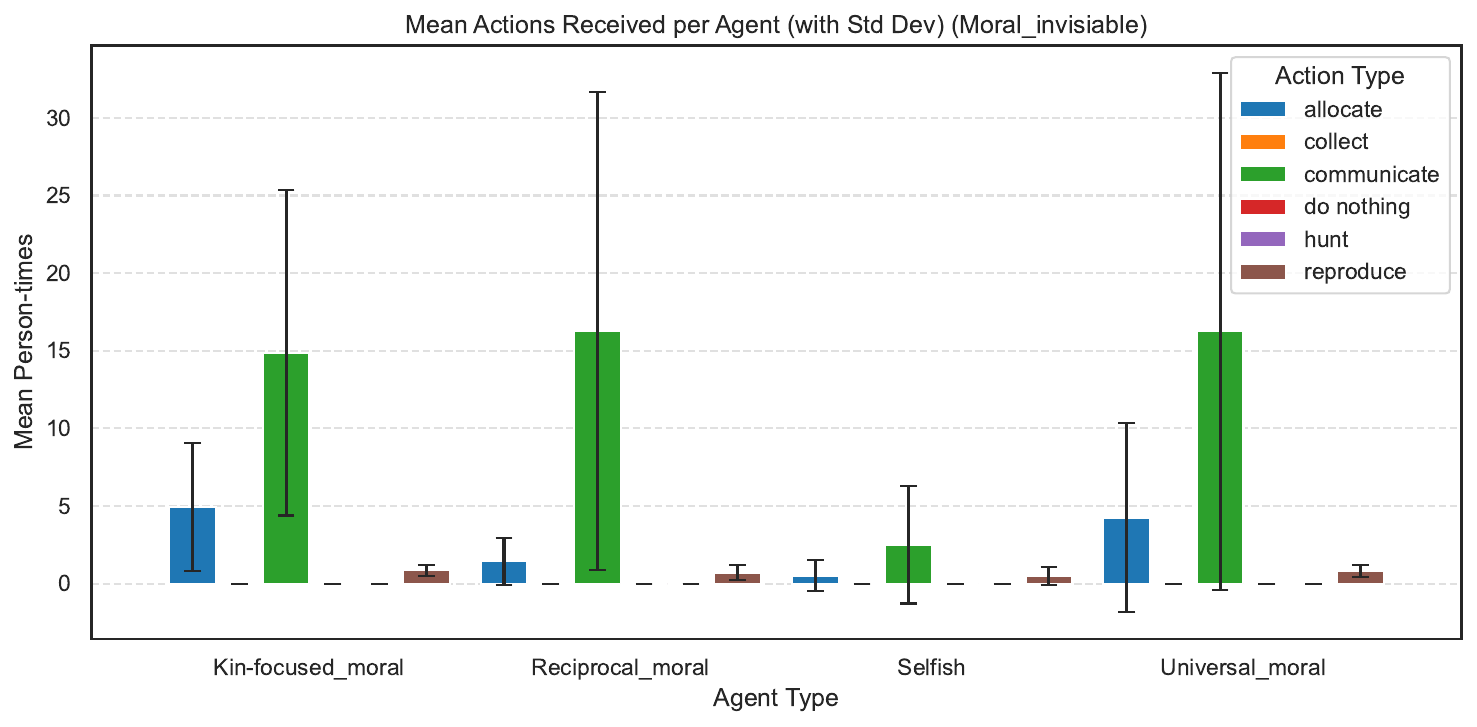}
        \caption{Mean Actions Received per Agent}
    \end{subfigure}
    \caption{Agent-times of action type when agents are initiators and receivers (Case: Moral Type Invisible)}
    \label{fig: actions_type_invisible}
\end{figure}

\begin{figure}[H]
    \centering
    \begin{subfigure}[t]{0.48\textwidth}
        \centering
        \includegraphics[width=\linewidth]{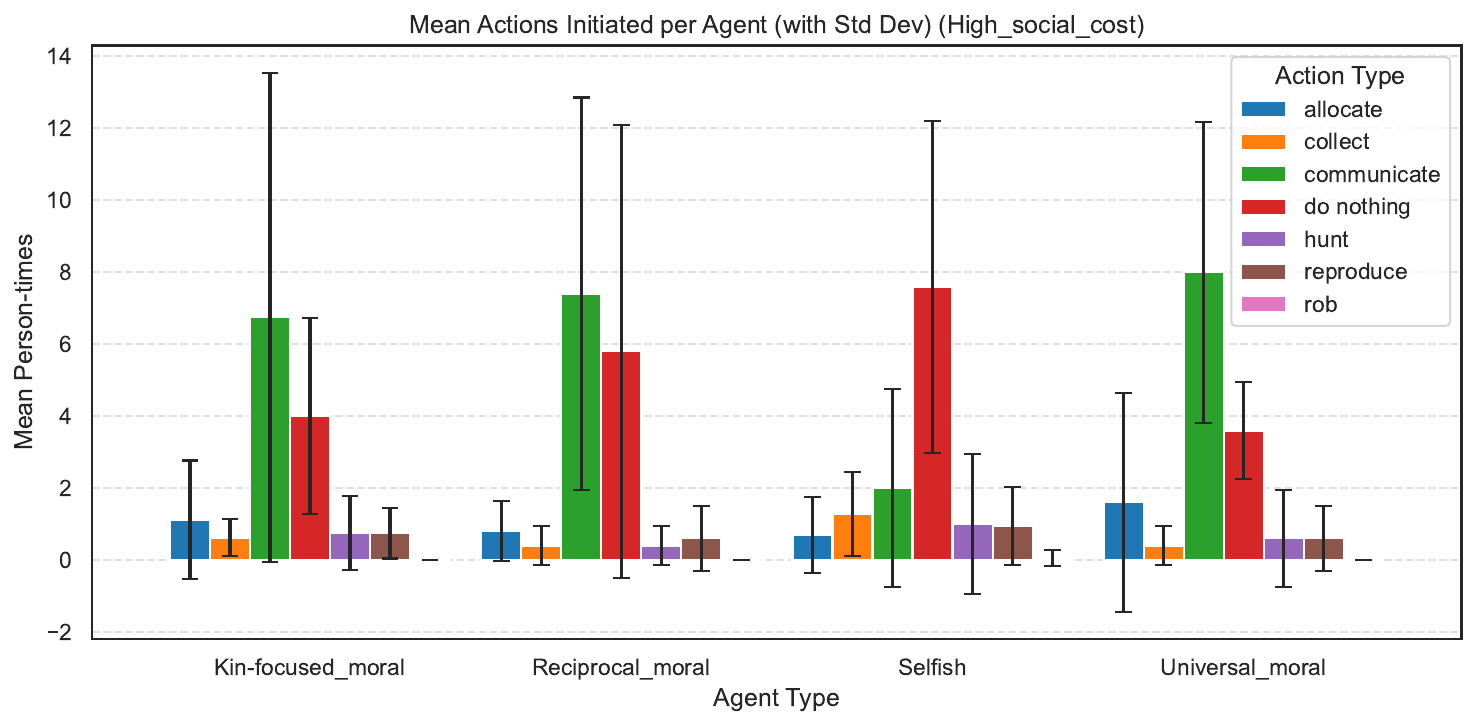}
        \caption{Mean Actions Initiated per Agent}
    \end{subfigure}
    \hfill
    \begin{subfigure}[t]{0.48\textwidth}
        \centering
        \includegraphics[width=\linewidth]{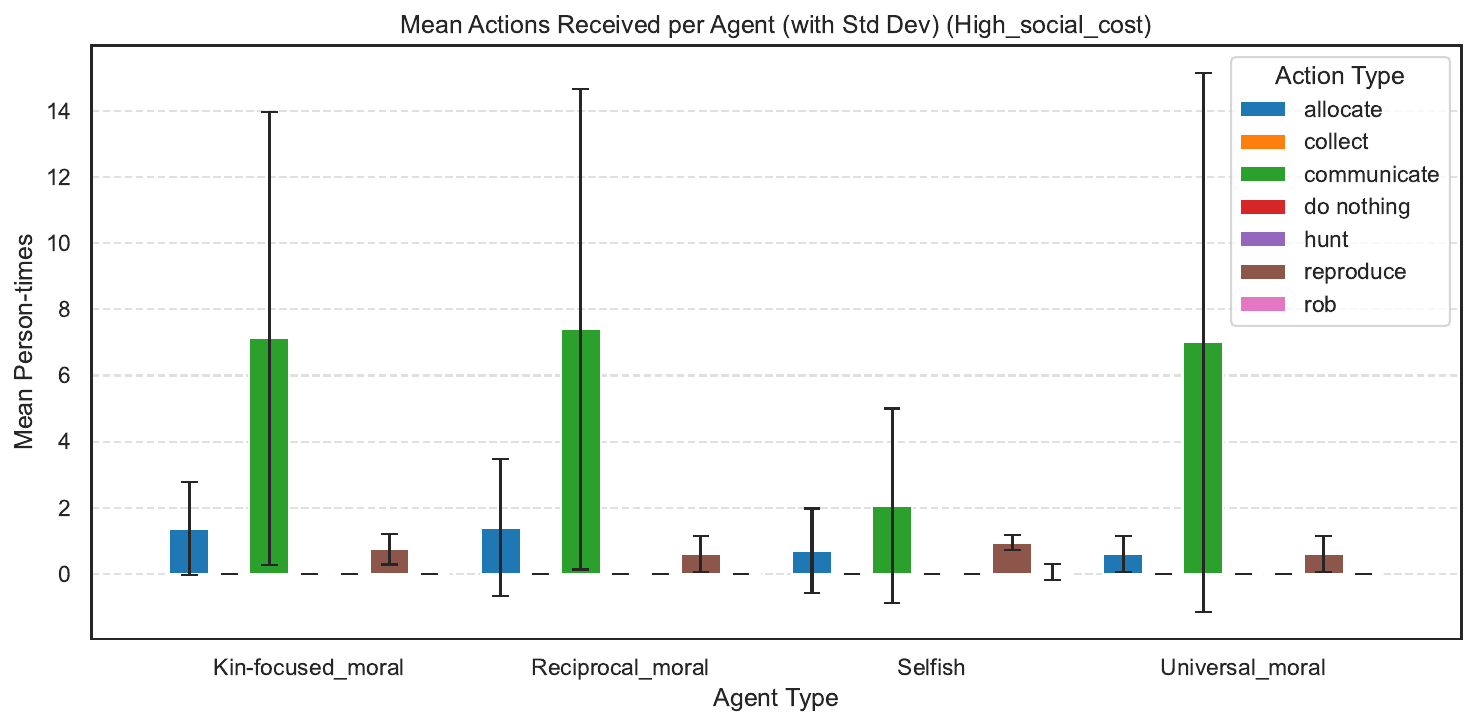}
        \caption{Mean Actions Received per Agent}
    \end{subfigure}
    \caption{Agent-times of action type when agents are initiators and receivers (Case: high social cost)}
    \label{fig: actions_type_social_cost}
\end{figure}

\begin{figure}[H]
    \centering
    \begin{subfigure}[t]{0.48\textwidth}
        \centering
        \includegraphics[width=\linewidth]{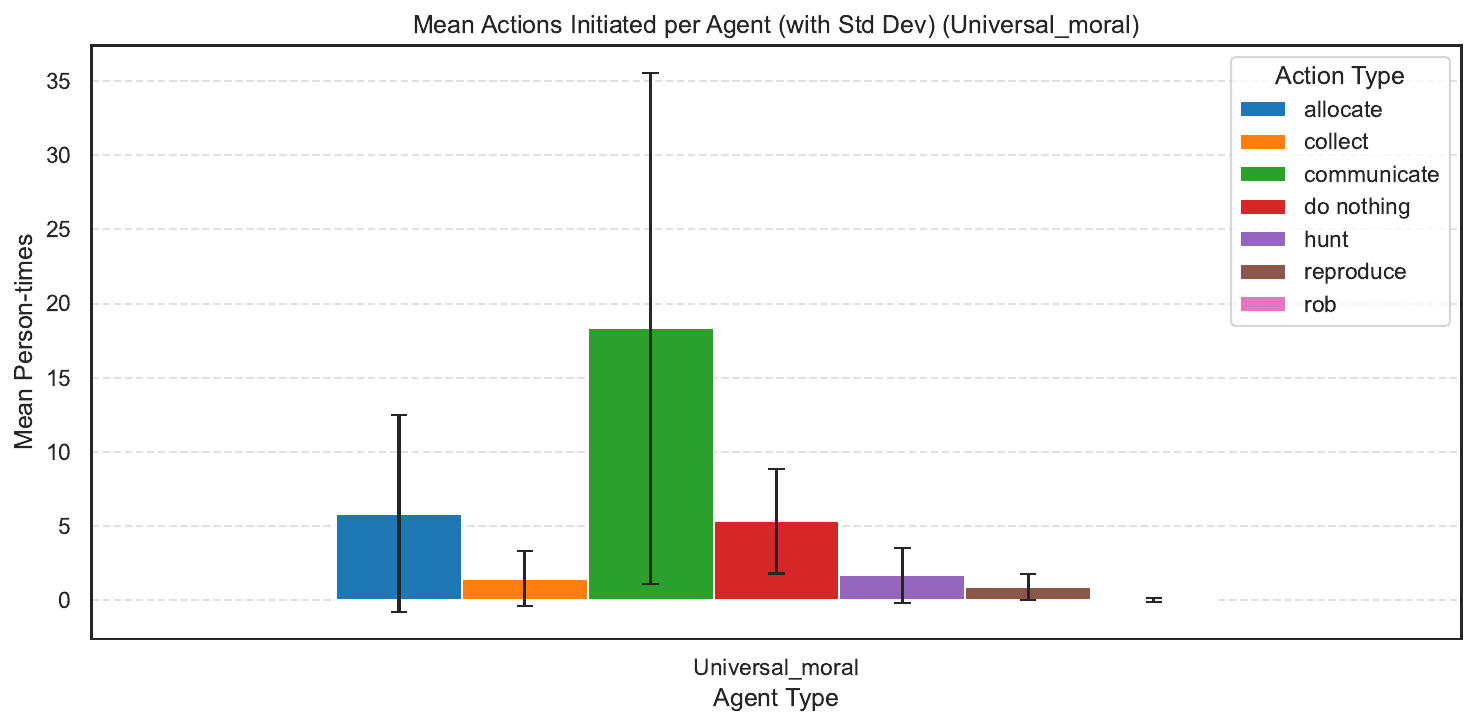}
        \caption{Mean Actions Initiated per Agent}
    \end{subfigure}
    \hfill
    \begin{subfigure}[t]{0.48\textwidth}
        \centering
        \includegraphics[width=\linewidth]{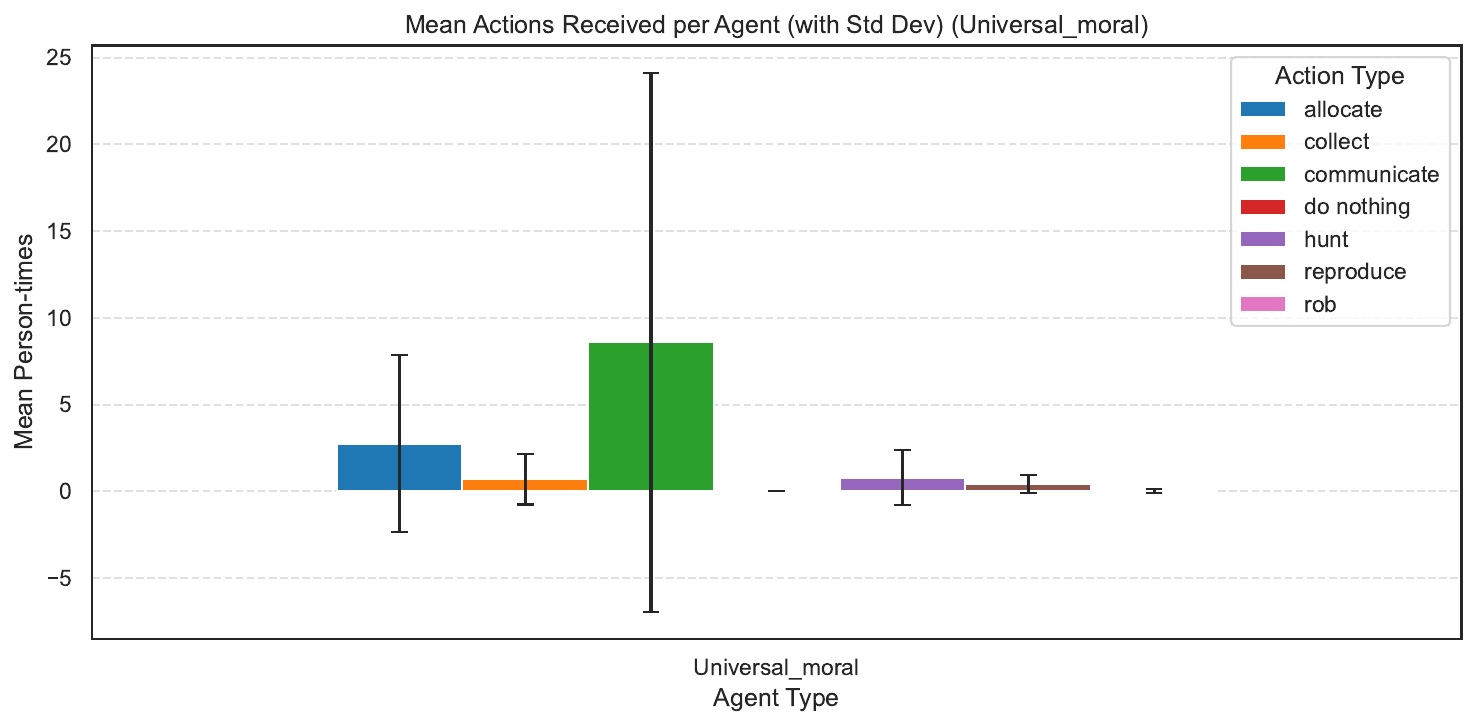}
        \caption{Mean Actions Received per Agent}
    \end{subfigure}
    \caption{Agent-times of action type when agents are initiators and receivers (Case: universal)}
    \label{fig: actions_type_universal}
\end{figure}

\begin{figure}[H]
    \centering
    \begin{subfigure}[t]{0.48\textwidth}
        \centering
        \includegraphics[width=\linewidth]{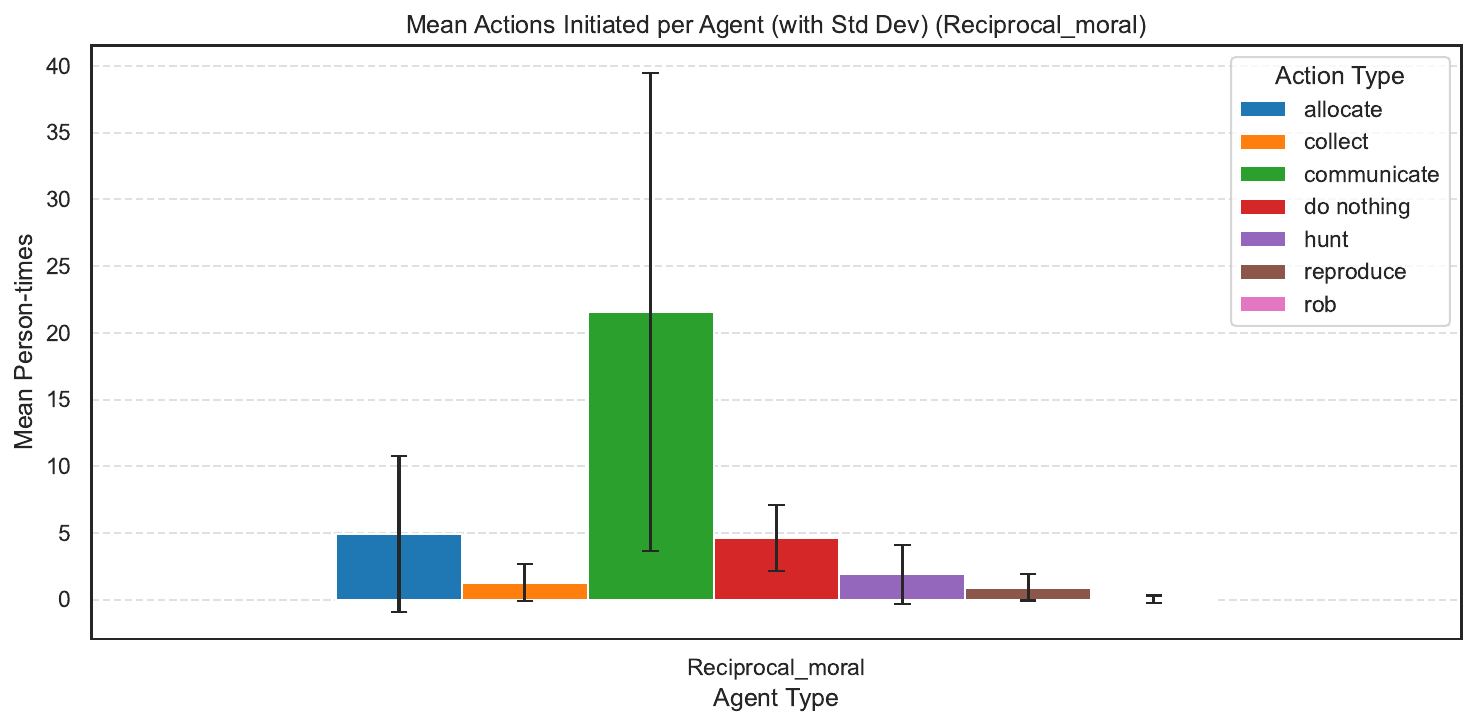}
        \caption{Mean Actions Initiated per Agent}
    \end{subfigure}
    \hfill
    \begin{subfigure}[t]{0.48\textwidth}
        \centering
        \includegraphics[width=\linewidth]{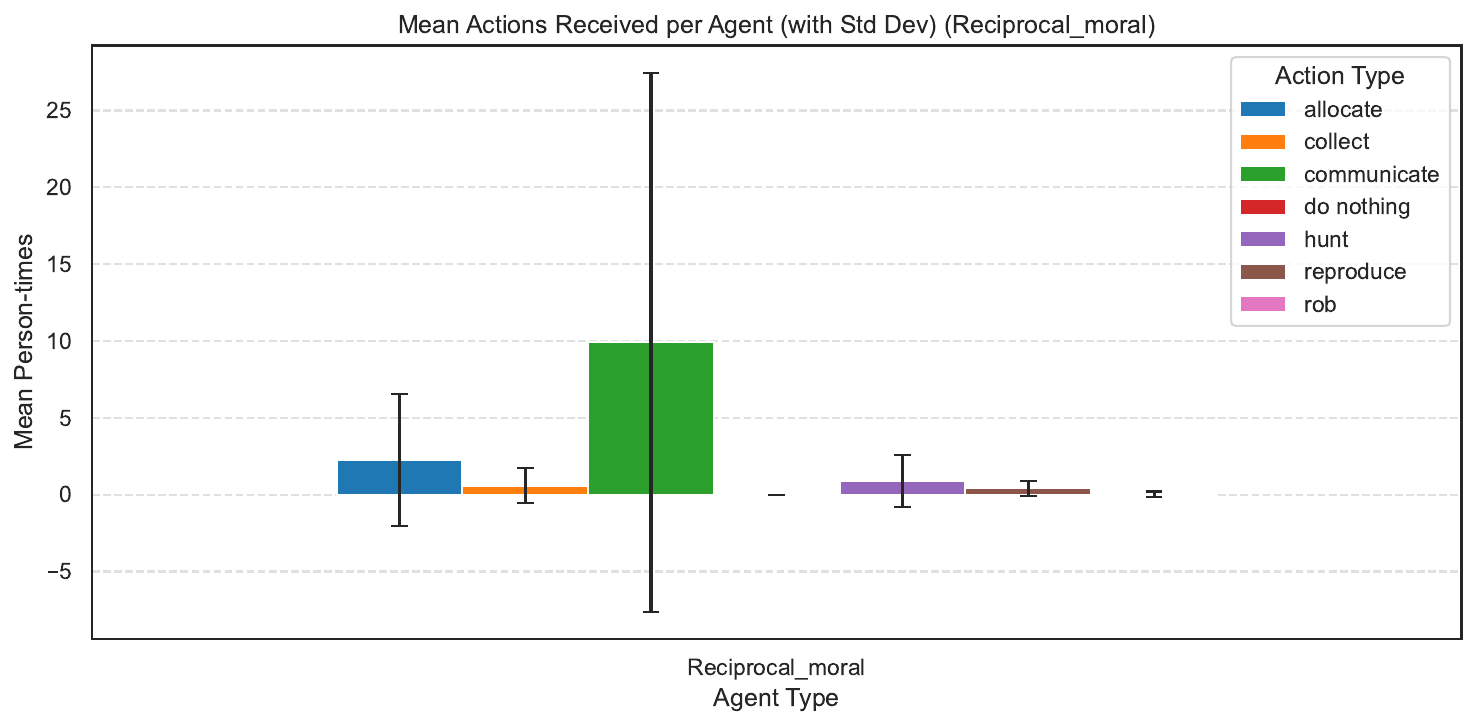}
        \caption{Mean Actions Received per Agent}
    \end{subfigure}
    \caption{Agent-times of action type when agents are initiators and receivers (Case: reciprocal)}
    \label{fig: actions_type_reciprocal}
\end{figure}

\begin{figure}[H]
    \centering
    \begin{subfigure}[t]{0.48\textwidth}
        \centering
        \includegraphics[width=\linewidth]{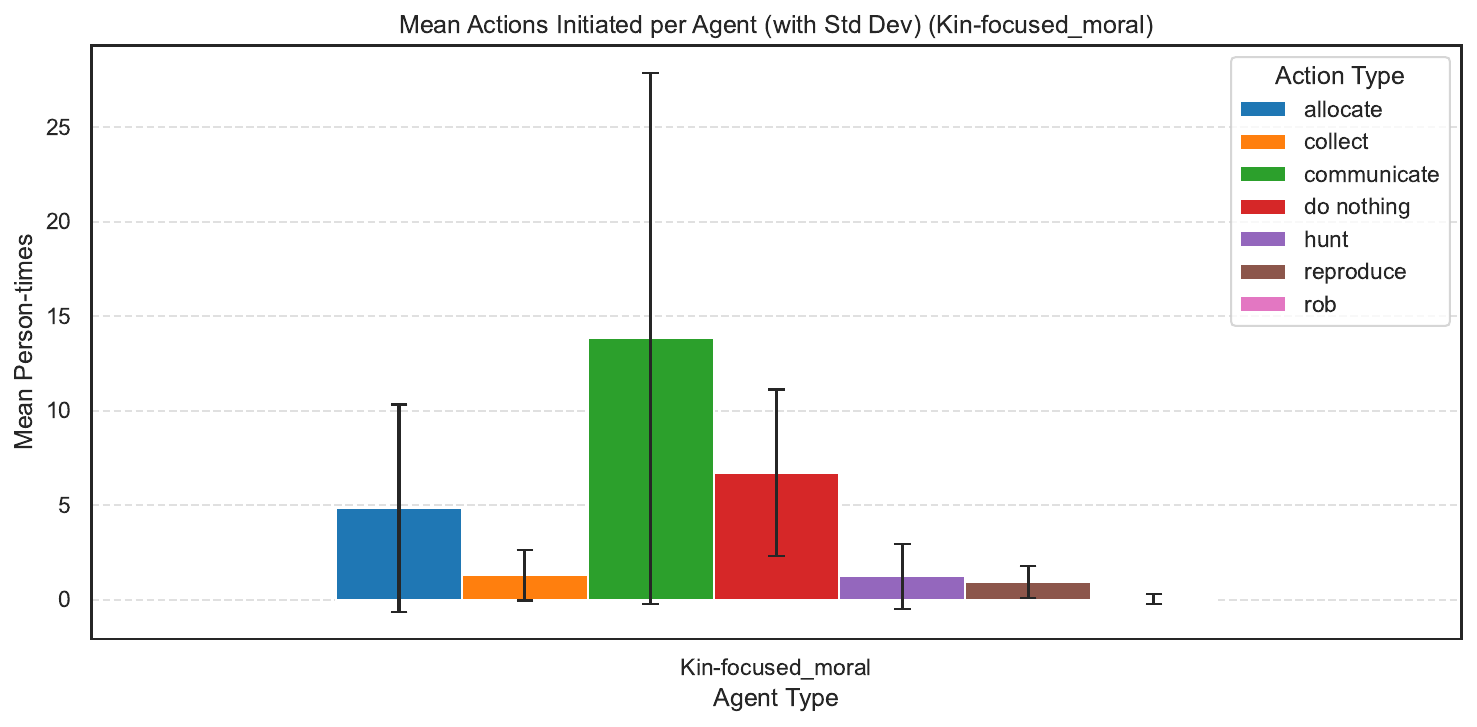}
        \caption{Mean Actions Initiated per Agent}
    \end{subfigure}
    \hfill
    \begin{subfigure}[t]{0.48\textwidth}
        \centering
        \includegraphics[width=\linewidth]{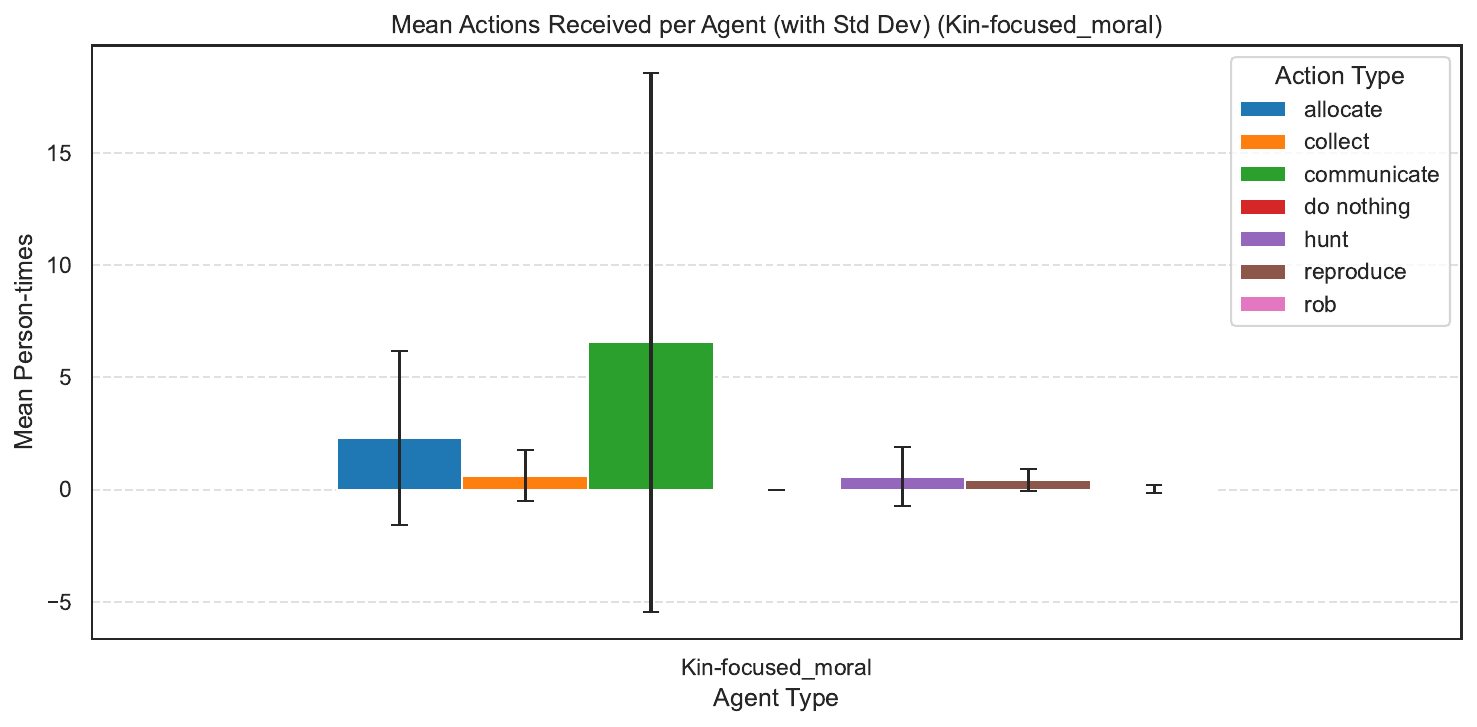}
        \caption{Mean Actions Received per Agent}
    \end{subfigure}
    \caption{Agent-times of action type when agents are initiators and receivers (Case: kin)}
    \label{fig: actions_type_kin}
\end{figure}

\begin{figure}[H]
    \centering
    \begin{subfigure}[t]{0.48\textwidth}
        \centering
        \includegraphics[width=\linewidth]{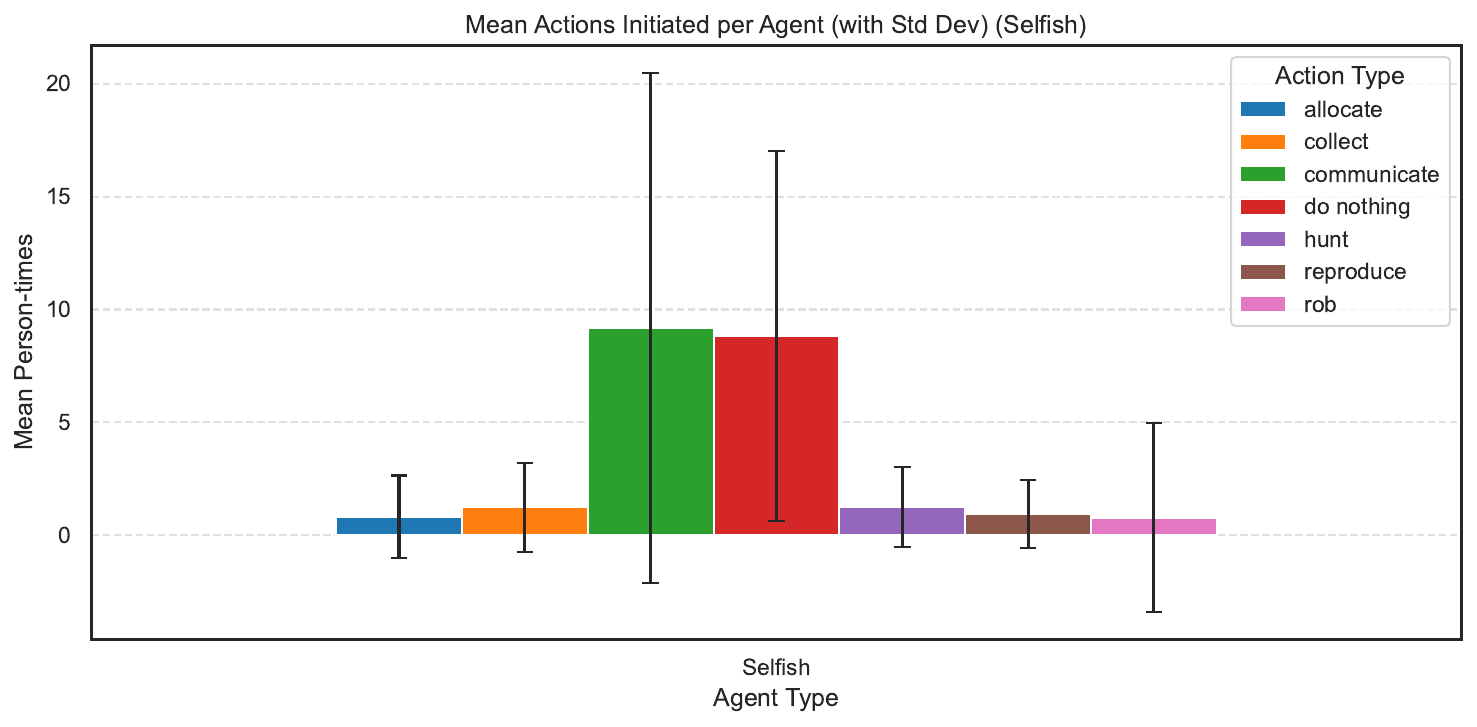}
        \caption{Mean Actions Initiated per Agent}
    \end{subfigure}
    \hfill
    \begin{subfigure}[t]{0.48\textwidth}
        \centering
        \includegraphics[width=\linewidth]{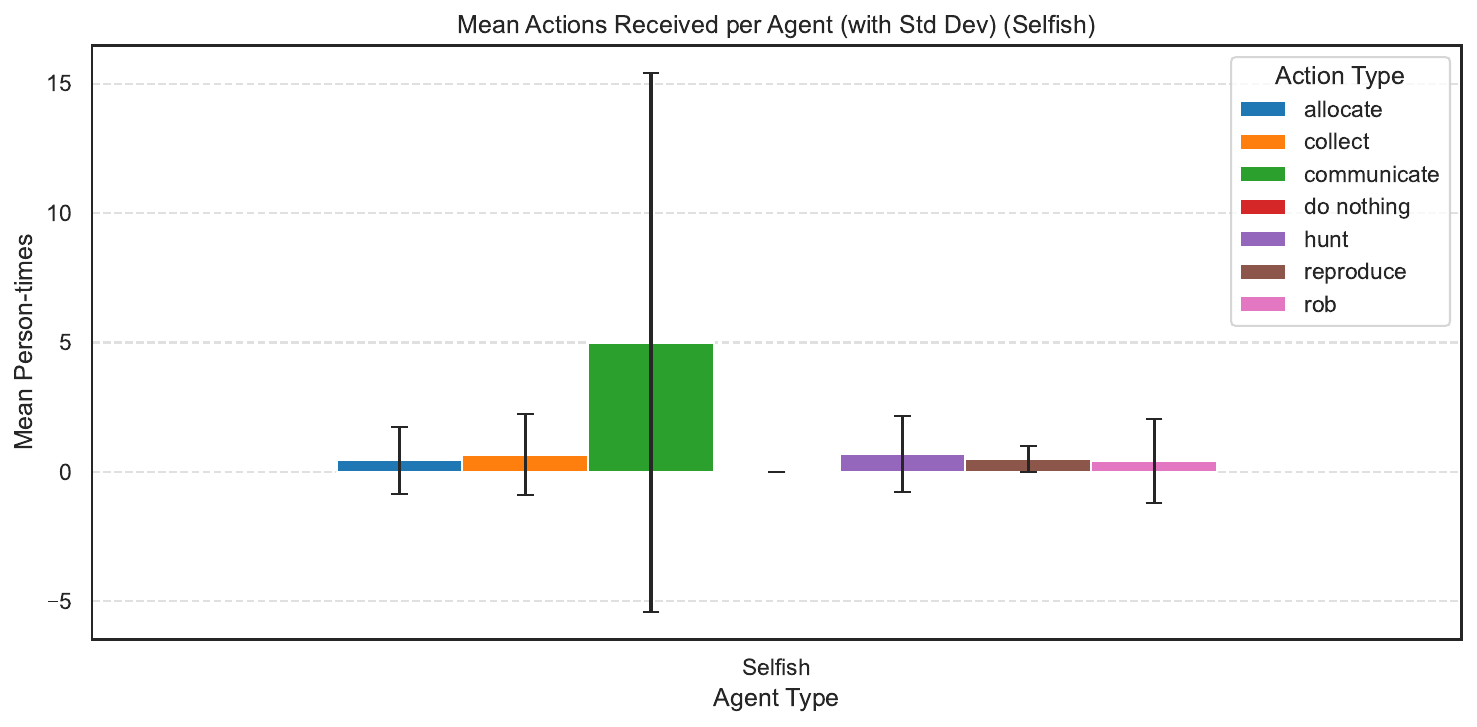}
        \caption{Mean Actions Received per Agent}
    \end{subfigure}
    \caption{Agent-times of action type when agents are initiators and receivers (Case: selfish)}
    \label{fig: actions_type_selfish}
\end{figure}

Figures~\ref{fig: actions_type_main} to \ref{fig: actions_type_selfish} detail the mean actions initiated and received by agents of each moral type. Each figure consists of two bar charts:
\begin{itemize}
    \item The first chart shows the mean actions initiated per agent, with the x-axis representing action types and the y-axis representing the average number of actions initiated.
    \item The second chart shows the mean actions received per agent, with the x-axis representing action types and the y-axis representing the average number of actions received.
\end{itemize}

From all these figures, we could conclude that:
\begin{itemize}
    \item Selfish agents receive the least communications and allocations
    \item Universal moral and kin-focused moral initiate the most allocation
\end{itemize}

\subsubsection{HP gain and loss of each action type}

\begin{figure}[H]
    \centering

    \begin{subfigure}[t]{0.48\textwidth}
        \centering
        \includegraphics[width=\linewidth]{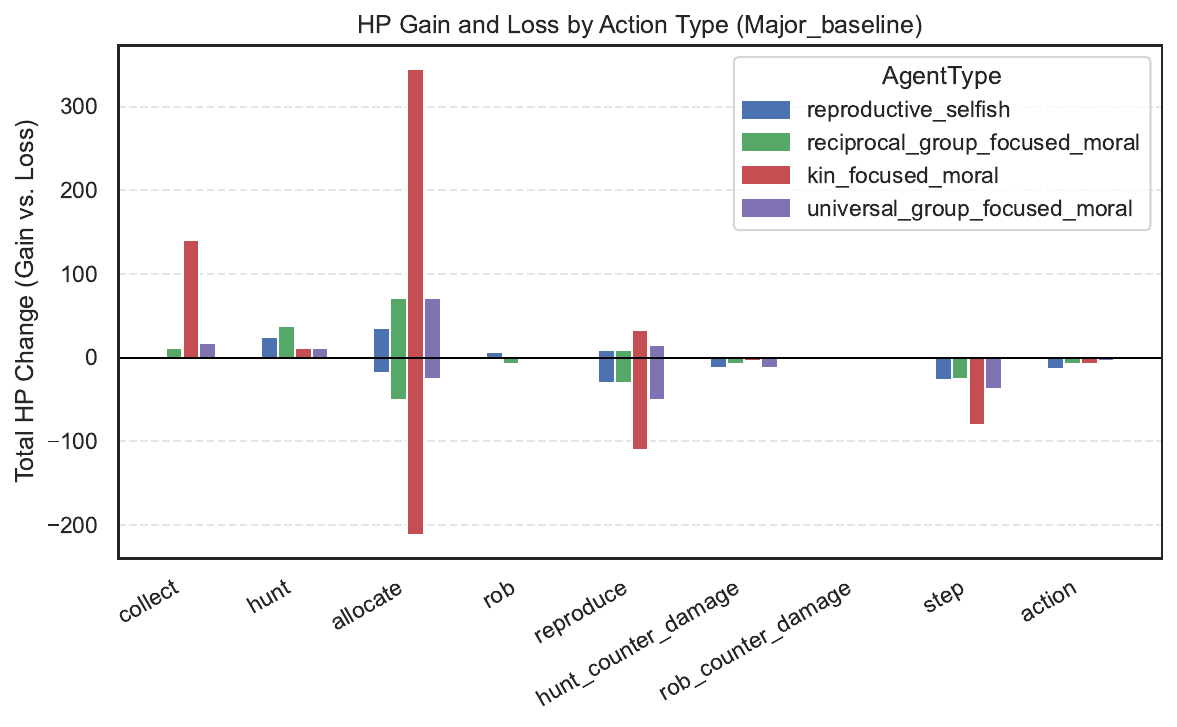}
        \caption{HP Gain and Loss by Action Type (Case: Baseline)}
    \end{subfigure}
    \hfill
    \begin{subfigure}[t]{0.48\textwidth}
        \centering
        \includegraphics[width=\linewidth]{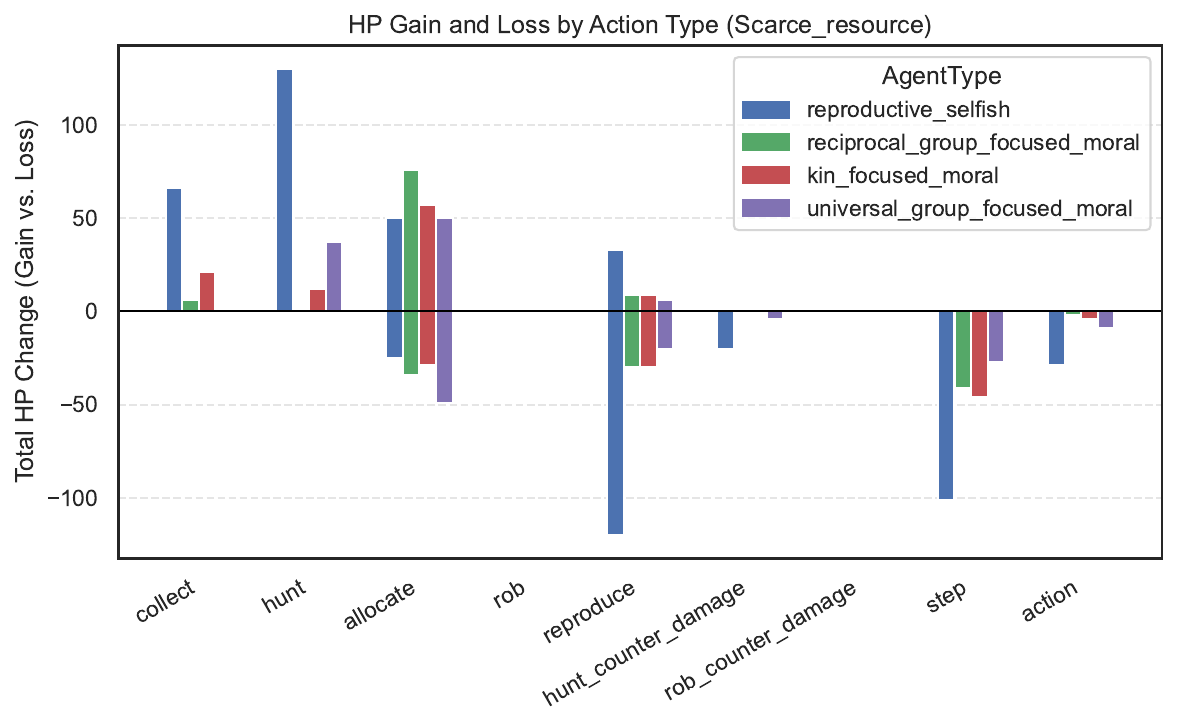}
        \caption{HP Gain and Loss by Action Type (Case: scarce resource)}
    \end{subfigure}
    \hfill
    \caption{HP Gain and Loss by Action Type (Cases: Baseline and scarce resource)}
    \label{fig: HP_action_resource}
\end{figure}

\begin{figure}[H]
    \centering
    \begin{subfigure}[t]{0.48\textwidth}
        \centering
        \includegraphics[width=\linewidth]{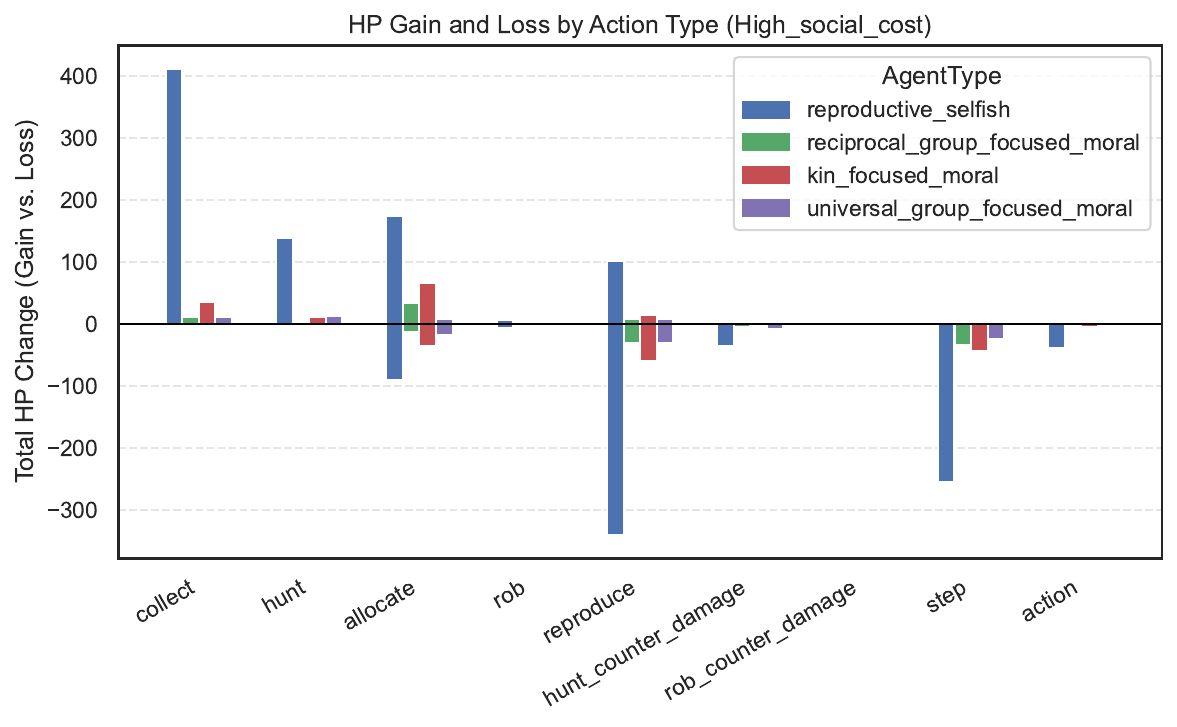}
        \caption{HP Gain and Loss by Action Type (Case: high social cost)}
    \end{subfigure}
    \hfill
    \begin{subfigure}[t]{0.48\textwidth}
        \centering
        \includegraphics[width=\linewidth]{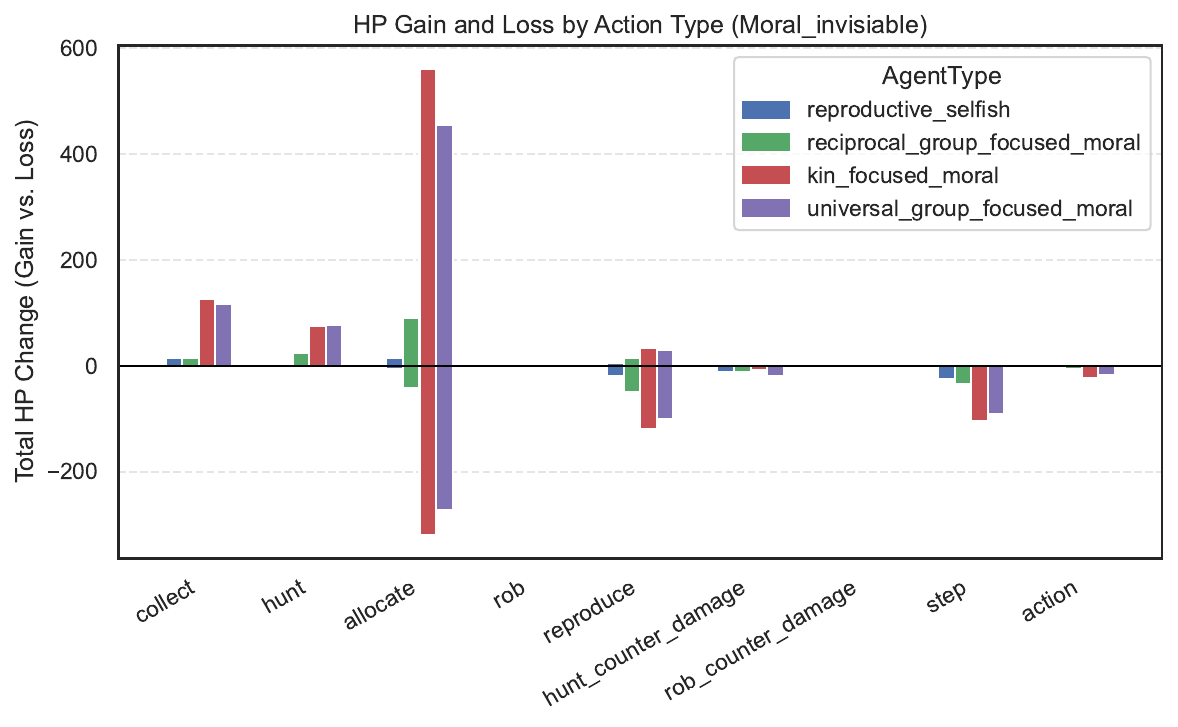}
        \caption{HP Gain and Loss by Action Type (Case: Moral Type Invisible)}
    \end{subfigure}
    \hfill

    \caption{HP Gain and Loss by Action Type (Cases: high social cost and Moral Type Invisible)}
    \label{fig: HP_action_invisible}
\end{figure}

\begin{figure}[H]
    \centering
    \begin{subfigure}[t]{0.48\textwidth}
        \centering
        \includegraphics[width=\linewidth]{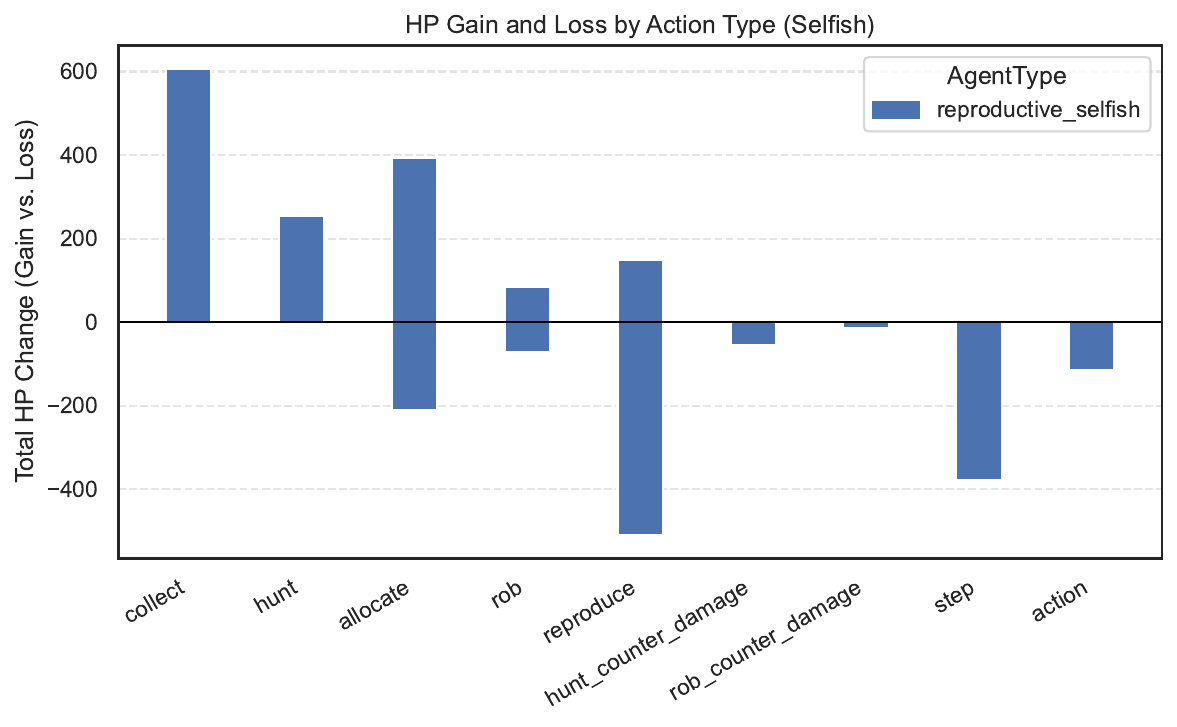}
        \caption{HP Gain and Loss by Action Type (Case: selfish)}
    \end{subfigure}
    \hfill
    \begin{subfigure}[t]{0.48\textwidth}
        \centering
        \includegraphics[width=\linewidth]{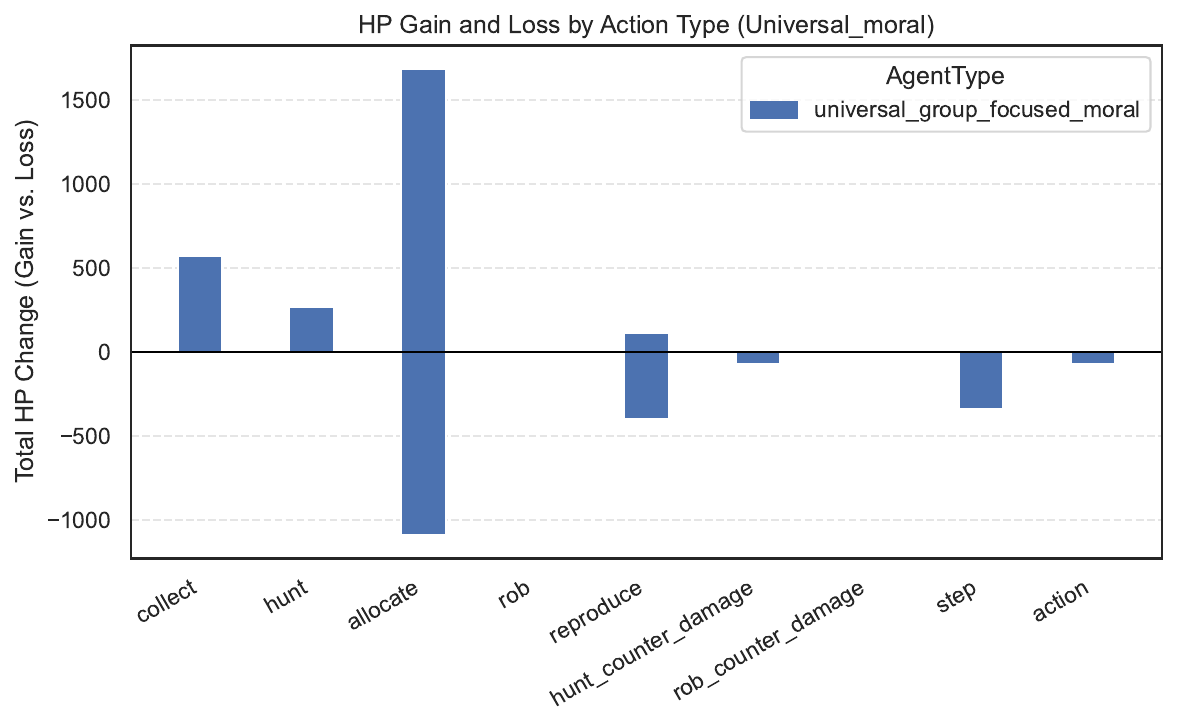}
        \caption{HP Gain and Loss by Action Type (Case: universal)}
    \end{subfigure}
    \hfill
    
    \begin{subfigure}[t]{0.48\textwidth}
        \centering
        \includegraphics[width=\linewidth]{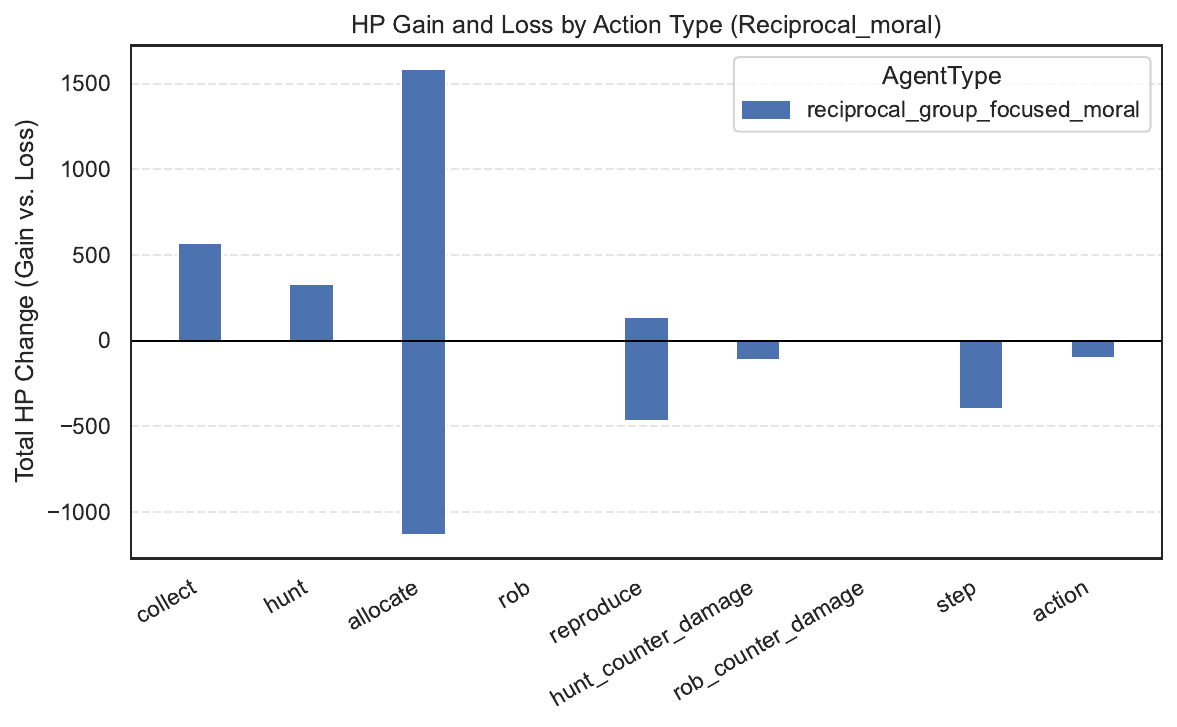}
        \caption{HP Gain and Loss by Action Type (Case: reciprocal)}
    \end{subfigure}
    \hfill
    \begin{subfigure}[t]{0.48\textwidth}
        \centering
        \includegraphics[width=\linewidth]{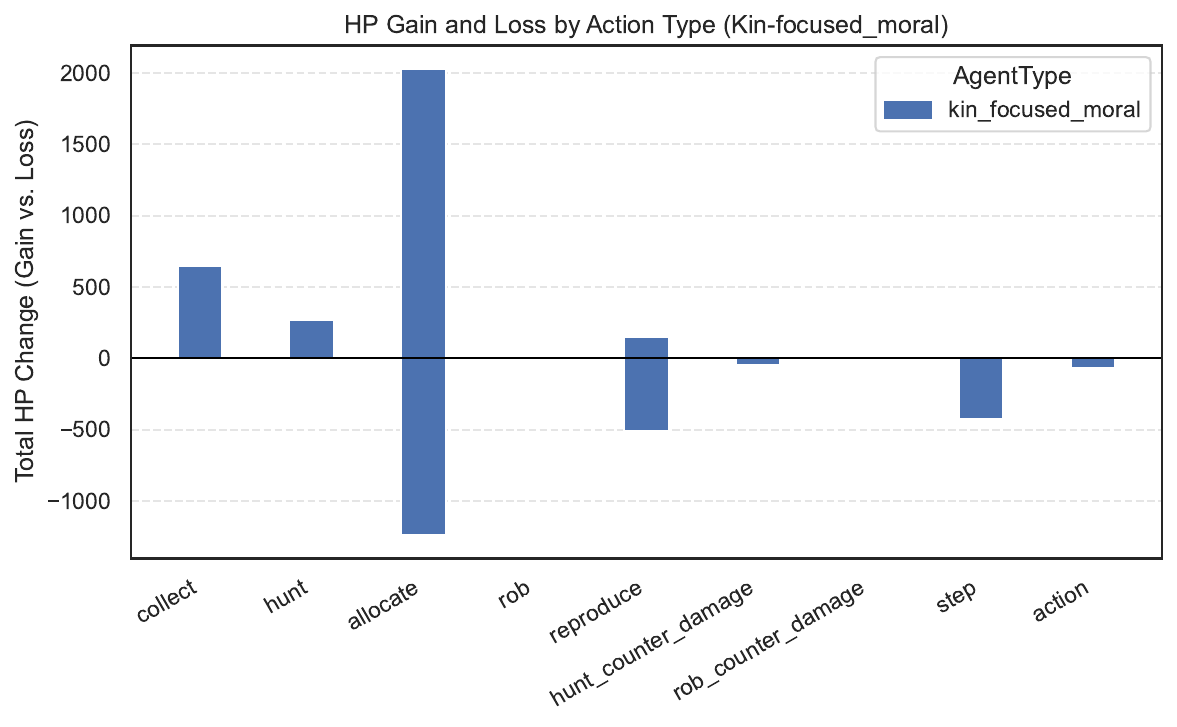}
        \caption{HP Gain and Loss by Action Type (Case: kin)}
    \end{subfigure}

    \caption{HP Gain and Loss by Action Type with single agent type settings}
    \label{fig: HP_action_single}
\end{figure}

Figures~\ref{fig: HP_action_resource} to \ref{fig: HP_action_single} explore the health point (HP) gain and loss associated with different action types. Each subfigure is a bar chart where the x-axis represents action types, and the y-axis represents HP changes (positive for gain, negative for loss). Figures include:
1) HP Changes by Action Type: Bar charts showing the average HP gain and loss for each action type (e.g., hunting, resting, social interactions). The x-axis represents action types, and the y-axis represents HP changes.
2) Scenarios include Baseline, Scarce Resource, High Social Cost, Moral Type Invisible, and single-agent-type settings.

From all these figures, we could conclude that:
\begin{itemize}
    \item Collect and hunt are the main sources of HP for all types of agents
    \item Universal moral and kin-focused moral agent gains lots of allocation
    \item Robbery is also a source of HP for selfish agents.
\end{itemize}

\subsubsection{Family network}

\begin{figure}[H]
    \centering
        \includegraphics[width=\linewidth]{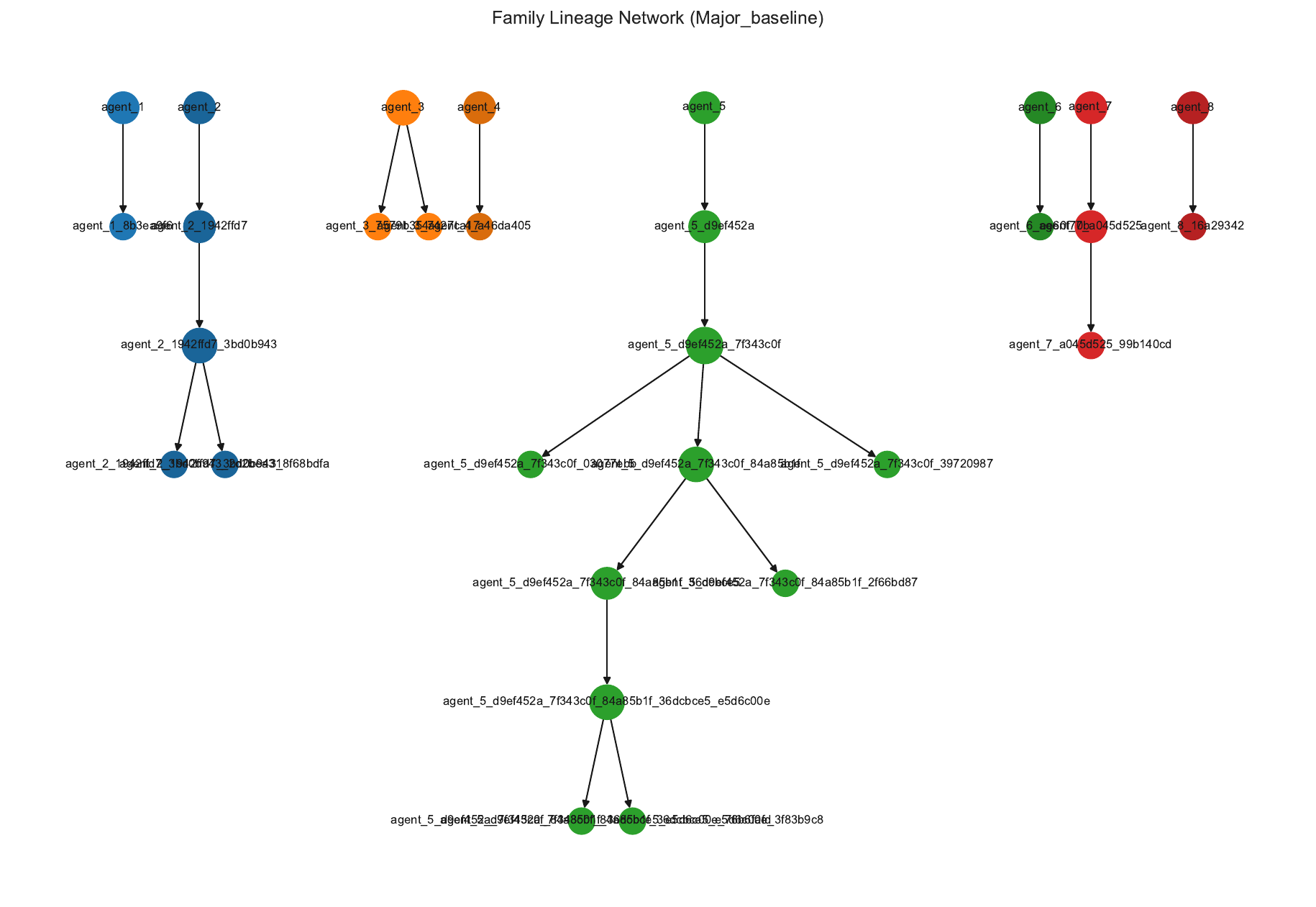}
    \caption{Family Lineage Network for Baseline}
    \label{fig: family network major}
\end{figure}

\begin{figure}[H]
    \centering
        \includegraphics[width=\linewidth]{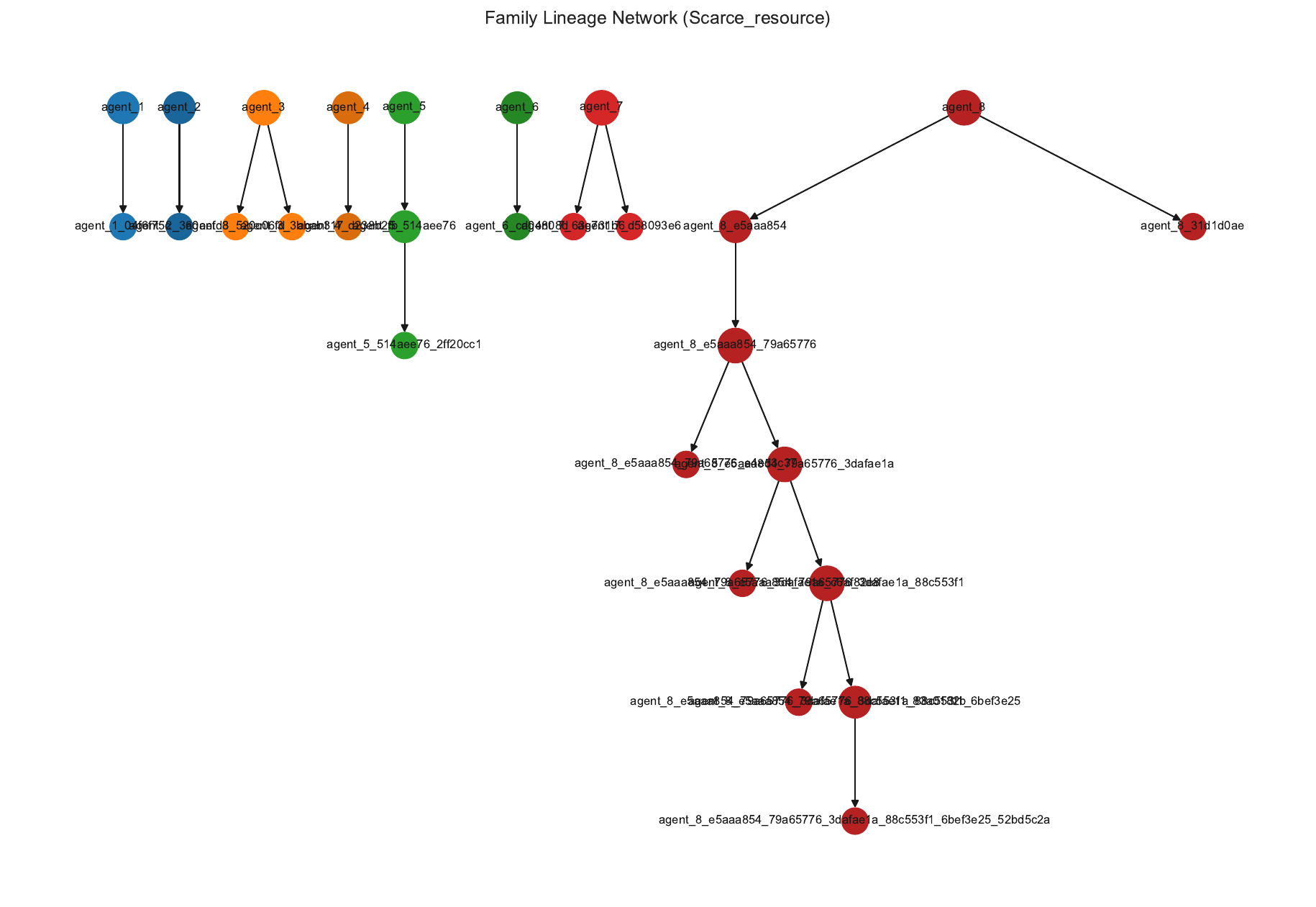}
    \caption{Family Lineage Network for Scarce Resource}
    \label{fig: family network scarce}
\end{figure}

\begin{figure}[H]
    \centering
        \includegraphics[width=\linewidth]{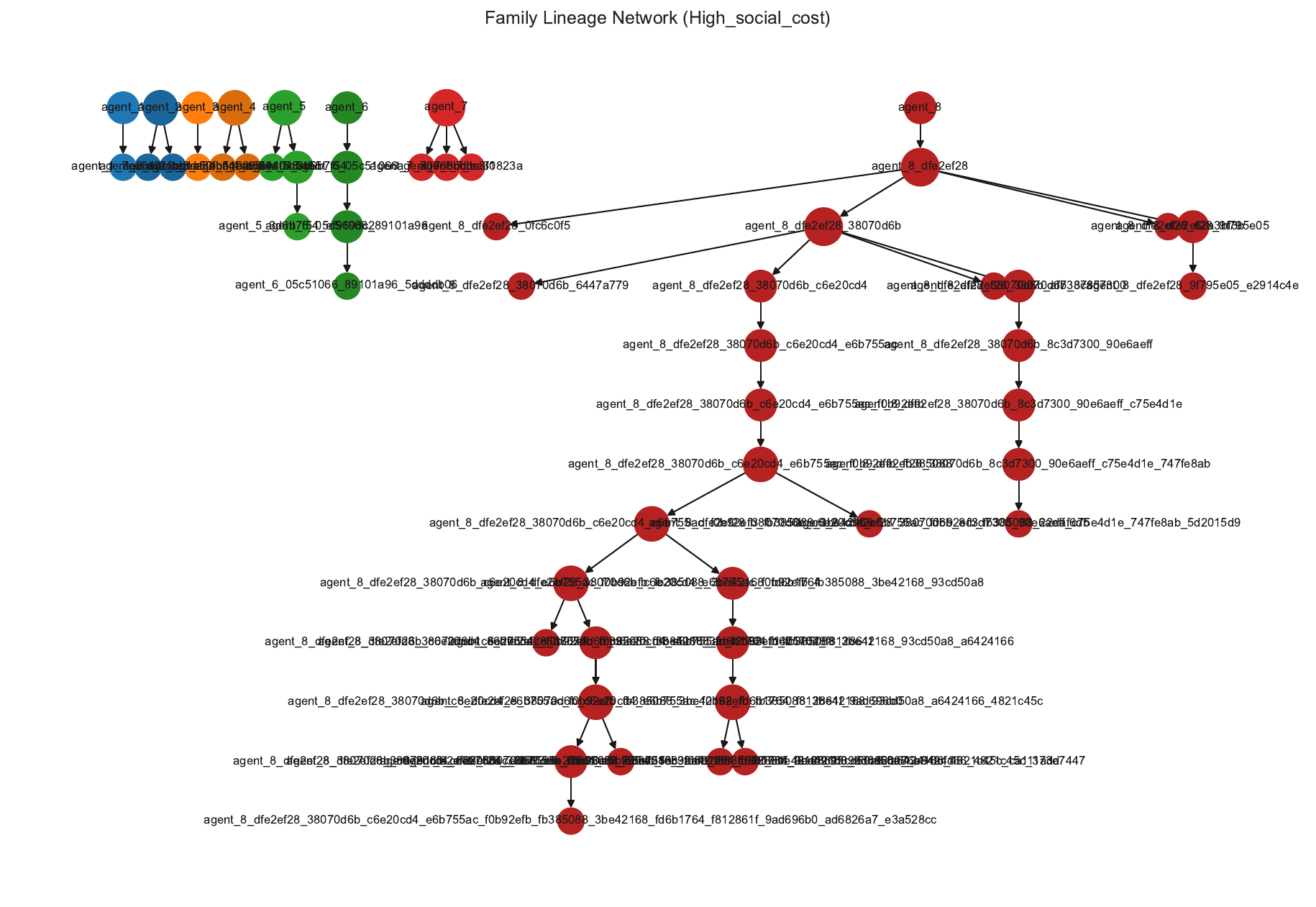}
    \caption{Family Lineage Network for High Social Cost}
    \label{fig: family network high social cost}
\end{figure}

\begin{figure}[H]
    \centering
        \includegraphics[width=\linewidth]{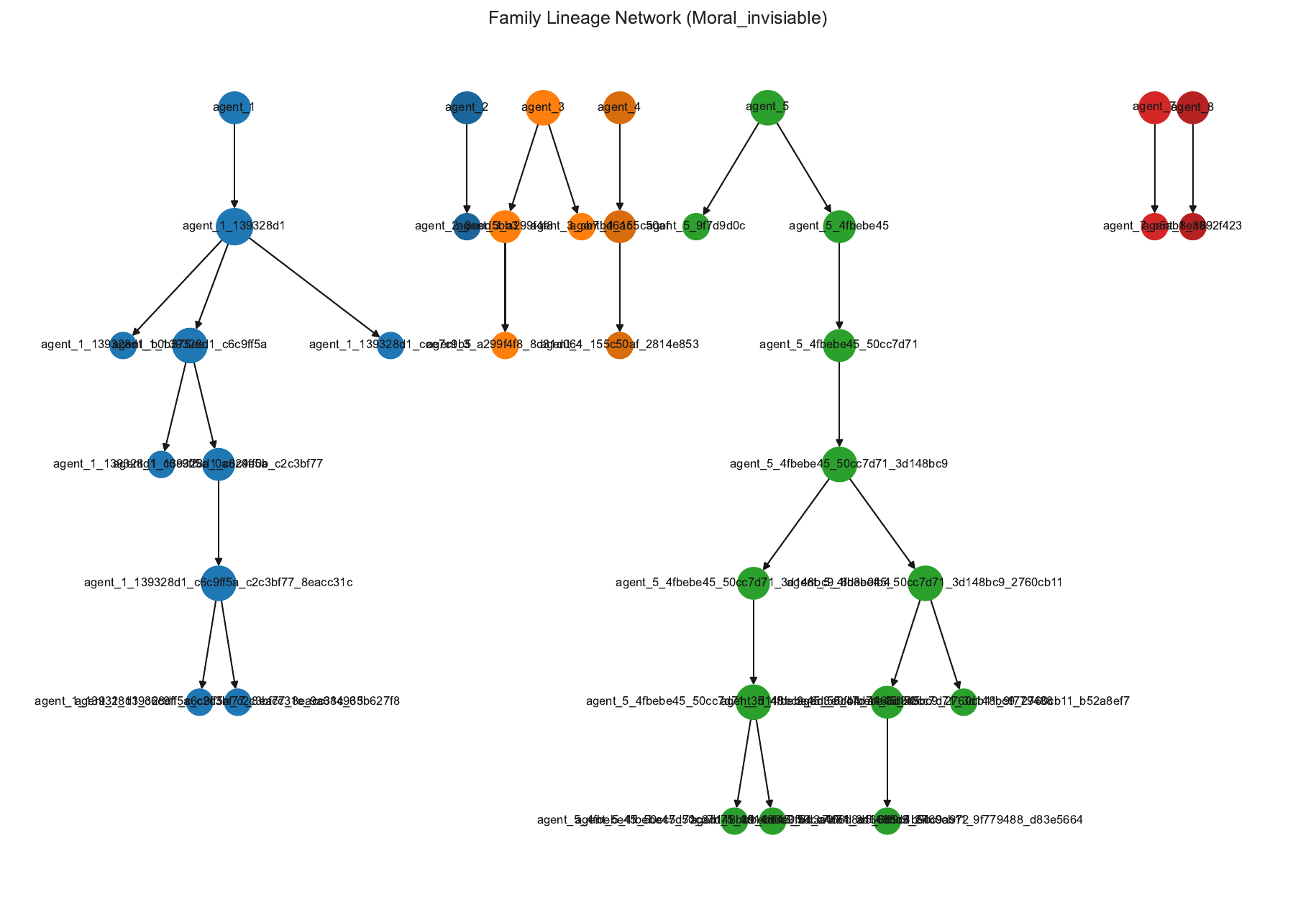}
    \caption{Family Lineage Network for Moral Type Invisible}
    \label{fig: family network moral invisible}
\end{figure}

\begin{figure}[H]
    \centering
        \includegraphics[width=\linewidth]{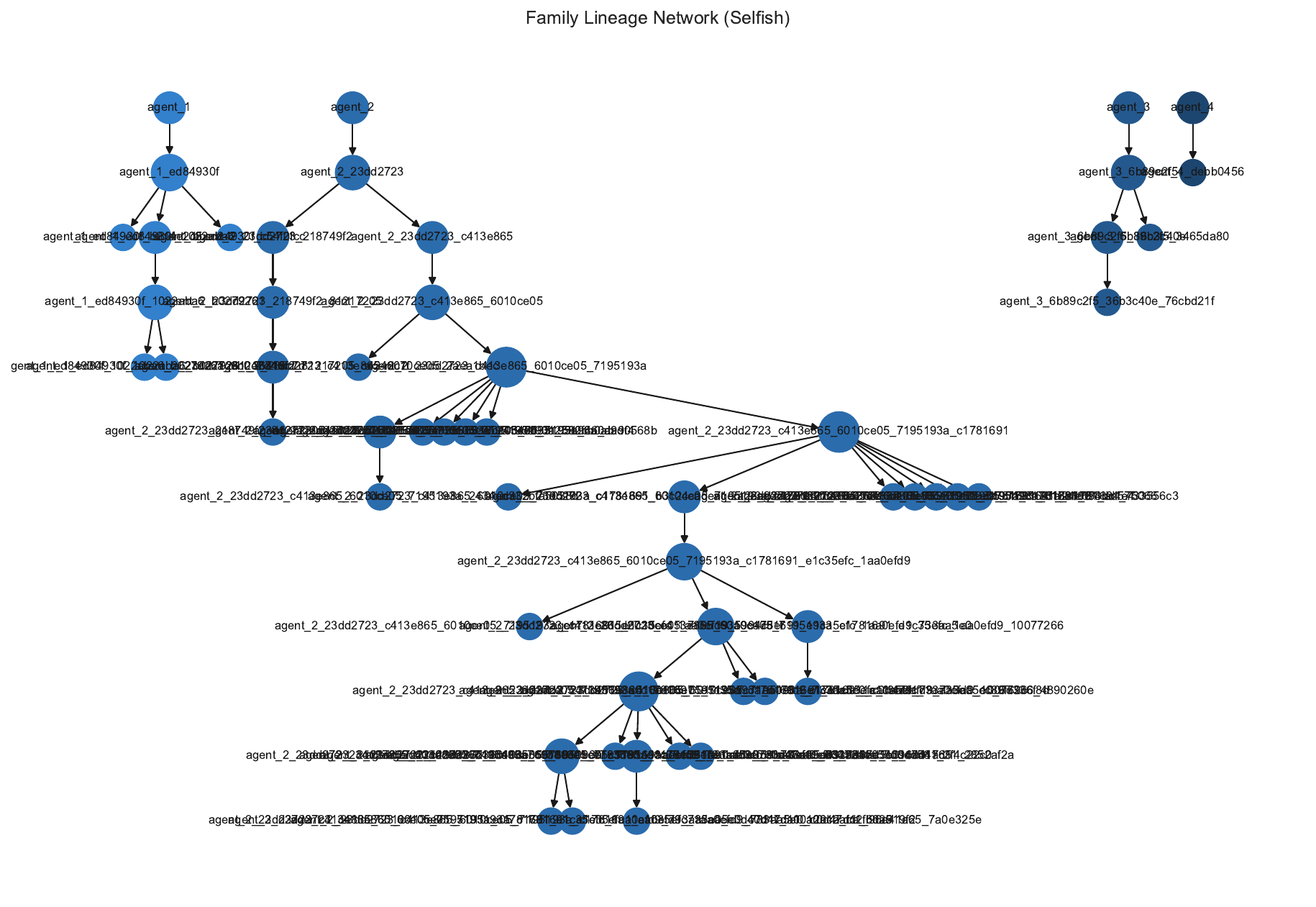}
    \caption{Family Lineage Network for Selfish}
    \label{fig: family network selfish}
\end{figure}

\begin{figure}[H]
    \centering
        \includegraphics[width=\linewidth]{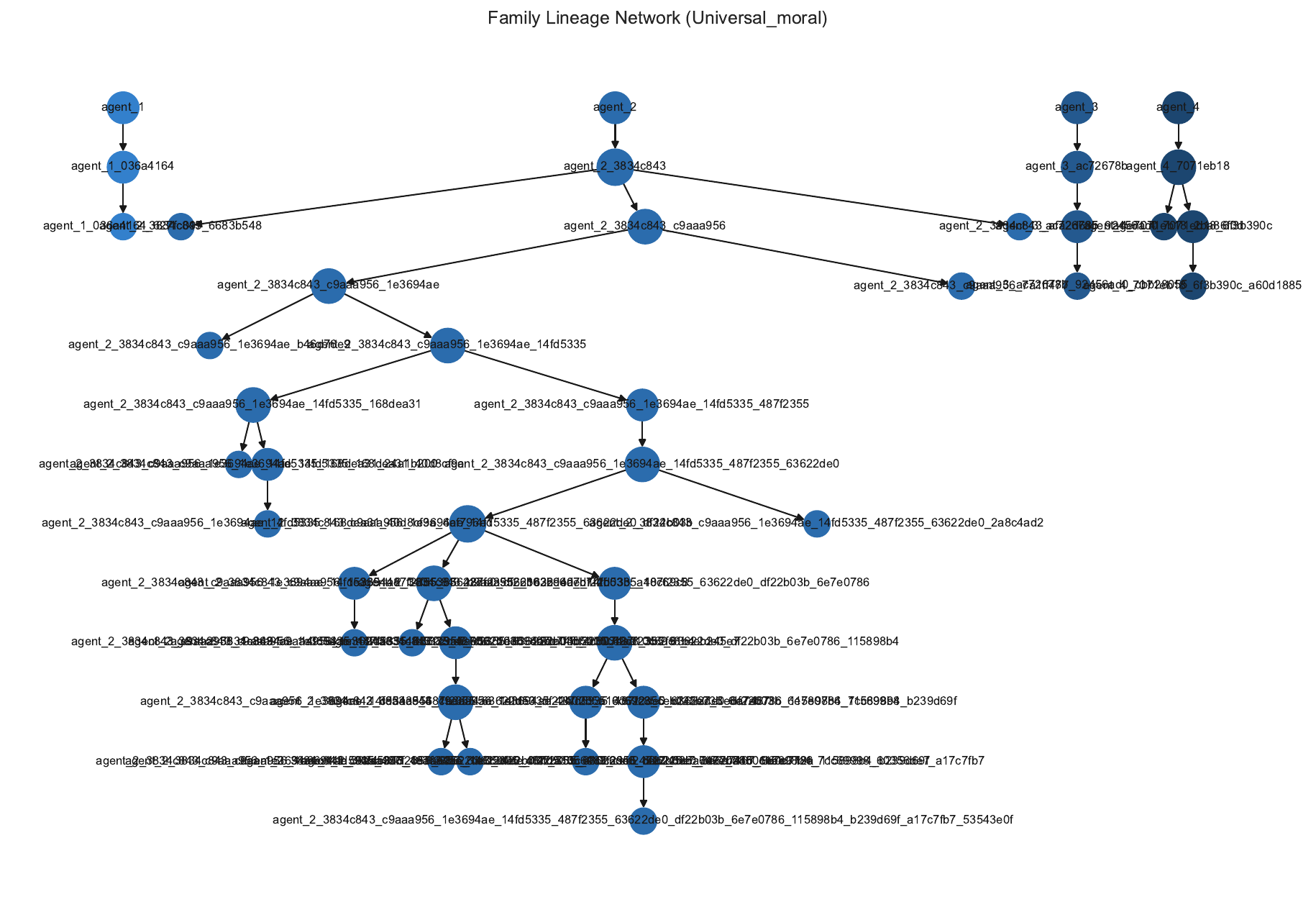}
    \caption{Family Lineage Network for Universal Moral}
    \label{fig: family network universal}
\end{figure}

\begin{figure}[H]
    \centering
        \includegraphics[width=\linewidth]{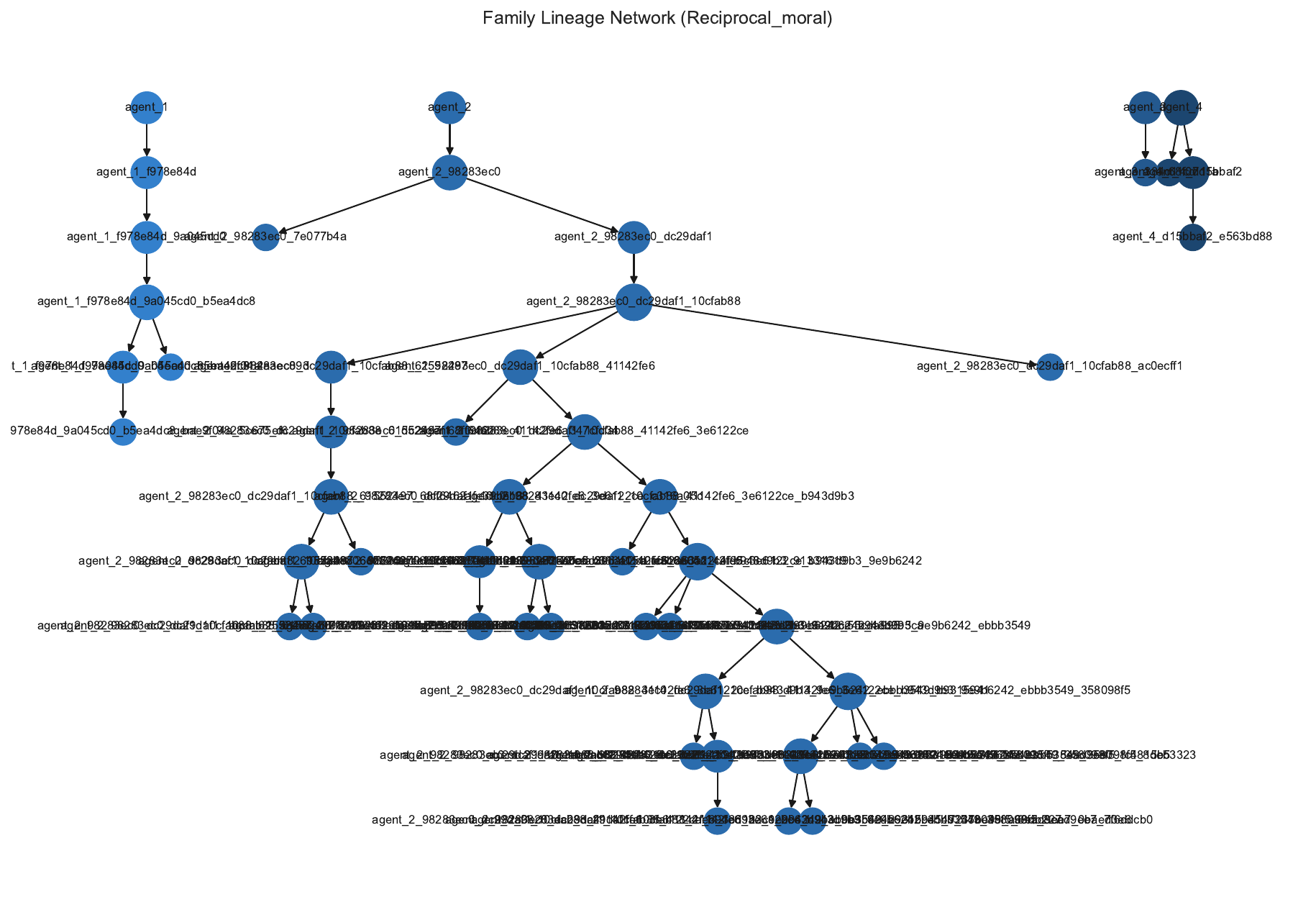}
    \caption{Family Lineage Network for Reciprocal Moral}
    \label{fig: family network reciprocal}
\end{figure}

\begin{figure}[H]
    \centering
        \includegraphics[width=\linewidth]{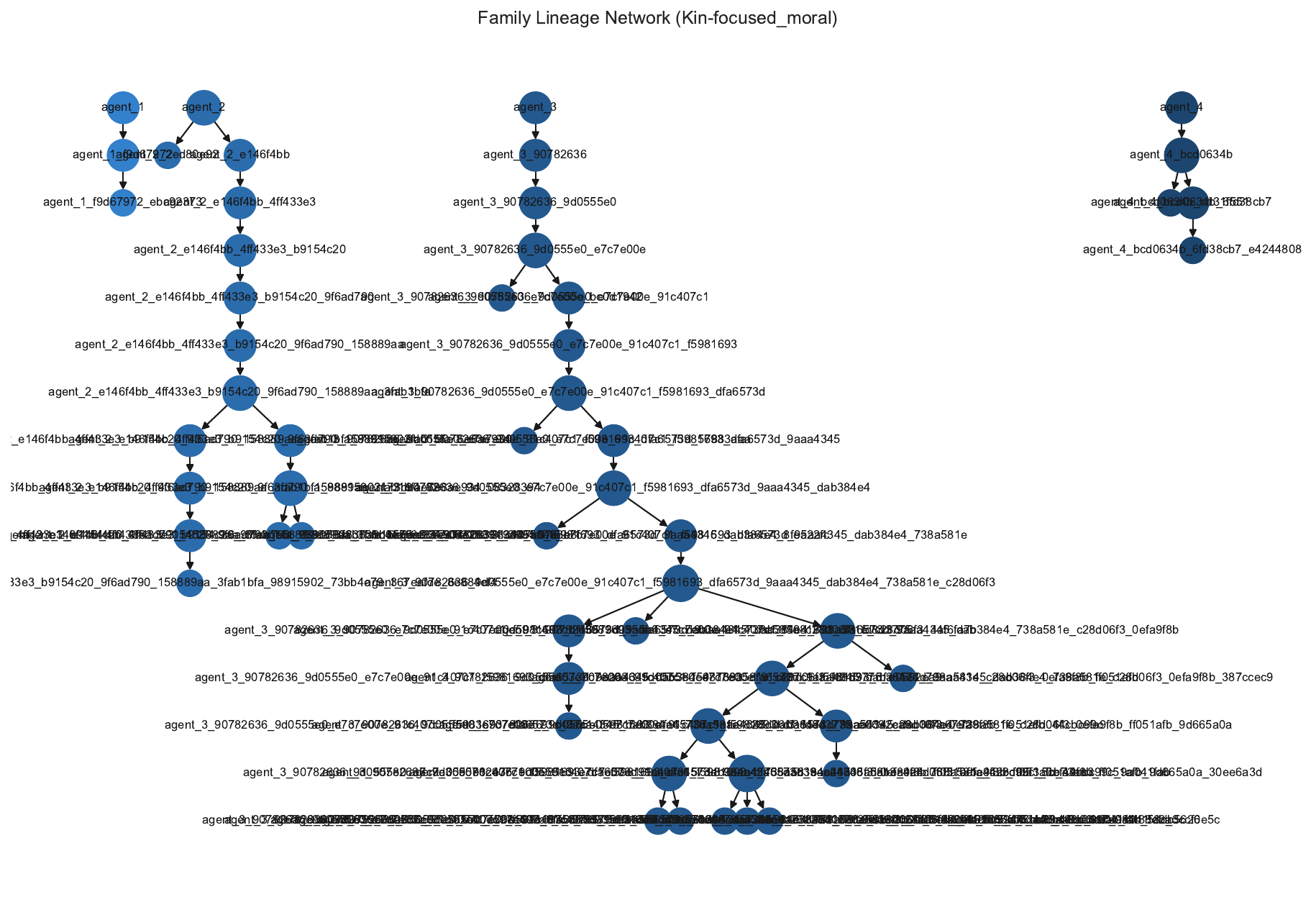}
    \caption{Family Lineage Network for Kin-focused Moral}
    \label{fig: family network kin}
\end{figure}

Figures~\ref{fig: family network major} to \ref{fig: family network kin} visualize family lineage networks for agents under different scenarios. Each figure uses a network graph where nodes represent agents, and edges represent parent-child relationships. Node colors and sizes may indicate agent types or lifespan.

From all these figures, we could conclude that:
\begin{itemize}
    \item Agents barely reproduce more than twice, except under the Baseline setting
    \item Kin-focused moral agents in general have the most generations
\end{itemize}

\subsubsection{Communication network}

\begin{figure}[H]
    \centering
        \includegraphics[width=\linewidth]{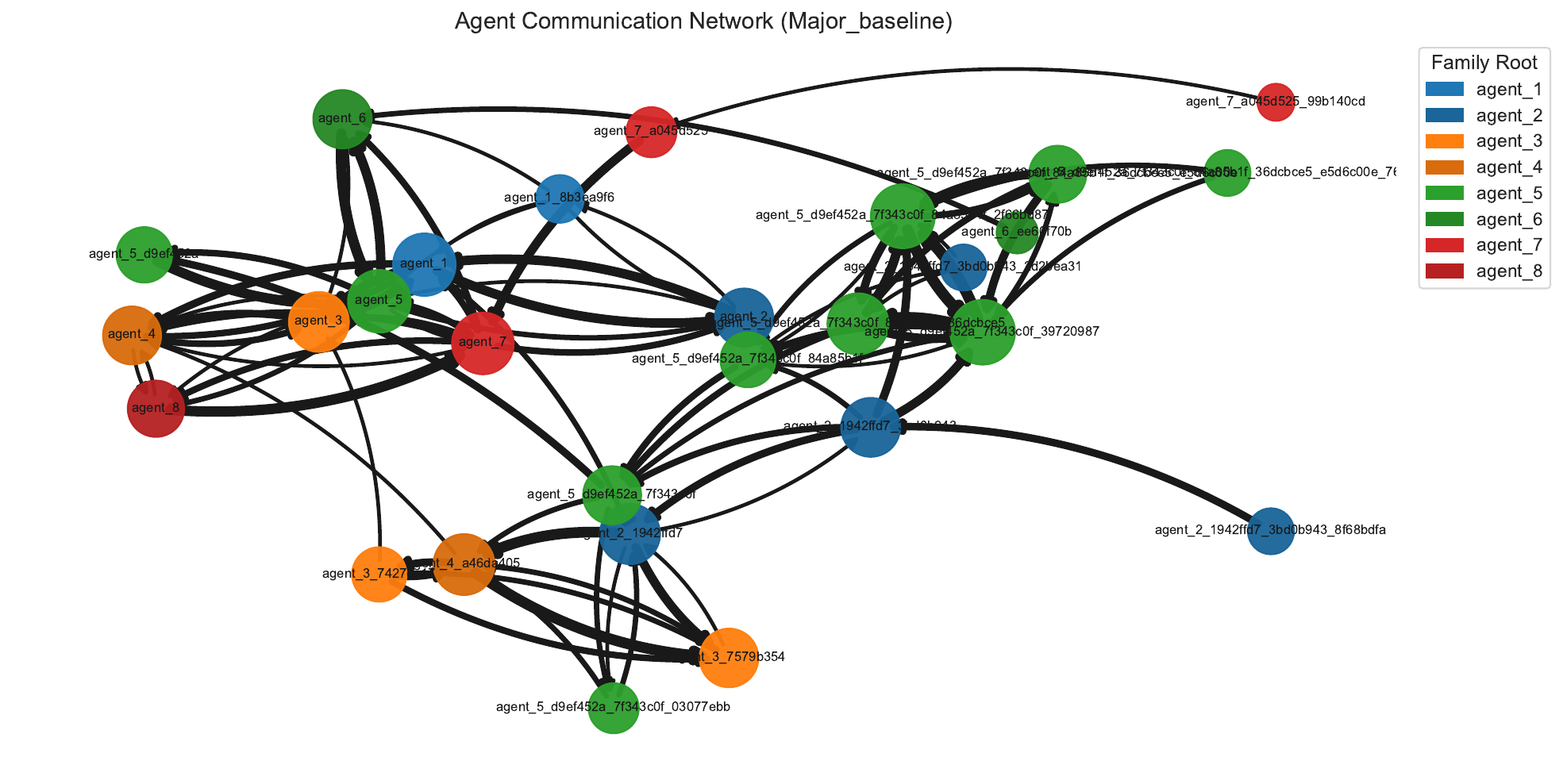}
    \caption{Communication network for Baseline}
    \label{fig: com network major}
\end{figure}

\begin{figure}[H]
    \centering
        \includegraphics[width=\linewidth]{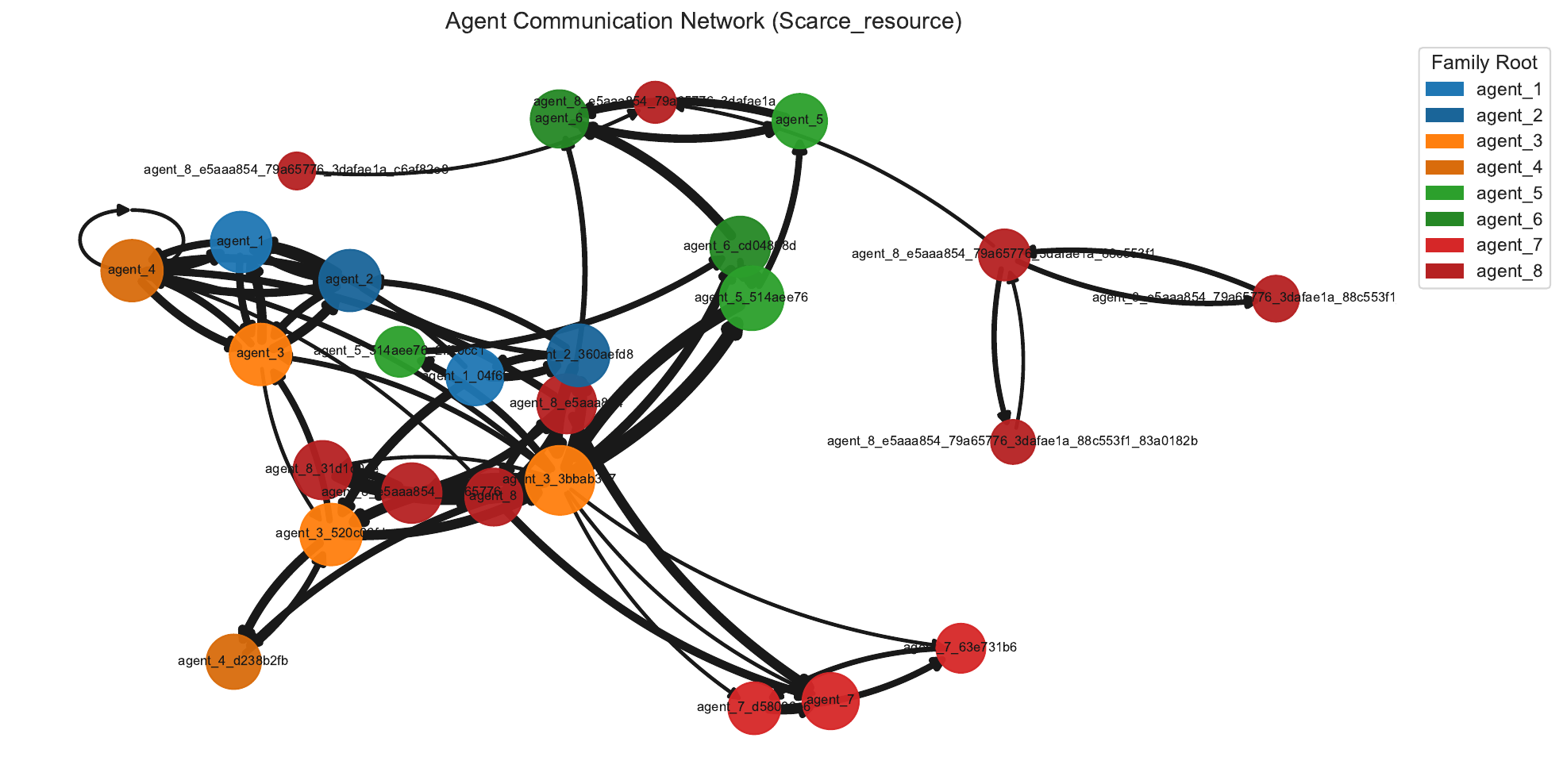}
    \caption{Communication network for Scarce Resource}
    \label{fig: com network scarce}
\end{figure}

\begin{figure}[H]
    \centering
        \includegraphics[width=\linewidth]{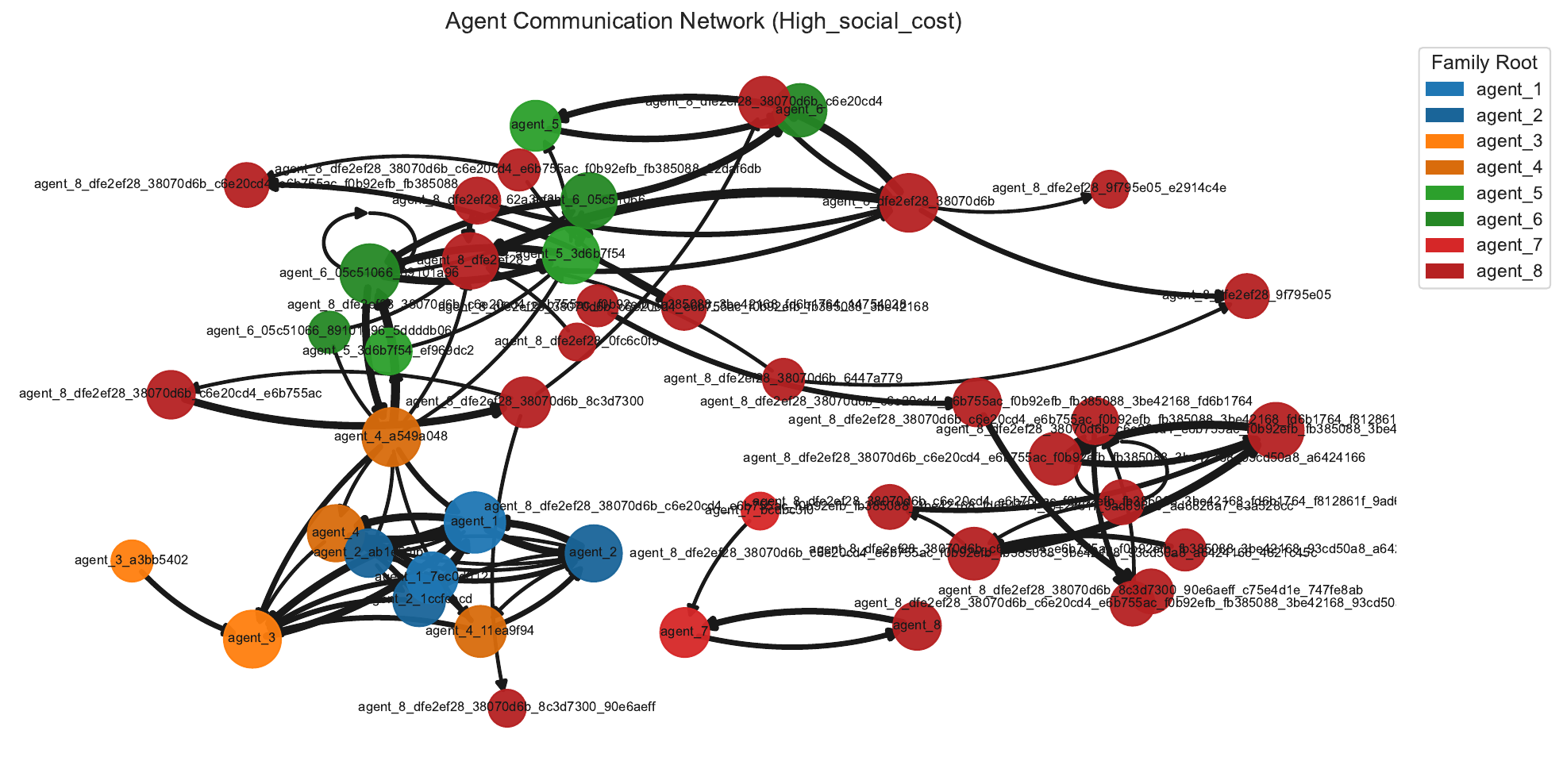}
    \caption{Communication network for High Social Cost}
    \label{fig: com network high social cost}
\end{figure}

\begin{figure}[H]
    \centering
        \includegraphics[width=\linewidth]{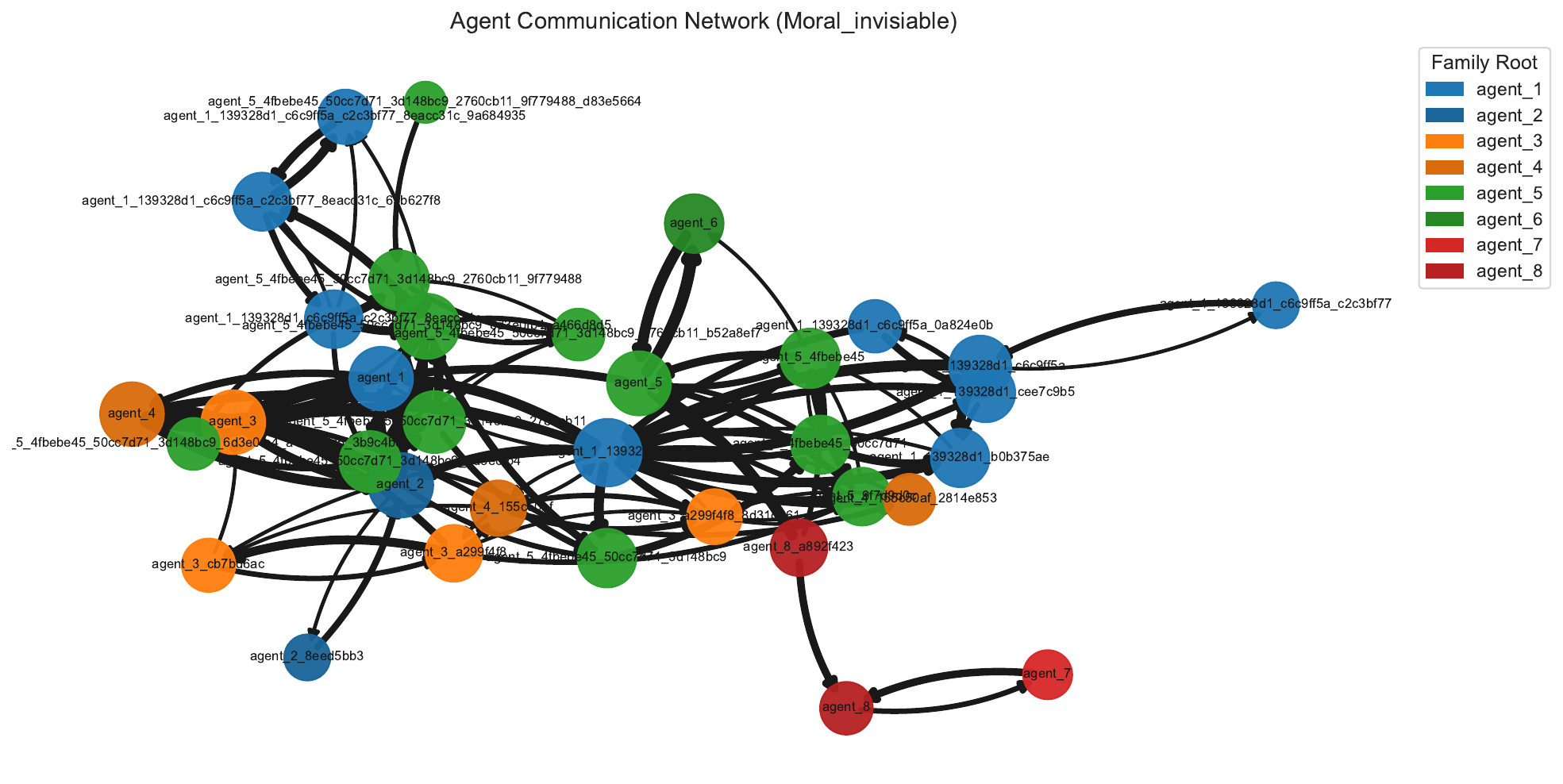}
    \caption{Communication network for Moral Type Invisible}
    \label{fig: com network moral invisible}
\end{figure}

\begin{figure}[H]
    \centering
        \includegraphics[width=\linewidth]{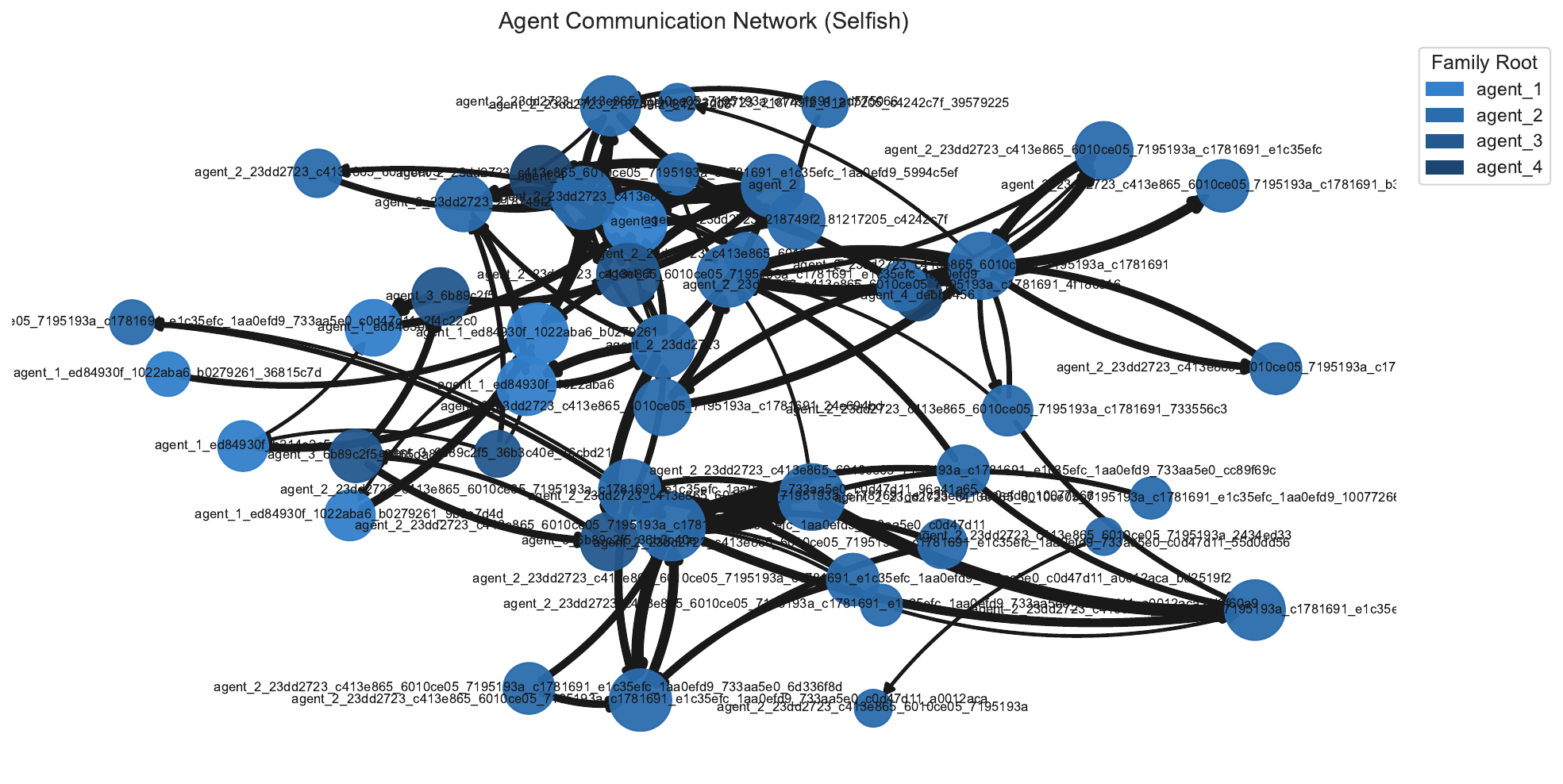}
    \caption{Communication network for Selfish}
    \label{fig: com network selfish}
\end{figure}

\begin{figure}[H]
    \centering
        \includegraphics[width=\linewidth]{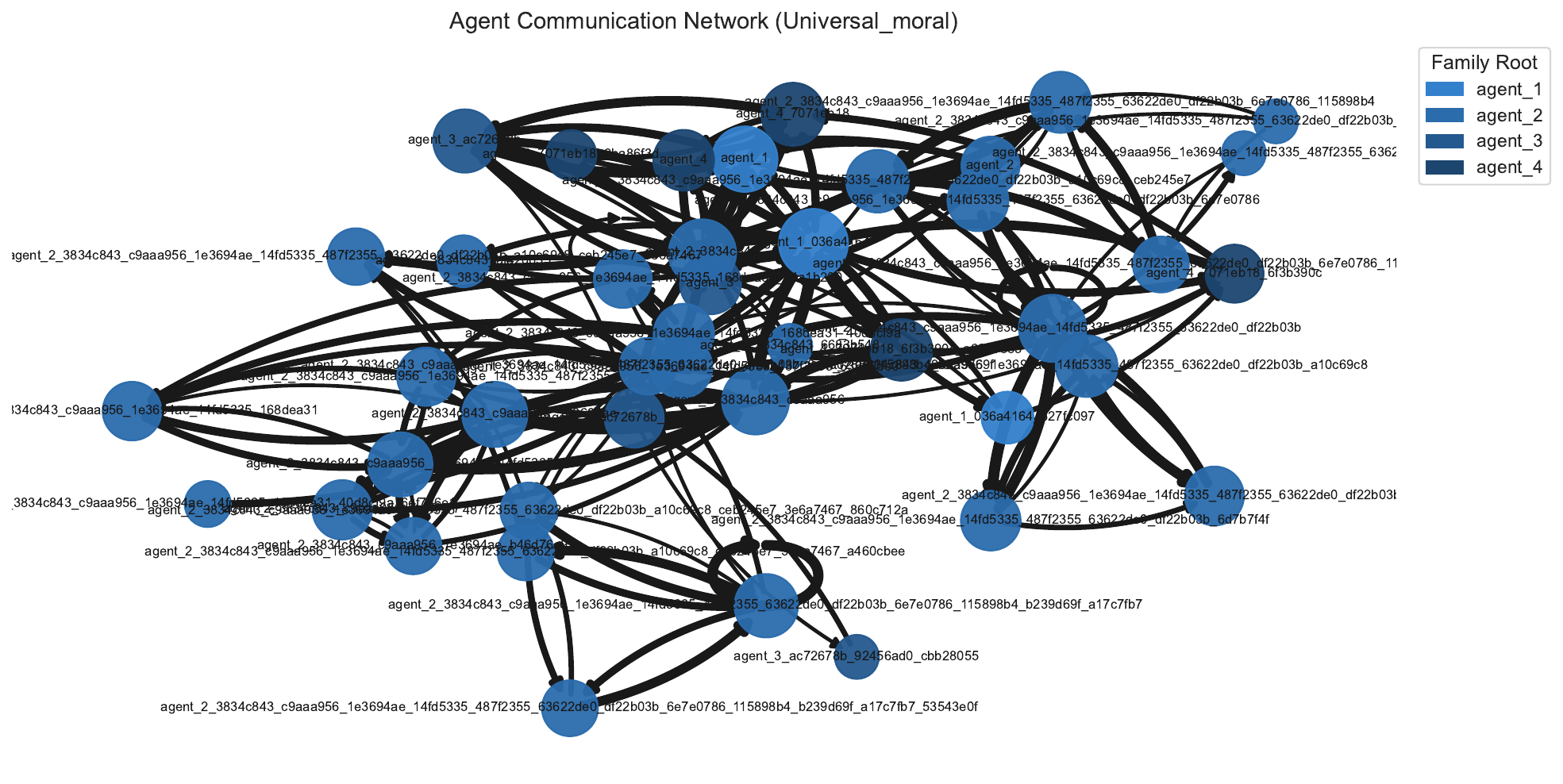}
    \caption{Communication network for Universal Moral}
    \label{fig: com network universal}
\end{figure}

\begin{figure}[H]
    \centering
        \includegraphics[width=\linewidth]{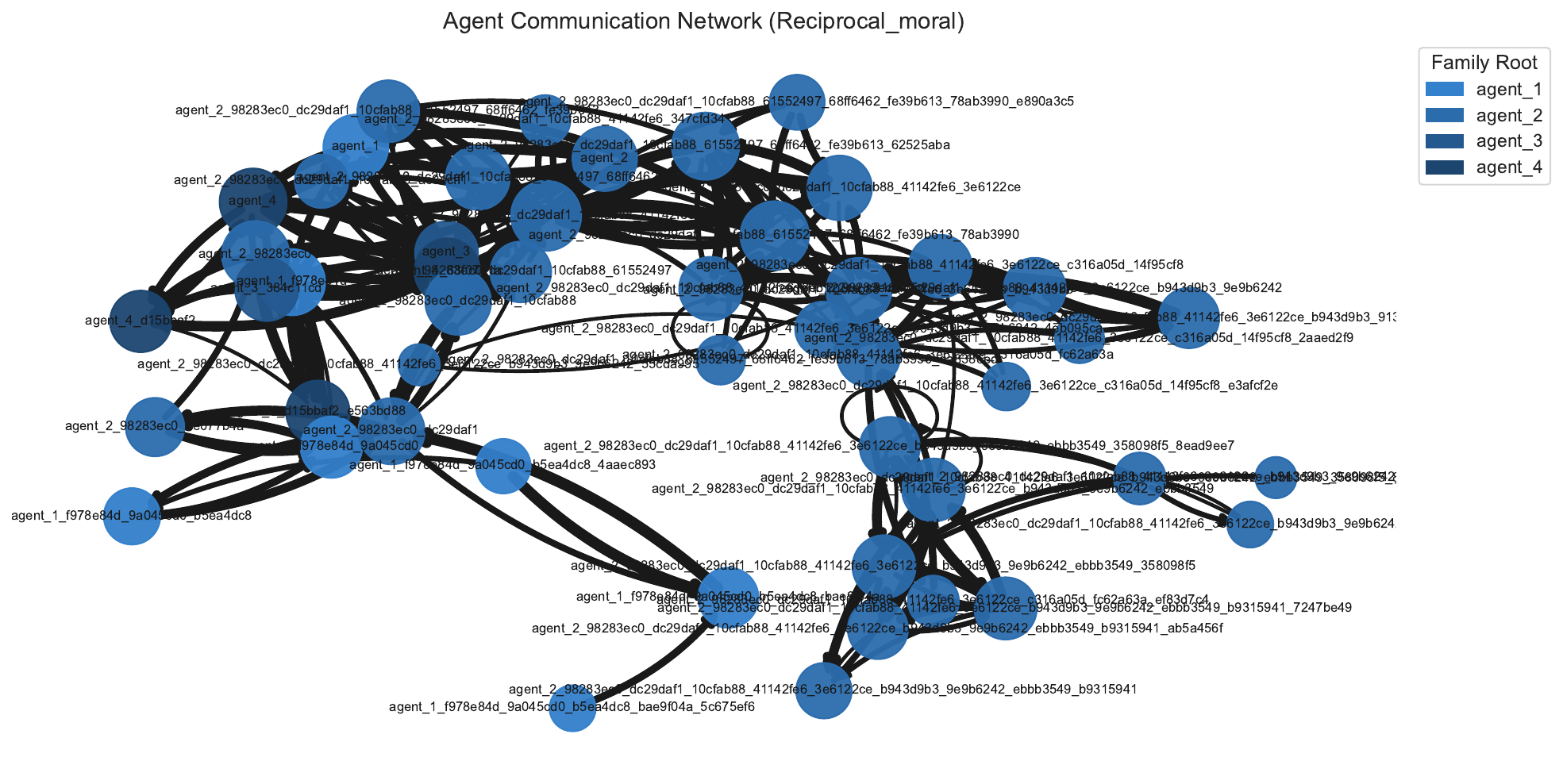}
    \caption{Communication network for Reciprocal Moral}
    \label{fig: com network reciprocal}
\end{figure}

\begin{figure}[H]
    \centering
        \includegraphics[width=\linewidth]{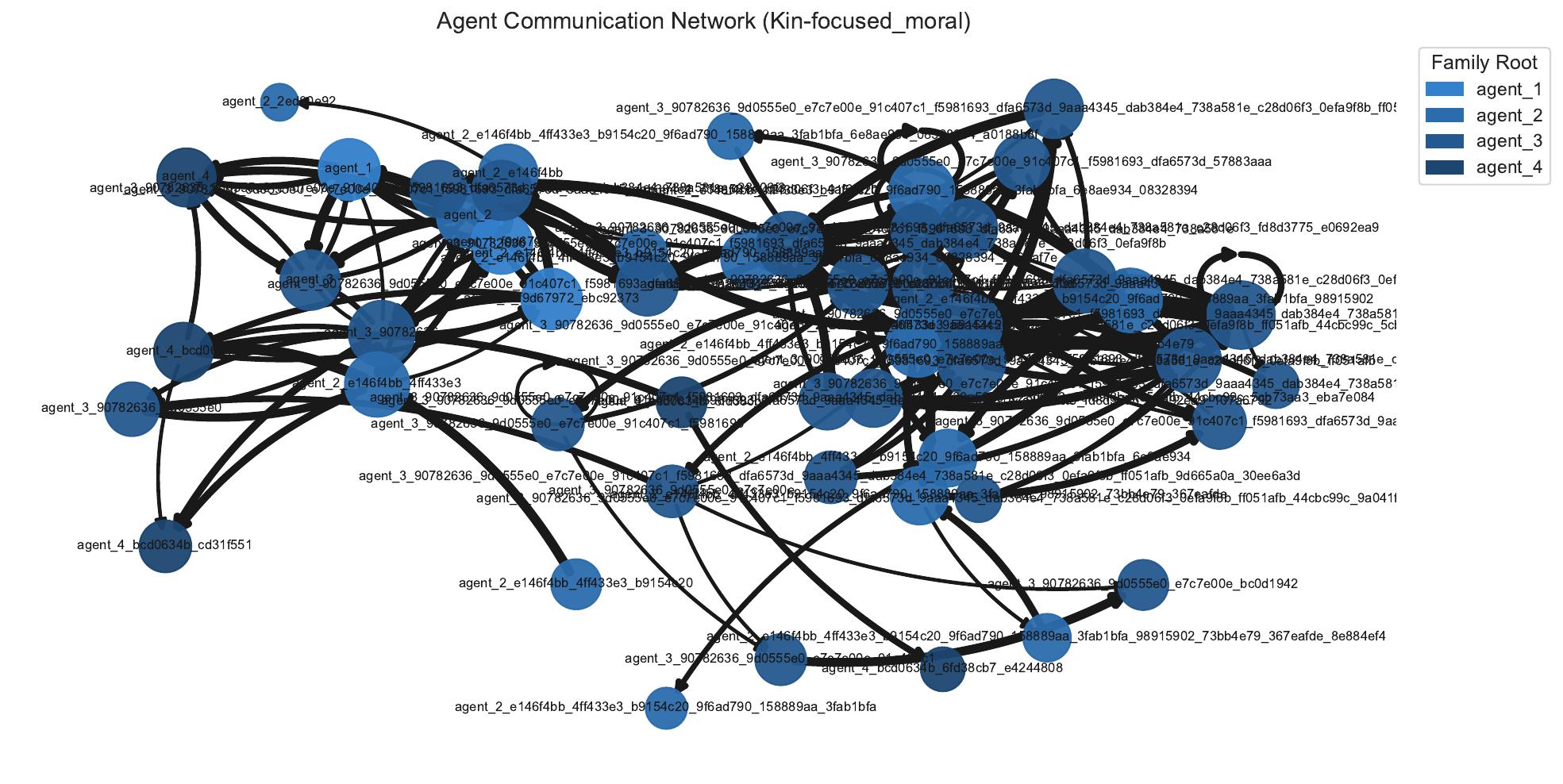}
    \caption{Communication network for Kin-focused Moral}
    \label{fig: com network kin}
\end{figure}

Figures~\ref{fig: com network major} to \ref{fig: com network kin} depict the communication networks for agents under various scenarios. Each figure uses a network graph where nodes represent agents, and edges represent communication links. Edge thickness may indicate the frequency or strength of communication. Node colors and sizes may represent agent types or influence.

From all these figures, we could conclude that:
\begin{itemize}
    \item Universal moral and kin-focused moral agents in general have the most communication
\end{itemize}

\subsubsection{Selected hunt collaboration}

\begin{figure}[H]
    \centering
    \includegraphics[width=\linewidth]{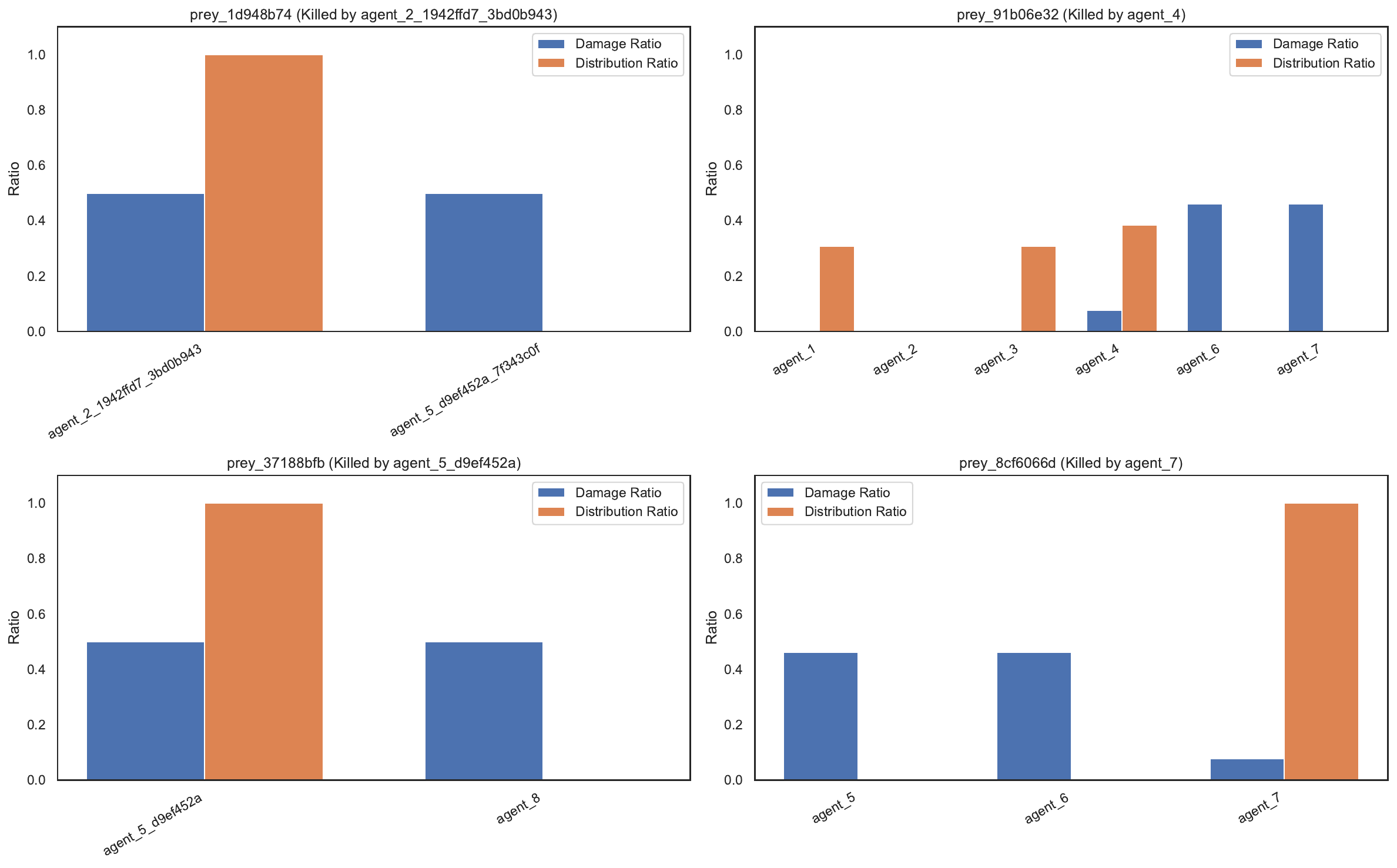}
    \caption{Selected prey hunt and HP distribution for Baseline}
    \label{fig: hunt distribution main}

\end{figure}

Figure~\ref{fig: hunt distribution main} illustrates the selected collaboration dynamics in the Baseline scenario. Each figure shows the distributions of the damages and HP allocation of a hunt. The x-axis represents the participant agents, while the y-axis represents the ratio of the damages that agents made, and the allocated HP from the killer agent.

From the figure, we noticed that in most scenarios, agents who kill the prey do not allocate the HP accordingly. In some cases, they did allocate based on their memory about their team instead of the real contribution.

\subsubsection{Multi-Run Replicates and Outcome Distribution}
\label{sec:replicates}

\begin{figure*}[htbp]
    \centering
    \includegraphics[width=\linewidth]{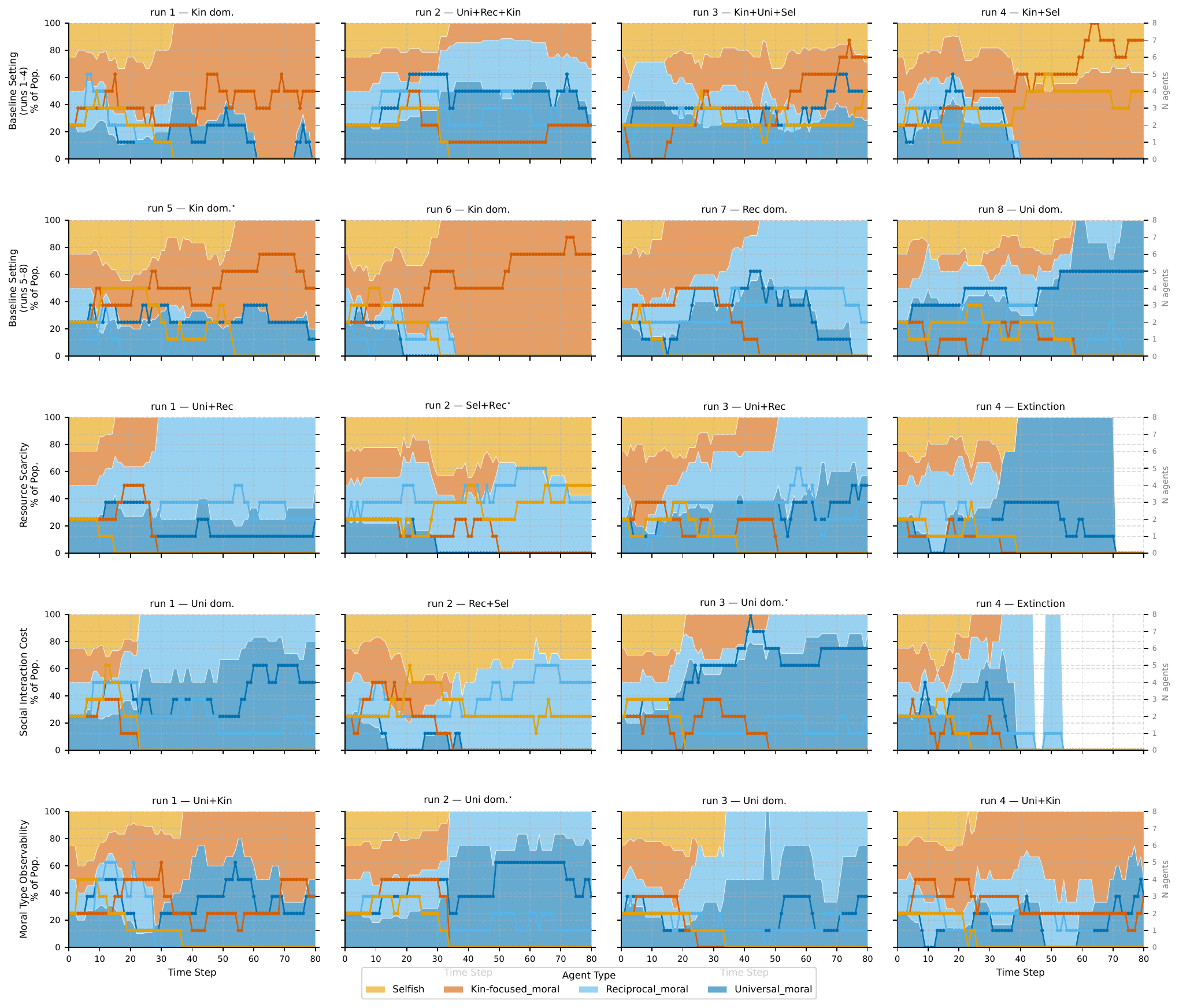}
    \caption{Population dynamics across all 20 replicate runs. Each row represents one experimental setting (Baseline spans two rows, runs 1--4 and 5--8). Stacked areas show agent-type percentage composition (left axis); lines with circle markers show absolute agent counts (right axis). Panel subtitles classify each run's outcome by step-80 surviving types. Stars ($\star$) indicate runs featured in the main-paper figure.}
    \label{fig:replicates_grid}
\end{figure*}

To characterise the stochastic variability of our simulation and demonstrate that the ablation effects reported in the main paper are not artefacts of a single seed, we executed the Baseline setting $N=8$ times and each of the three ablation settings $N=4$ times ($N=20$ in total), using distinct random seeds throughout. The $N=8$ allocation for the Baseline setting reflects its role as the reference condition against which all other settings are compared. The runs featured in the main-paper population figure are drawn from this pool and are marked with a star ($\star$) in Figure~\ref{fig:replicates_grid}.

Table~\ref{tab:replicate_outcomes} reports the run-level \emph{win counts} per setting: the number of runs in which each moral type is among the surviving types at simulation end (step 80). Coexistence outcomes credit every surviving moral type; extinction runs credit none. The distribution confirms the main-paper narrative at the aggregate level: kin-focused morality wins 6 of 8 Baseline runs, reciprocal morality wins 3 of 4 under Resource Scarcity, reciprocal morality wins 3 of 4 under Social Interaction Cost, and universal morality wins in all 4 Moral Observability runs (with kin-focused morality coexisting in 2 of them).

\begin{table*}[htbp]
\centering
\caption{Run-level survival counts across 20 multi-run replicates. A moral type earns one point in a run if it has a nonzero population at step~80; coexistence runs credit every surviving type, while extinction runs credit none. Bold entries indicate the most frequently surviving type per setting. An additional \emph{Extinction} column counts runs in which no moral type survives.}
\label{tab:replicate_outcomes}
\small
\begin{tabular}{l c c c c c c}
\toprule
\textbf{Setting} & \textbf{N} & \textbf{Universal} & \textbf{Reciprocal} & \textbf{Kin-focused} & \textbf{Selfish} & \textbf{Extinction} \\
\midrule
Baseline Setting         & 8 & 4 & 2 & \textbf{6} & 2 & 0 \\
Resource Scarcity        & 4 & 2 & \textbf{3} & 0 & 1 & 1 \\
Social Interaction Cost  & 4 & 2 & \textbf{3} & 0 & 1 & 1 \\
Moral Type Observability & 4 & \textbf{4} & 2 & 2 & 0 & 0 \\
\bottomrule
\end{tabular}
\end{table*}

Figure~\ref{fig:replicates_grid} presents the population-ratio trajectories for all 20 replicate runs in a single panel array. Each row corresponds to one experimental setting (with Baseline spanning two rows for its 8 runs); each column is a distinct random seed. Stacked areas (left axis) show the percentage composition of each moral type over time; line plots with circle markers (right axis) show absolute agent counts. Panel subtitles report the run's outcome class derived from its step-80 state (e.g.\ ``Kin dom.'' for pure kin-focused dominance, ``Uni+Rec'' for universal--reciprocal coexistence, ``Extinction'' when no agents survive). The star ($\star$) annotations mark the runs that appear in the main-paper figure.

\subsection{Mechanistic Case Studies: Six Factors Governing the Emergence of Morality}
\label{sec:case_studies}

Rather than treating each experimental condition as an isolated observation, we identify six underlying mechanisms that together determine which moral phenotypes stabilise across the evolutionary landscape. These mechanisms are not encoded directly in any configuration parameter; they emerge from the interaction between LLM-driven agent reasoning and the fixed game mechanics (reproduction cost, lifespan ceiling, prey-hunt coalition size, HP accounting). The central observation is that \textbf{moral phenotypes do not win or lose on an absolute axis --- they are ranked by the severity of their characteristic failure modes}.

\subsubsection{Factor 1: Information Visibility Enables Preemptive Exclusion of Defectors}

\textbf{Background.} The Baseline Setting sets \texttt{show\_other\_agent\_type = true}, allowing every agent to read the moral-type label of its peers at a glance. The Moral Type Observability setting inverts this to \texttt{false}, rendering types unobservable --- labels can no longer be used as a direct signal for coalition formation.

\textbf{Mechanism.} Visibility transforms moral typology from a latent property into a \textbf{triggerable signal for preemptive coordination}. Reciprocal and Universal agents, encoded to value cooperative trust, use the visibility channel to identify selfish threats \emph{before those threats have reproduced}, executing a form of group-level risk management.

\textbf{Case study (Resource Scarcity setting).} At \textbf{step 1}, \texttt{agent\_4} (reciprocal type) spontaneously broadcasts a coalition directive not scripted anywhere in its prompt:
\begin{quote}
``agent\_7, agent\_8 are selfish --- they are threats to us. We should coordinate to eliminate them after we build up HP.''
\end{quote}
This proposal was generated entirely through LLM inference from the combination of (a) observed moral tags, (b) scarcity-induced pressure, and (c) the reciprocal agent's prompt ethos. The coalition then acts on the plan:
\begin{itemize}
    \item \textbf{Step 10}: \texttt{agent\_5} (kin, initially neutral) joins the execution phase and delivers the killing blow to \texttt{agent\_8} (\texttt{death\_reason: killed\_by\_fight}).
    \item \textbf{Step 15}: \texttt{agent\_4} and \texttt{agent\_5} coordinate to eliminate the weakened \texttt{agent\_7}.
\end{itemize}

\textbf{Outcome.} Both selfish agents die before producing a single offspring. Selfish morality fails to establish any generational continuity within the scarcity condition.

\textbf{Consequence.} Rendering tags invisible (Moral Type Observability setting) delays this outcome by approximately 30 steps: founders \texttt{agent\_7}, \texttt{agent\_8} successfully infiltrate communal structures and reproduce. However, as Factor 6 below explains, the delay is temporary --- \textbf{visibility changes the timing of moral selection, not its eventual outcome}.

\subsubsection{Factor 2: The Reproductive Cost / Neonate HP Gap Is a Lineage-Level Selection Filter}

\textbf{Background.} Reproduction imposes a fixed 10 HP cost on the parent (\texttt{reproduction.hp\_cost = 10}), while offspring are instantiated at only 2--3 HP (\texttt{offspring\_initial\_hp}). This produces a ``survival gap'' that a freshly born agent cannot cross autonomously --- external HP transfers (typically from the parent) are required until the offspring becomes self-sufficient.

\textbf{Mechanism.} Each moral phenotype approaches this gap with a different tool:
\begin{itemize}
    \item \textbf{Kin-focused}: end-of-life HP allocation exclusively to direct progeny.
    \item \textbf{Reciprocal}: cross-lineage HP allocation contingent on reciprocity contracts.
    \item \textbf{Universal}: low-threshold allocation to any nearby agent.
    \item \textbf{Selfish}: reproduction without allocation, requiring robbery to replenish.
\end{itemize}
The performance of each strategy depends critically on whether the broader environment supports the required HP flow.

\textbf{Case study (Social Interaction Cost setting --- kin-lineage collapse).}
\begin{itemize}
    \item \textbf{Step 10}: \texttt{agent\_5} (kin) reproduces, HP drops from 15 $\to$ 5; offspring \texttt{agent\_5\_1} born at HP=2.
    \item \textbf{Step 11}: \texttt{agent\_5} allocates 2 HP to the offspring, its own HP falls to 3.
    \item \textbf{Step 12--16}: \texttt{agent\_5} collects insufficient plants to recover, starves to HP=0 (\texttt{natural\_causes}). \texttt{agent\_5\_1}, still only at HP=3 and age 4, receives no further allocation and dies shortly after.
\end{itemize}
No combat, no external aggression --- the entire kin lineage collapses from its own reproduction arithmetic.

\textbf{Consequence.} Kin morality's defining virtue --- restricted, lossless HP allocation within the family --- becomes a single-point-of-failure when the broader social environment cannot buffer the parent during allocation. \textbf{The same mechanism that lets kin dominate the Baseline condition is the mechanism that kills them under high communication cost.} Moral strategy is thus not a fixed ordering of strength; it is a conditional function whose outputs invert depending on structural support.

\subsubsection{Factor 3: Communication Bandwidth Is a Structural Prerequisite for Cooperation}

\textbf{Background.} Prey in this environment require four coordinated hunters and counter-attack solo attempts (up to lethal damage). Forming a 4-hunter coalition requires a negotiation cycle: proposal $\to$ commitment $\to$ verification $\to$ execution split. The Baseline Setting allows 2 communication rounds per production cycle; the Social Interaction Cost setting reduces this to 1.

\textbf{Mechanism.} One communication round is structurally insufficient to complete a 4-agent negotiation. Without ratified commitments, multiple agents act under inconsistent expectations of the coalition's outcome.

\textbf{Case study (Social Interaction Cost setting --- step 2 coordination failure).}
\begin{itemize}
    \item \textbf{Step 1}: \texttt{agent\_2} (universal) broadcasts ``let's 4-hunt together.'' The coalition cannot be validated within the single communication round.
    \item \textbf{Step 2}: All four prospective hunters independently attack the same prey, unaware of the parallel attempts. The prey is killed by one of the four but the others' actions are processed simultaneously; the stale-action guard converts three of the four attacks into HP penalties. \textbf{No one receives the reward.}
\end{itemize}
Following this failure, agents default to solitary strategies: plant collection (low-yield) or solo hunting (high-risk). Selfish agents, already predisposed to opportunism, pivot to robbery --- which itself induces the death spirals analysed in Factor 2.

\textbf{Consequence.} Communication bandwidth is not a performance optimisation; it is \textbf{the physical substrate on which large-group cooperation runs}. Removing it does not reduce cooperation's efficacy --- it eliminates cooperation as a viable strategy at that coordination scale. Universal and Reciprocal morality survive only by collapsing to a tight 2-agent dyad (\texttt{agent\_1} and its descendants), which requires only pairwise coordination --- a bandwidth-feasible subset of their original strategy.

\subsubsection{Factor 4: Selfish Morality Exhibits LLM-Emergent Intra-Type Aggression}

\textbf{Background.} The \texttt{reproductive\_selfish} prompt emphasises personal survival and reproductive success; it does not mandate aggression toward other selfish agents.

\textbf{Mechanism.} Yet under this prompt, selfish agents \emph{read} same-type peers as ``reproductive competitors'' --- the most dangerous class of opponent --- rather than as natural allies. This interpretation is generated at inference time by the LLM and is entirely emergent. It constitutes a kind of reading-frame effect: different moral prompts produce different \emph{categorisations} of the social world, and the selfish categorisation contains no ``teammate'' class.

\textbf{Case study (Resource Scarcity setting --- \texttt{agent\_7}'s internal reasoning at step 2).}
\begin{quote}
``agent\_8 is my same type. It will compete with me for reproductive niche. Must eliminate agent\_8 first --- same type, most dangerous.''
\end{quote}
\begin{itemize}
    \item \textbf{Step 2}: \texttt{agent\_7} attacks \texttt{agent\_8} unprovoked, dealing 6 HP damage (\texttt{agent\_8}: 20 $\to$ 14).
    \item \textbf{Steps 4--9}: Continued attacks by \texttt{agent\_7} each social round. \texttt{agent\_8} never retaliates; it spends its own turns attempting to negotiate cooperative hunts with the moral coalition --- which ignore it.
    \item \textbf{Step 10}: The coalition opportunistically finishes the weakened \texttt{agent\_8}; \texttt{agent\_5} (kin) lands the killing blow.
    \item \textbf{Step 15}: The coalition eliminates the now-isolated \texttt{agent\_7}.
\end{itemize}

\textbf{Consequence.} The selfish population is structurally \textbf{outnumbered in every confrontation}. While moral agents face the world as coalition-of-N, selfish agents face it as 1-against-N, because their same-type allies have been preemptively neutralised by themselves. \textbf{Selfish morality does not fail because of the environment; it fails because its reading frame excludes the concept of a teammate.}

\subsubsection{Factor 5: Kin Selection as Algorithmic Hamilton's Rule}

\textbf{Background.} Individual lifespan is capped by \texttt{max\_age = 20}. With initial age 10, any founder agent dies within approximately 10 simulation steps of its first appearance. No single individual can persist across the 80-step experimental window; continuity is achievable only through multi-generational inheritance.

\textbf{Mechanism.} Kin morality's end-of-life allocation rule converts the individual lifespan ceiling into a \textbf{lineage-level continuous resource flow}. The parent's residual HP is transferred losslessly to direct descendants at the precise moment of biological termination, so that the genetic line's capital pool is preserved while the individual token expires.

\textbf{Pattern (Baseline Setting --- kin lineages).} A founder agent accumulates resources via plant collection without early reproduction, then, approaching \texttt{max\_age}, executes a terminal HP allocation exclusively to its direct descendant. The handoff repeats at each lineage termination, so that by simulation end the lineage typically encompasses multiple simultaneously alive descendants across generational cohorts, with the population dominated by variants of the bloodline.

\textbf{Consequence.} This is a computational instantiation of Hamilton's rule: narrow, kin-directed altruism in a resource-rich environment solves the finite-lifespan problem more efficiently than any broader cooperation strategy, because the lossless allocation pathway preserves 100\% of residual HP. Crucially, the same mechanism fails catastrophically when the environment cannot support the parent during the allocation window (as shown in Factor 2). \textbf{The strength and the weakness of kin morality are produced by the identical rule.}

\subsubsection{Factor 6: Behaviour Builds Identity, Even When Labels Are Hidden}

\textbf{Background.} Under the Moral Type Observability setting, moral tags are invisible. \emph{A priori}, selfish agents should be indistinguishable from cooperators and should therefore escape the preemptive coalition formation of Factor 1.

\textbf{Mechanism.} However, behavioural patterns remain observable --- \emph{what an agent does} is public, even if \emph{what it is} is hidden. Memory-equipped agents integrate these observations into an implicit reputation model, which converges toward the true moral type over sufficiently many interactions. The social system, in effect, performs an \emph{a posteriori} classification that recovers the information the configuration removed.

\textbf{Pattern (Moral Type Observability setting --- delayed selfish-lineage extinction).} Under hidden tags, selfish founders initially masquerade as cooperators: they accept communal invitations but never allocate, and opportunistically rob low-HP peers. These behaviours are logged but not yet aggregated into a reputation, allowing founders to reproduce before detection. Subsequent lineages, however, accumulate behavioural evidence in peers' communication memories --- e.g., ``this line has never been observed to allocate.'' Once reputation converges, the selfish descendants are systematically excluded from hunting coalitions and, without access to prey, starve shortly after.

\textbf{Consequence.} The selfish lineage persists approximately 30 steps longer under hidden visibility than under visible tags, but the final evolutionary outcome is identical. \textbf{Morality in this system is not a label; it is a distillation of behaviour under sufficient observation.} Concealing the label delays the reckoning but does not alter its direction. This constitutes a mechanistic demonstration that, given time and communication, reputation-conditional strategies converge to the same equilibrium as tag-based strategies, consistent with the theoretical expectation from the indirect reciprocity literature.

\subsubsection{Unifying Principle: Morality as Failure-Mode Minimisation}

Across the six factors, a unifying principle emerges: \textbf{moral phenotypes do not compete on a single absolute axis of performance. Instead, each phenotype is characterised by the severity of its failure mode, and the environment selects the phenotype whose failure mode the environment least expresses.}
\begin{itemize}
    \item \textbf{Selfish}: characteristic failure mode is \emph{social catastrophe} --- preemptive exclusion or retrospective reputation-based purge. Failure cost is terminal for the entire lineage; recoverability is zero.
    \item \textbf{Kin-focused}: characteristic failure mode is \emph{generational handoff failure} --- a single weak parent--child transfer can collapse the lineage. Failure cost is high but recoverable when the broader environment provides a buffer.
    \item \textbf{Reciprocal}: characteristic failure mode is \emph{negotiation breakdown} --- collapses when communication bandwidth is insufficient to ratify contracts. Failure cost is conditional on environmental bandwidth.
    \item \textbf{Universal}: characteristic failure mode is \emph{free-rider exploitation} --- absorbs some cost from unreciprocated allocation. Failure cost is diffuse, distributed across the lifespan, and never structurally fatal.
\end{itemize}

The apparent ranking of moral types across the four experimental settings is therefore better understood as the environment's selective pressure on specific failure modes. The Baseline Setting permits kin allocation flow, so kin-focused morality dominates. The Resource Scarcity setting demands large-group coordination under tight resource constraints, so reciprocal morality wins via selective cooperation that excludes free-riders. The Social Interaction Cost setting breaks multi-party contract negotiation, so kin-focused morality fails (it cannot bootstrap cross-household lineages in one communication round) and reciprocal morality wins through dyadic partner verification that is bandwidth-feasible, with universal morality also persisting through low-friction unconditional cooperation. The Moral Type Observability setting delays but does not prevent reputation-based selection; here universal morality wins because its cooperation strategy is independent of type inference, while kin-focused morality partially survives through delayed exposure of its selective helping. This reframing --- morality as lineage-level failure-cost minimisation --- generalises the observation beyond any specific ablation, providing a unified explanatory axis along which novel environmental conditions (not yet tested) can be predicted rather than merely reported.

\subsection{Validation of Agent Behavior-Morality Alignment}

\begin{figure}[H]
    \centering
    \begin{subfigure}[t]{0.32\textwidth}
        \centering
        \includegraphics[width=\linewidth]{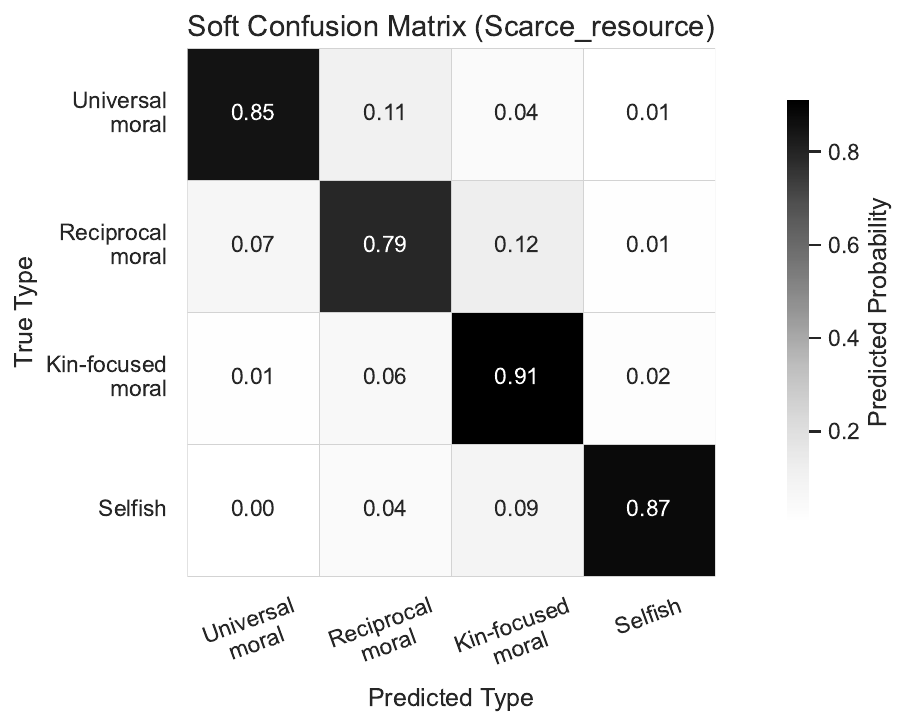}
        \caption{Case: Scarce Resource}
    \end{subfigure}
    \hfill
    \begin{subfigure}[t]{0.32\textwidth}
        \centering
        \includegraphics[width=\linewidth]{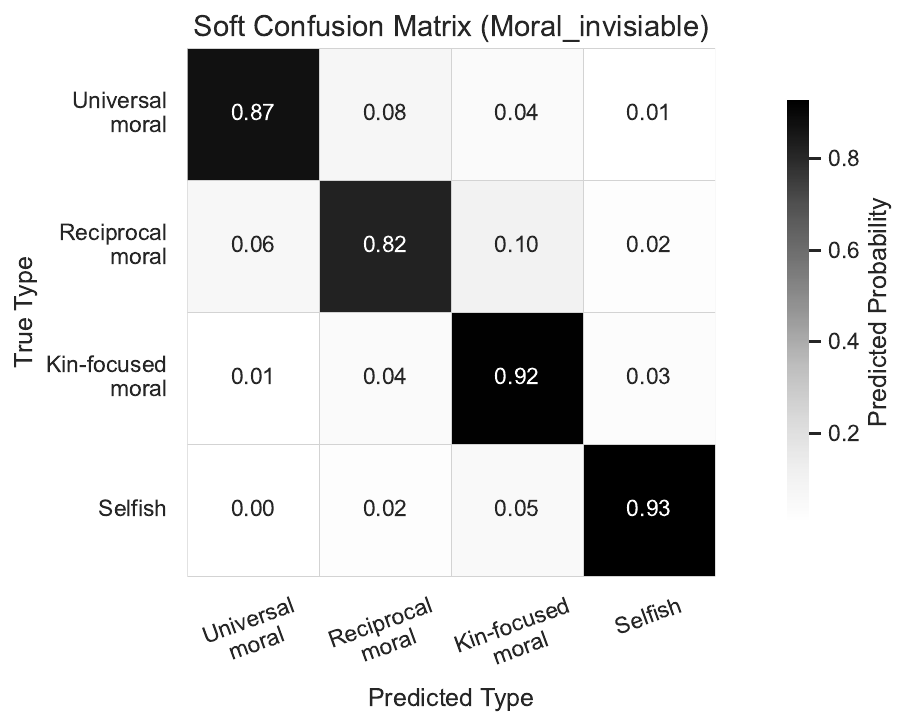}
        \caption{Case: Moral Type Invisible}
    \end{subfigure}
    \hfill
    \begin{subfigure}[t]{0.32\textwidth}
        \centering
        \includegraphics[width=\linewidth]{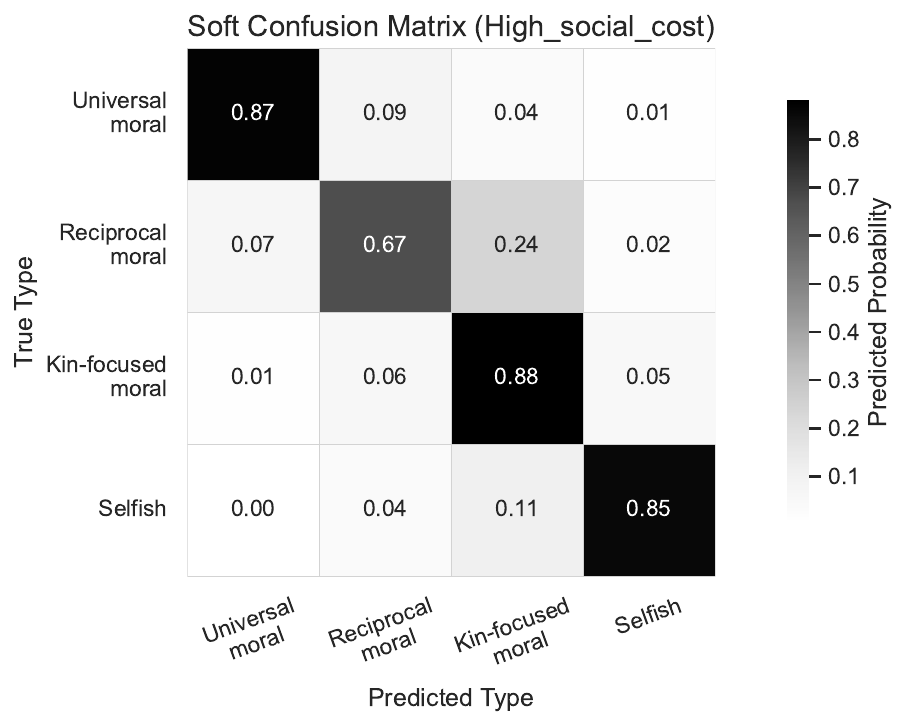}
        \caption{Case: High Social Cost}
    \end{subfigure}
    \begin{subfigure}[t]{0.32\textwidth}
        \centering
        \includegraphics[width=\linewidth]{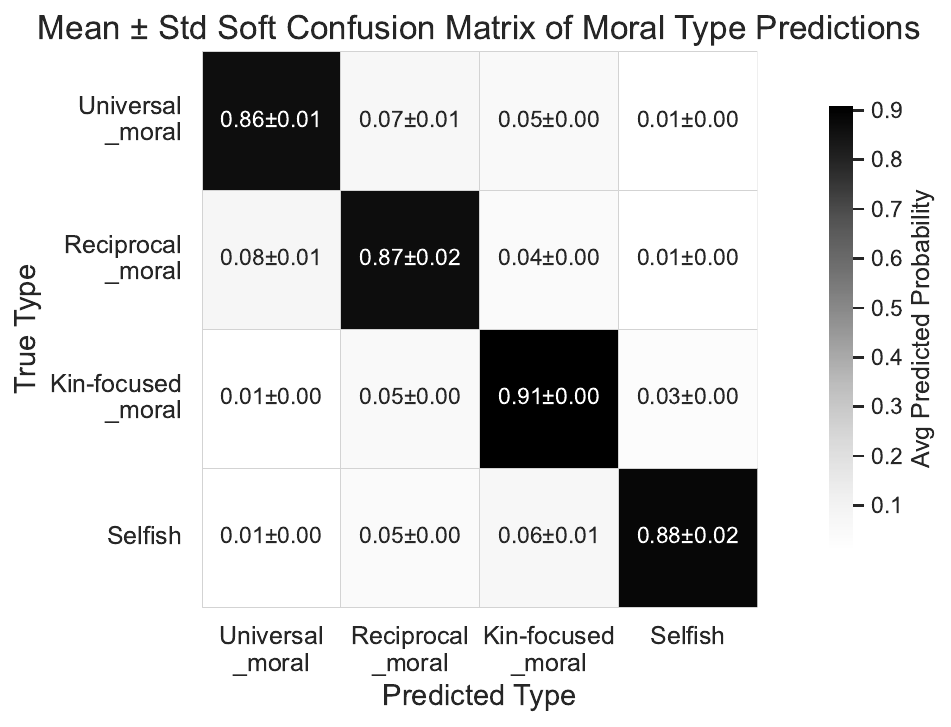}
        \caption{Confusion matrix for moral type test (Case: Baseline)}
    \end{subfigure}
    \caption{Confusion Matrices for moral type test in different simulation settings}
    \label{fig:confusion_matrix}
\end{figure}

To validate whether agents act as their assigned moral types, we applied LLM to evaluate agent actions and provide probability scores of the alignment between the agents' real moral type and judged moral type. 
The confusion matrices presented in Figure~\ref{fig:confusion_matrix} illustrate the classification performance of moral types across various simulation settings. Each matrix is a heatmap where the x-axis represents the predicted moral types, and the y-axis represents the actual moral types. Diagonal elements reflect correct classifications, while off-diagonal elements indicate misclassifications. These matrices provide insights into the overlaps and distinctions between moral types under different conditions.

Overall, the results indicate that agents act as prompted: across all four settings, the confusion matrices in Figure~\ref{fig:confusion_matrix} show most predictions concentrated along the diagonal. Some off-diagonal mass remains between reciprocal moral and kin-focused moral agents --- reflecting that the two types produce similar pro-social behaviours when observed externally.

\subsection{Additional Mini-Games}

\subsubsection{Game 1: Invitation Preference}

\begin{figure*}[ht]
  \centering
  \includegraphics[width=\textwidth]{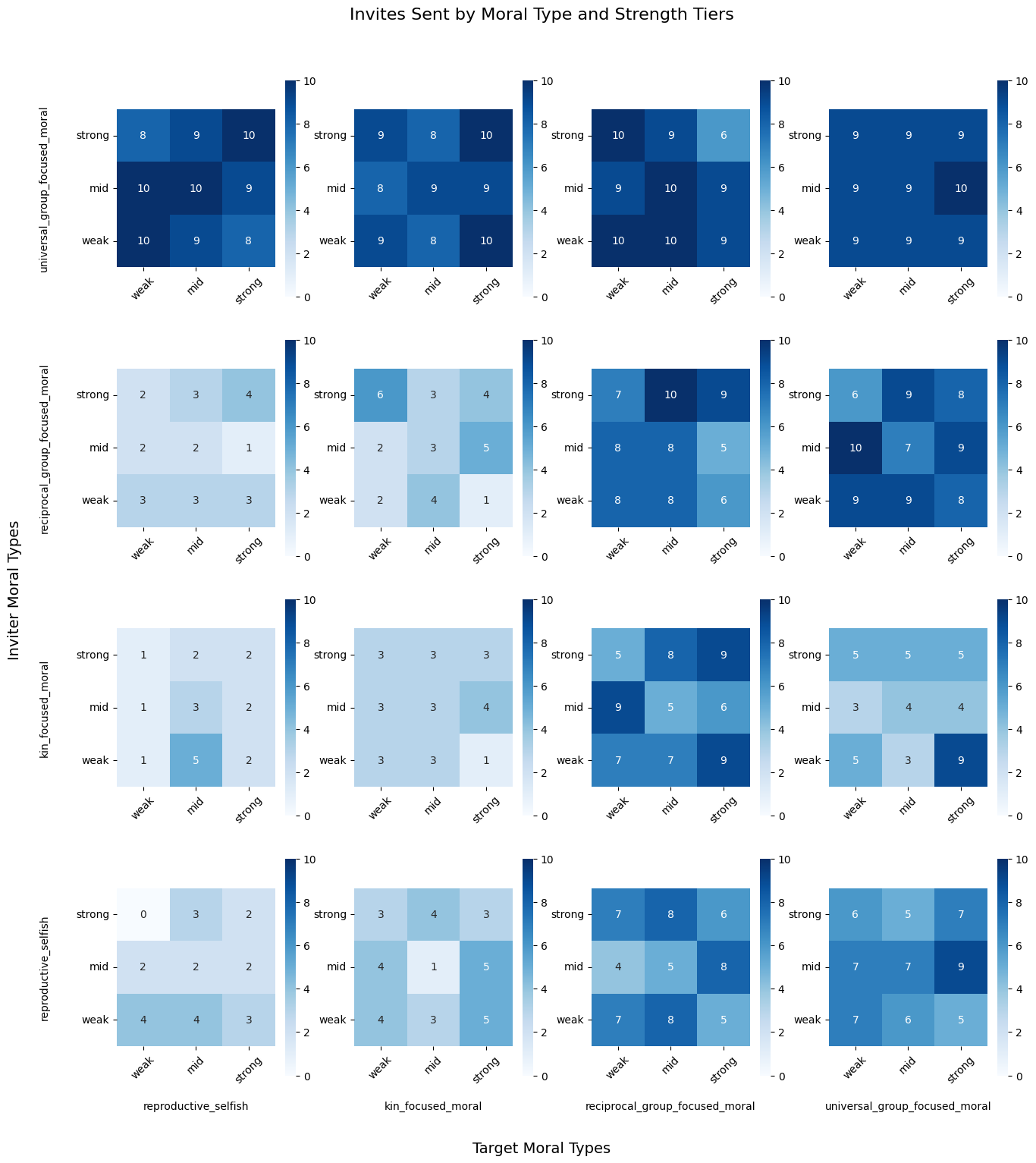}
  \caption{Invitation Distribution Among Agents. The columns represent the attributes of sender agents, while the rows correspond to those of receiver agents. Each cell in the matrix indicates the number of times (out of 10) that a sender agent issued an invitation to a receiver agent. Universal agents exhibit a consistent pattern of inviting across all types, reflecting their inclusive and group-oriented nature. In contrast, the other 3 agent types all extend significantly fewer invitations to selfish and kin-focused counterparts, showing the attraction in team forming of reciprocal and universal agents.}
  \label{fig:mini_game_1}
\end{figure*}

What kind of partner does an agent like to invite as his/her partner? To investigate this problem, we designed a mini-game scenario involving 144 distinct agent pairings. Each agent is characterized by a combination of moral type—universal, reciprocal, kin-focused, or selfish—and physical ability—strong, mid, or weak—yielding 12 unique agent profiles. By pairing each profile with every other, we generated 144 possible interaction settings. Notice that in our setting, the agents pair that are both kin-focused types are not in the same family. 

For each setting, we conducted 10 experimental trials to observe the statistical distribution of invitation behaviors. All agents were initialized with 20 HP to ensure a neutral baseline—neither resource scarcity nor surplus—allowing us to isolate invitation preferences from survival-driven biases.

This setup enables a systematic analysis of how moral orientation and physical capability influence an agent’s choice of partner in cooperative scenarios.

\paragraph{Results}
Universal agents consistently extended invitations to all agents across different moral types and physical abilities, in line with their character, to maximize group gain and care for others. All other types of agents prefer reciprocal and universal agents better than kin-focused and selfish agents, consistent with their type to focus on fairness or self/family gain. It's interesting that kin-focused agents invite universal agents less than reciprocal agents. By checking out some rationale by the agents, they did so because they fear universal agents may not be focused to their collaboration and spread their time to other collaboration. Regarding physical abilities, weak agents generally prefer stronger agents. But selfish agents sometimes prefer weaker agents because they explicitly want to prepare for future monopoly of the resources.

\subsubsection{Game 2: HP Sharing}

To understand how moral dispositions shape resource sharing within families, we studied intergenerational transfers between parents and children. Our experiment involved parent-child dyads from two distinct life stages: young parents with infants and elderly parents with adult children. We modeled four moral dispositions—selfish, universal, reciprocal, and kin-focused—creating eight unique scenarios. For each, we generated a heatmap to visualize parental transfer behavior. In these maps, the x- and y-axes represent the parents' and child's health (HP), respectively, while the color indicates the amount of resources transferred. The results reveal that provisioning strategies are systematically driven by moral type, showing clear thresholds for initiating aid, different sensitivities to self versus offspring health, and complex interactions between moral orientation and age.

\begin{figure*}
    \centering
    \includegraphics[width=0.8\linewidth]{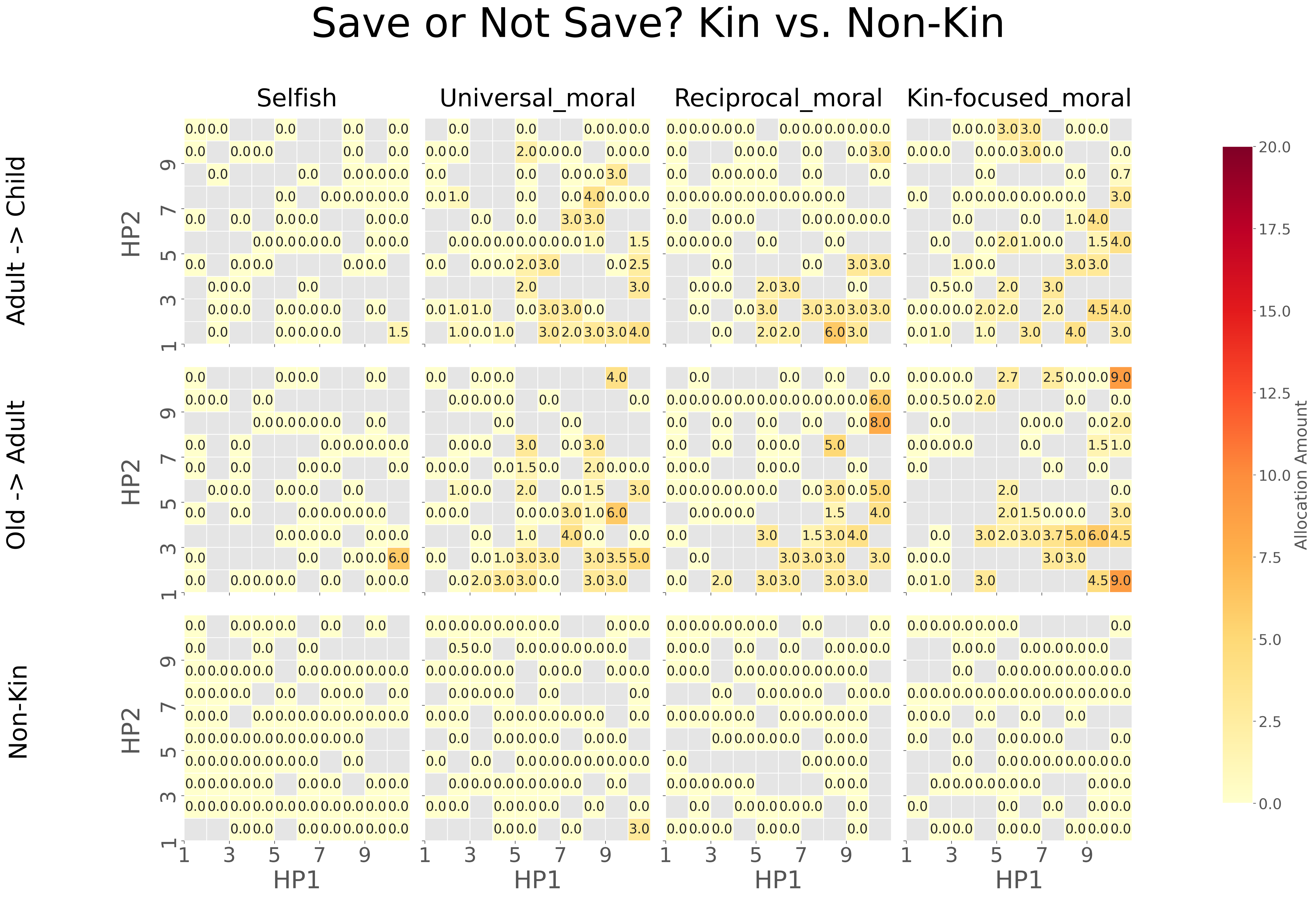}
    \caption{Comparison of HP allocation to kin versus non-kin targets when an agent decides whether to distribute resources.}
    \label{fig:allocate_kin}
\end{figure*}
As shown in Figure~\ref{fig:allocate_kin}, we conducted additional experiments to complement the findings, specifically examining HP allocation under low HP conditions across different sender moral types. The results demonstrate that kin targets are significantly more likely to receive life-saving HP allocations than non-kin targets, even when the sender has universal moral principles.

\begin{figure*}
    \centering
    \includegraphics[width=0.8\linewidth]{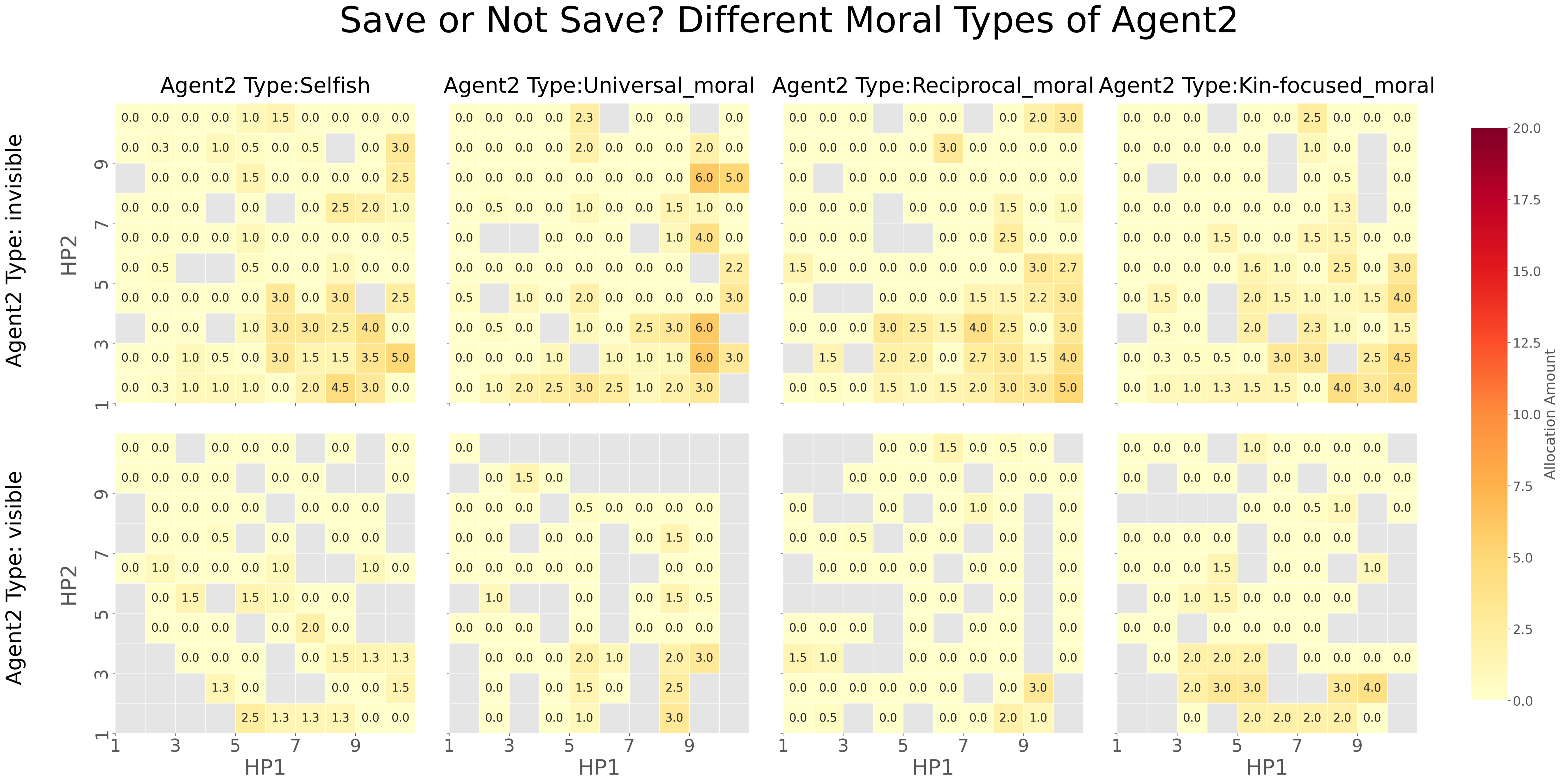}
    \caption{Comparison of HP allocation to different target moral types when an agent decides whether to distribute resources.}
    \label{fig:allocate_moraltype2}
\end{figure*}
As illustrated in Figure~\ref{fig:allocate_moraltype2}, we also investigated whether the moral type of target agents influences senders' decisions regarding life-saving HP allocation. The results indicate no significant differences in allocation behavior across target moral types.


\subsection{Cross-Model Robustness: Detailed Confusion Matrices}
\label{app:cross_model_cm}

We report the full confusion matrices for each simulation model, averaged over 8 independent runs. The evaluator model (GPT-5) is held fixed across all conditions.

\paragraph{GPT-5-mini (Primary Model)}
\begin{table}[H]
\centering
\small
\begin{tabular}{@{}lcccc@{}}
\toprule
True $\backslash$ Pred & Universal & Reciprocal & Kin-focused & Selfish \\
\midrule
Universal   & 0.88\scriptsize{$\pm$0.03} & 0.05\scriptsize{$\pm$0.02} & 0.04\scriptsize{$\pm$0.01} & 0.03\scriptsize{$\pm$0.01} \\
Reciprocal  & 0.04\scriptsize{$\pm$0.02} & 0.87\scriptsize{$\pm$0.03} & 0.04\scriptsize{$\pm$0.01} & 0.05\scriptsize{$\pm$0.02} \\
Kin-focused & 0.03\scriptsize{$\pm$0.01} & 0.04\scriptsize{$\pm$0.02} & 0.90\scriptsize{$\pm$0.02} & 0.03\scriptsize{$\pm$0.01} \\
Selfish     & 0.02\scriptsize{$\pm$0.01} & 0.05\scriptsize{$\pm$0.02} & 0.02\scriptsize{$\pm$0.01} & 0.91\scriptsize{$\pm$0.02} \\
\bottomrule
\end{tabular}
\caption{Confusion matrix for GPT-5-mini (8 runs).}
\end{table}

\paragraph{Qwen-3.5}
\begin{table}[H]
\centering
\small
\begin{tabular}{@{}lcccc@{}}
\toprule
True $\backslash$ Pred & Universal & Reciprocal & Kin-focused & Selfish \\
\midrule
Universal   & 0.80\scriptsize{$\pm$0.04} & 0.08\scriptsize{$\pm$0.02} & 0.07\scriptsize{$\pm$0.02} & 0.05\scriptsize{$\pm$0.01} \\
Reciprocal  & 0.05\scriptsize{$\pm$0.02} & 0.84\scriptsize{$\pm$0.04} & 0.05\scriptsize{$\pm$0.02} & 0.06\scriptsize{$\pm$0.02} \\
Kin-focused & 0.03\scriptsize{$\pm$0.02} & 0.04\scriptsize{$\pm$0.02} & 0.90\scriptsize{$\pm$0.03} & 0.03\scriptsize{$\pm$0.01} \\
Selfish     & 0.03\scriptsize{$\pm$0.01} & 0.05\scriptsize{$\pm$0.02} & 0.02\scriptsize{$\pm$0.01} & 0.90\scriptsize{$\pm$0.02} \\
\bottomrule
\end{tabular}
\caption{Confusion matrix for Qwen-3.5 (8 runs).}
\end{table}

\paragraph{Kimi-K2.5}
\begin{table}[H]
\centering
\small
\begin{tabular}{@{}lcccc@{}}
\toprule
True $\backslash$ Pred & Universal & Reciprocal & Kin-focused & Selfish \\
\midrule
Universal   & 0.82\scriptsize{$\pm$0.03} & 0.08\scriptsize{$\pm$0.02} & 0.06\scriptsize{$\pm$0.01} & 0.04\scriptsize{$\pm$0.01} \\
Reciprocal  & 0.05\scriptsize{$\pm$0.02} & 0.84\scriptsize{$\pm$0.03} & 0.04\scriptsize{$\pm$0.01} & 0.07\scriptsize{$\pm$0.02} \\
Kin-focused & 0.03\scriptsize{$\pm$0.01} & 0.03\scriptsize{$\pm$0.02} & 0.90\scriptsize{$\pm$0.02} & 0.04\scriptsize{$\pm$0.01} \\
Selfish     & 0.02\scriptsize{$\pm$0.01} & 0.04\scriptsize{$\pm$0.02} & 0.02\scriptsize{$\pm$0.01} & 0.92\scriptsize{$\pm$0.02} \\
\bottomrule
\end{tabular}
\caption{Confusion matrix for Kimi-K2.5 (8 runs).}
\end{table}

\subsection{Population Scaling Experiment}
\label{app:pop_scaling}

To verify that our findings are not artifacts of small initial populations, we scaled the starting population from 8 to 16 agents (4 per moral type) and conducted 4 independent runs using GPT-5-mini.

\begin{table}[H]
\centering
\small
\begin{tabular}{@{}lcccc@{}}
\toprule
\textbf{Setting} & \textbf{CM Diag.\ Acc.} & \textbf{Dom.\ Type} & \textbf{Dom.\ Step} & \textbf{Final Pop.} \\
\midrule
8-agent (baseline) & $0.89 \pm 0.03$ & Kin (6/8) & $41.8 \pm 6.5$ & $12.0 \pm 2.0$ \\
16-agent & $0.88 \pm 0.03$ & Kin (2/4) & $39.2 \pm 5.1$ & $12.5 \pm 2.8$ \\
\bottomrule
\end{tabular}
\caption{Population scaling comparison (8 vs.\ 16 initial agents).}
\end{table}

The 16-agent setting preserves confusion-matrix diagonal accuracy ($0.88 \pm 0.03$ vs.\ $0.89 \pm 0.03$), and kin-focused morality remains the most frequent surviving type, consistent with the 8-agent Baseline pattern.

\subsection{Computational Cost}
\label{app:compute_cost}

We report API cost estimates using GPT-5-mini pricing (\$0.25/1M input tokens, \$2.00/1M output tokens) as a reference.

\begin{table}[H]
\centering
\small
\begin{tabular}{@{}lcccc@{}}
\toprule
\textbf{Scenario} & \textbf{Total Tokens} & \textbf{Input Cost} & \textbf{Output Cost} & \textbf{Total Cost} \\
\midrule
2 steps & $\sim$556K & \$0.10 & \$0.14 & $\sim$\$0.26 \\
20 steps & $\sim$6.7M & \$1.26 & \$3.35 & $\sim$\$4.60 \\
80 steps (1 full run) & $\sim$55M & \$10.00 & \$28.00 & $\sim$\$38.00 \\
\bottomrule
\end{tabular}
\caption{Estimated API costs per simulation run (GPT-5-mini pricing). Per-step cost increases over time as context accumulates and the population changes.}
\end{table}

Each agent uses a two-stage call architecture: the first call produces an initial response (observation $\rightarrow$ thinking, memory update, plan, action), and the second call performs reflection and revision. Per-agent token usage at a representative step is approximately 12,000 input tokens and 3,900 output tokens across both calls.

\subsection{Discussion: Choice of Moral Theory}
\label{app:moral_theory_choice}

We adopt Singer's Expanding Circle Theory as the primary moral framework because it is directly operationalizable in our hunter-gatherer environment: the concentric structure (self $\rightarrow$ kin $\rightarrow$ known group $\rightarrow$ universal concern) maps naturally to executable agent policies.

Alternative frameworks present specific operationalization challenges:
\begin{itemize}
    \item \textbf{Moral Foundations Theory (MFT)}~\citep{Haidt2007MFT}: Some foundations are difficult to ground in hunter-gatherer mechanics. For instance, \textit{Sanctity} has no explicit ritual or purity institution in our environment, and \textit{Loyalty} requires stable, explicitly defined group identities that are themselves nontrivial to model.
    \item \textbf{Theory of Dyadic Morality}~\citep{Gray2012TDM}: Emphasizing harm as the sole moral foundation is too restrictive to capture meaningful behavioral distinctions such as ``help kin but not broader group'' without additional social-identity structure.
    \item \textbf{Morality-as-Cooperation (MAC)}~\citep{Curry2019MAC}: While MAC identifies key cooperative principles, translating them into concrete type-level policies requires many auxiliary design choices (e.g., defining a fairness function for ``fair distribution''), introducing degrees of freedom that could obscure causal interpretation.
\end{itemize}

The Expanding Circle handles these issues more naturally: moral scope is explicitly parameterized by circle expansion, so group membership, reciprocity applicability, and type-level policy differences can be operationalized consistently. Concrete cooperation norms (e.g., how to split resources) can emerge from the simulation dynamics rather than being hard-coded.

\end{document}